\documentclass[12pt]{article}
\usepackage{amsmath}
\usepackage{mathtools}
\usepackage{amssymb}
\usepackage[T1]{fontenc}
\usepackage{amsfonts}
\usepackage{amsthm}
\usepackage{mathrsfs}
\usepackage[all]{xy}
\usepackage{tensor}
\usepackage{comment}
\usepackage{stmaryrd}
\usepackage{bm}
\usepackage{physics}
\usepackage{bbm}
\usepackage{subcaption}
\usepackage[normalem]{ulem}
\usepackage[usenames,dvipsnames]{xcolor}
\usepackage[colorlinks=true, citecolor=Blue, linkcolor=Blue]{hyperref}
\usepackage{ragged2e}
\usepackage{enumitem}
\usepackage{slashed}

\usepackage{geometry}
\newsavebox{\vertbox}

\usepackage{setspace}
\usepackage[labelfont={small, bf, sf}, textfont={small,sf}]{caption}
\numberwithin{equation}{section} %section numbers in equations
\usepackage{authblk}
\usepackage[in]{fullpage}
\usepackage{cite}

\newcommand{\lie}{\pounds}

\newcommand{\ccol}[1]{\omit\hfill #1\hfill}

\newcommand{\tone}{\text{I}}
\newcommand{\ttwo}{\text{II}}
\newcommand{\tthr}{\text{III}}
\newcommand{\Aobs}{\mathcal{A}_\text{obs}}
\newcommand{\alg}{\mathcal{A}}
\newcommand{\hb}{\frac{h_\Psi}{\beta}}
\renewcommand{\vev}[1]{\langle #1\rangle}
\newcommand{\gc}{\varkappa}
\newcommand{\hs}{\mathcal{H}}
\newcommand{\hqft}{\hs_\text{QFT}}
\newcommand{\aqft}{\alg_\text{QFT}}
\newcommand{\sr}{\mathcal{S}}
\newcommand{\Hxig}{H_\xi^g}
\newcommand{\aproj}{\widetilde{\alg}}  % projected algebra
\newcommand{\csfull}{\Sigma_c}
\newcommand{\Pio}{\Pi_\text{o}}
\newcommand{\Piadm}{\Pi_\text{ADM}}
\newcommand{\aout}{\alg_\text{out}}
\newcommand{\atr}{\wh{\tr}} % algebra trace
\newcommand{\ac}{\alg^\mathcal{C}}
\newcommand{\cloc}{\mathcal{C}}
\newcommand{\fbdy}{\tilde{\partial}}

\newcommand{\msf}{\mathsf}
\newcommand{\wh}{\widehat}
\renewcommand{\tr}{\operatorname{Tr}}
\newcommand{\ho}[1]{\wh{\msf{#1}}}

\newcommand{\beq}{\begin{equation}}
\newcommand{\eeq}{\end{equation}}

\title{
Generalized entropy for general subregions in \\ quantum gravity 
}

\author[1]{Kristan Jensen\thanks{kristanj@uvic.ca}}

\author[2]{Jonathan Sorce\thanks{jsorce@mit.edu}}

\author[3]{Antony J. Speranza\thanks{asperanz@gmail.com}}

\affil[1]{\small\it Department of Physics and Astronomy, University of Victoria, Victoria, BC V8W 3P6, Canada}

\affil[2]{\small\it Center for Theoretical Physics, Massachusetts Institute of Technology, 182 Memorial Dr, Cambridge, MA 02142, U.S.A.}

\affil[3]{\small \it Department of Physics, University of Illinois, Urbana-Champaign, Urbana IL 61801, USA}

\date{June 21, 2023}

\begin{document}
\maketitle

\begin{abstract}

We consider quantum algebras of observables associated with subregions in theories of Einstein gravity coupled to matter in the $G_N\rightarrow 0$ limit. When the subregion  is spatially compact or encompasses an asymptotic boundary, we argue that the algebra is a type $\ttwo$ von Neumann factor. To do so in the former case we introduce a model of an observer living in the region; in the latter, the ADM Hamiltonian effectively serves as an observer. In both cases the entropy of states on which this algebra acts is UV finite, and we find that it agrees, up to a state-independent constant, with the generalized entropy. For spatially compact regions the algebra is type $\text{II}_1$, implying the existence of an entropy maximizing state, which realizes a version of Jacobson's entanglement equilibrium hypothesis. The construction relies on the existence of well-motivated but conjectural states whose modular flow is geometric at an instant in time. Our results generalize the recent work of Chandrasekaran, Longo, Penington, and Witten on an algebra of operators for the static patch of de Sitter space.
\end{abstract}

\flushbottom

\vspace{3em}
\begin{flushright}\footnotesize{Preprint MIT-CTP/5571}\end{flushright}

\newpage

\tableofcontents

\section{Introduction}

A major lesson of modern physics is that it is fruitful to study entanglement measures in many-body quantum mechanical systems. These measures have many applications across physics, from quantum computing to the emergence of spacetime from conformal field theory.
To discuss such measures in quantum mechanical systems, the usual starting point is a tensor product decomposition of the Hilbert space, from which one can obtain reduced density matrices.
However, if one wishes to study entanglement measures in local quantum field theory by dividing the degrees of freedom according to where they live in space, there is an obstruction.
While the Hilbert space of local quantum field theory has a tensor product decomposition in a lattice regularization, and thereby has reduced density matrices associated with a subregion, those reduced density matrices are ill-defined in the continuum limit.
Said another way, subregions do not carry renormalized density matrices, but they do carry local operators and their algebras.
Despite the absence of reduced density matrices, and correspondingly the absence of well-defined entanglement entropies, these local algebras possess relative entanglement measures such as mutual information and relative entropy~
\cite{Haag1992, Petz1993}.

In this work we are interested not in quantum field theory, but in quantum gravity coupled to matter in the $G_N\to 0$ limit. In this regime there is an effective field theory description at low energies~\cite{Burgess2004, Donoghue:2012zc}, where one can still divide space into subregions and associate local operators and thus their algebra with a region.
Remarkably, coupling to gravity is expected to strengthen the tools of information theory by providing a renormalized notion of entanglement entropy for subregions given via the generalized entropy
\beq \label{eqn:Sgen}
S_\text{gen} = \left\langle \frac{A}{4 \hbar G_N} \right\rangle + S_\text{EE},
\eeq
consisting of a Bekenstein-Hawking-like term involving 
the area
of the  entangling surface
and a term representing the  entanglement entropy
of the state of quantum fields restricted to the subregion.  
Each term in (\ref{eqn:Sgen}) is separately UV divergent---the second due
to infinite vacuum entanglement in quantum field theory, the first 
due to loop effects that renormalize the gravitational
coupling $G_N$---but a number of arguments suggest that these divergences
cancel in their contributions to $S_\text{gen}$ in order to make
it UV-finite and regulator-independent 
\cite{Susskind:1994sm, Jacobson:1994iw, Larsen:1995ax,
Solodukhin:2011gn, Cooperman:2013iqr, Bousso:2015mna}.  
This hints that $S_\text{gen}$ may represent the true entropy
of the fundamental quantum gravitational degrees of freedom, which
organizes into a sum of the two terms in \eqref{eqn:Sgen} when working
within  the low energy effective theory.

This
interpretation of $S_\text{gen}$  underlies many of the connections 
that have been discovered
between quantum information and quantum gravity.  These  have 
their origin in 
black hole thermodynamics \cite{Bekenstein1973a, Bardeen1973},
which first motivated the introduction of the generalized entropy 
in order to 
make sense of the second law of thermodynamics in the presence of 
black holes \cite{Bekenstein1972, Bekenstein1973a}.  
The resulting generalized second law, which states 
that $S_\text{gen}$ increases under evolution to the future along the 
black hole horizon, provides a semiclassical upgrade of the classical
area theorem of general relativity \cite{Hawking1971, Hawking1972}.
This procedure of replacing areas with generalized entropies has 
been applied in several other contexts \cite{Wall:2010jtc, Bousso:2015mna, Bousso:2015eda}, 
leading to various semiclassical
generalizations of classical theorems of general relativity, while at the same time
providing  information-theoretic explanations for why the theorems are true.
Foremost among these is the quantum focusing conjecture
\cite{Bousso:2015mna}, a 
semiclassical generalization of the classical focusing 
theorem that implies a number of other interesting statements about quantum field theory and semiclassical general relativity, such as the quantum null energy condition
\cite{Bousso:2015mna, Bousso:2015wca, Balakrishnan:2017bjg,
Ceyhan:2018zfg}
and the generalized
second law for causal horizons \cite{Jacobson:1999mi, Jacobson:2003wv}.

In holographic contexts, the generalized entropy features prominently
in the Ryu-Taka\-yanagi (RT) formula and its quantum generalizations
\cite{Ryu:2006bv, Ryu:2006ef, Hubeny:2007xt, Faulkner:2013ana, Engelhardt:2014gca}.  
The quantum-corrected formula states that the entanglement
entropy of a subregion in the boundary conformal field theory is 
equal to the generalized entropy of a specific subregion in the dual
bulk spacetime.  The bulk subregion is selected by extremizing
the generalized entropy over all choices of subregions in the bulk
whose asymptotic boundary is the boundary subregion.  The resulting bulk region is called an entanglement wedge, and its spatial boundary is known as
a quantum extremal surface (QES). 
Considerations of entanglement entropies computed via the 
RT formula and quantum extremal surfaces have
led to a wealth of ideas in holography and 
quantum gravity, including bulk reconstruction \cite{Czech:2012bh, Jafferis:2015del,
Dong:2016eik}, connections between
holography and quantum error correction
\cite{Almheiri:2014lwa, Pastawski:2015qua, Harlow:2016vwg, Kang:2019dfi,
Faulkner:2020hzi, Akers:2022qdl,
Faulkner:2022ada}, and  the black
hole information problem and the island formula 
\cite{Almheiri:2019psf, Penington:2019npb, Almheiri:2019hni, Almheiri:2019qdq, Penington:2019kki, Almheiri:2020cfm}.

In fact, it has been 
shown
that the semiclassical bulk dynamics are largely
determined by demanding that the bulk geometry be consistent with 
the RT and QES formulas \cite{Lashkari:2013koa, Faulkner:2013ica,
Swingle:2014uza, Faulkner:2017tkh, Lewkowycz:2018sgn}, 
allowing one to postulate that the bulk
geometry arises entirely from the entanglement structure of the 
dual conformal field theory 
\cite{VanRaamsdonk:2009ar, VanRaamsdonk:2010pw}.  
The arguments leading to the derivation of bulk dynamics from the 
RT formula bear a close resemblance to previous works deriving the 
Einstein equation from horizon thermodynamics
\cite{Jacobson:1995ab}.  This connection
is most explicit in Jacobson's recent entanglement equilibrium 
conjecture, where the Einstein equation is argued to follow
purely from bulk quantum gravity arguments and an assumption that the vacuum state restricted to a subregion has maximal entropy
\cite{Jacobson2015}.

Given the wide range of applications and insights that rely on
a notion of entanglement entropy for local subregions in quantum
gravity, it is  unsettling that such subregions are 
at the same time problematic.  The culprit
is diffeomorphism invariance, which tends to forbid the existence 
of  localized gauge-invariant 
observables in both classical and quantum gravitational
theories \cite{Torre:1993fq, Giddings:2015lla, 
Donnelly:2016rvo, Giddings2022}.  
The fact that diffeomorphisms can change the location
of a subregion requires that the subregion be specified in an
invariant manner; doing so leads to gravitational dressing of 
observables that can interfere with local algebraic properties
such as microcausality \cite{Heemskerk:2012np, Donnelly:2015hta,
Donnelly:2016rvo, Giddings2022}.  More generally,
introducing a boundary gives rise to gravitational edge modes 
from diffeomorphisms acting near the boundary,
leading to the concept of an extended phase space for quasilocal
gravitational charges \cite{Balachandran:1994up, Carlip:1995cd, Donnelly2016, Speranza2017, Freidel:2020xyx, 
Freidel:2020svx, 
Chandrasekaran2020, Donnelly:2020xgu, Ciambelli:2021vnn, Ciambelli2021,
Speranza2022, Donnelly:2022kfs}.

Despite the challenges posed by diffeomorphism invariance, the numerous 
applications of generalized entropy  detailed above 
suggest 
it should be well defined for generic subregions in semiclassical
gravity 
\cite{Bianchi:2012ev}.
Ideally it would 
arise as an entropy of a
quasilocal
operator algebra associated with the subregion,
and this algebra  would  enable rigorous discussions
of entanglement entropy and other quantum information theoretic quantities.
A further desideratum of such an algebraic description 
is that it
would make manifest the finiteness of the generalized entropy, 
demonstrating that the split into an area and entanglement entropy term
as in (\ref{eqn:Sgen}) should simply be viewed as a choice of 
renormalization scheme.  
Doing so would bolster  existing arguments in favor 
of finiteness of generalized entropy by providing 
an independent justification 
that does not rely on Euclidean methods, symmetry, or specific
field content.

The goal of the present paper is to offer a proposal for 
such a quasilocal algebra of observables for subregions in 
semiclassical quantum gravity.  This algebra is constructed 
in the limit of small gravitational coupling $G_N\rightarrow 0$,
in which gravitational backreaction is suppressed.  
In this limit, the description in terms of quantum field theory
in a fixed background is expected to capture the 
leading behavior, which can be further corrected order by order in
the $G_N$ expansion.  Since gravity can be treated as a low-energy
effective field theory in this limit, one expects the language 
of local quantum field theory and 
von Neumann algebras \cite{Haag1992, Borchers:2000pv, Witten:2018zxz} to be applicable in order to provide a
description of the subregion algebras.
In constructing such algebras, we will find that gravitational
constraints arising from diffeomorphism invariance enter the description in a crucial way.
Imposing these constraints results in an algebra in which entanglement entropy can be uniquely defined up to an overall additive ambiguity.
Under regularization, this entropy agrees with generalized entropy up to the additive ambiguity, which can be thought of as a universal entanglement divergence.

Our construction of local gravitational algebras relies heavily
on recent insights that have been made on
strict large-$N$ limits in holography.  
These began with the works of Leutheusser and Liu
\cite{Leutheusser:2021qhd, Leutheusser:2021frk},
which noted that
the large $N$ limit of a holographic CFT above the Hawking-Page phase 
transition produces an emergent type $\tthr_1$ von Neumann algebra, indicating
the 
presence of a black hole horizon in the bulk gravitational theory.\footnote{See also \cite{Papadodimas:2015jra, Banerjee:2016mhh, Jefferson:2018ksk} for related earlier work.}
Type $\tthr_1$ algebras are ubiquitous in quantum field theories 
when restricting to subregions \cite{fredenhagen1985modular, buchholz1995scaling}, and the emergent holographic
algebra is naturally interpreted as the algebra of bulk quantum
fields restricted to the black hole exterior.
Subsequent work  argued that generic causally complete subregions
in the bulk theory should be associated with emergent type $\tthr_1$
algebras in the boundary CFT \cite{Leutheusser:2022bgi, Bahiru:2022oas,
Bahiru:2023zlc}. An important further development
was made by Witten, who demonstrated that the inclusion of $\frac{1}{N}$ 
corrections significantly changes the properties of the emergent
algebras, resulting in algebras of type $\ttwo$ 
\cite{Witten2021}.  
Unlike their type $\tthr$ counterparts, type $\ttwo$ 
von Neumann algebras possess well-defined notions of density matrices and 
traces \cite{murray1936rings, murray1937rings}, and hence allow for 
renormalized entanglement entropies to be defined \cite{segal1960,
Longo:2022lod, sorce2023notes}.  The renormalized entropy was then shown to 
agree, up to a state-independent constant, with the generalized entropy in the cases of the static patch of 
de Sitter space and the AdS black hole \cite{Chandrasekaran2022a,
Chandrasekaran2022b}. 
Further investigations into algebraic constructions in JT gravity also yielded emergent type $\ttwo$ algebras \cite{Penington:2023dql, Kolchmeyer:2023gwa}, suggesting that such algebras appear generically in gravitational theories.

We will argue here that the same mechanism leading to 
type 
$\ttwo$ algebras in 
the dS static patch and the AdS black hole applies
to arbitrary 
subregions in quantum gravity.  Thus, the appropriate algebraic
formulation of gravitational subregions is in terms of type $\ttwo$ 
von Neumann algebras.  This represents a substantial generalization
of the constructions presented in 
\cite{Witten2021, Chandrasekaran2022a, Chandrasekaran2022b}, which 
all involved symmetric configurations in which
the subregion is bounded by a Killing horizon.  
Making the generalization to generic subregions requires two key
modifications of the original arguments.

\begin{enumerate}
\item First, we will show that treating perturbative gravity carefully beyond linear order requires imposing gravitational constraints associated with subregion-preserving diffeomorphisms even when these diffeomorphisms are not isometries.
\item Second, we will argue that there are states on the subregion algebra whose modular flow generates boost-like diffeomorphisms in an infinitesimal neighborhood of a Cauchy slice, even when there is no global boost symmetry.
\end{enumerate}

The details of our construction of a type $\ttwo$ algebra for subregions 
in quantum gravity will closely follow the  construction
for the de Sitter static patch given by 
Chandrasekaran, Longo, Penington, and Witten (CLPW) in 
\cite{Chandrasekaran2022a}.  
Most notably, this procedure involves the introduction
of an observer degree of freedom within the subregion to serve as an
anchor for gravitationally dressing operators in the subregion algebra.
Rather than arguing for the existence of such an observer degree of 
freedom from first principles, we will show that introducing the 
observer has the desired effect of producing a local 
gravitational algebra in which the renormalized entropy agrees 
with the subregion generalized entropy.  Additional arguments in
favor of the existence of the observer come from considerations 
of the algebra for a region which extends out to infinity,
discussed in section \ref{sec:asymp}.
In this case, the asymptotic boundary can be used as the observer, but
since the resulting type $\ttwo$ algebra must have a nontrivial
commutant, we end up concluding that 
the local algebra associated with the causal complement
must be associated with a type $\ttwo$ algebra constructed with
an observer degree of freedom.  Further speculations on the 
nature of the observer are given in the discussion,
section \ref{sec:observer}.

The final step in the construction of the algebra concerns energy
conditions imposed on the observer.  Just as in the 
CLPW construction, the observer can be restricted to have positive (or bounded
below) energy, which is implemented via a projection in the crossed 
product algebra.  For local gravitational subregions, this projection
results in an algebra of type $\ttwo_1$, which, in particular, possesses
a maximal entropy state. 
Intriguingly,
the existence of a maximal entropy state for the gravitational 
subregion immediately implies a version of Jacobson's entanglement
equilibrium hypothesis \cite{Jacobson2015}.  When applied to the asymptotic
boundary, the positive energy projection coincides with the positivity
of the ADM energy, but due to certain sign differences, produces
a type $\ttwo_\infty$ algebra, similar to the case of the AdS
black hole \cite{Witten2021, Chandrasekaran2022b}.\footnote{Here 
we explicitly exclude subregions that divide
an asymptotic such as entanglement wedges of boundary subregions 
in AdS;
such regions are associated with type $\tthr_1$ algebras
in the dual CFT for any value of $N$.  We speculate 
how these algebras should be handled in more detail
in section \ref{sec:holoapp}. \label{ftn:rindler}}
This suggests that bounded subregions
in quantum gravity are associated with type $\ttwo_1$ algebras, while
subregions that
encompass an asymptotic boundary are type $\ttwo_\infty$.
More succinctly, type $\ttwo$ algebras arise
for gravitational subregions  with compact 
entangling surfaces.

The basic argument leading to the type $\ttwo$ gravitational
algebras is straightforward to state, 
and so we begin in
section \ref{sec:outline} with an overview of the argument.
This section serves to clarify the logic of the paper
and to emphasize the major results.  The assumptions
entering into the argument are then listed in 
section \ref{sec:assumptions} to provide
an easy reference for later discussions in the paper.
A reader interested in understanding the main
claims of this work is encouraged to read section
\ref{sec:outline} and then also section \ref{sec:discussion}
which discusses numerous possible applications of the 
present work.  The remaining sections  
provide further justifications and explanations
of the assumptions listed in section \ref{sec:assumptions}
and describe in greater detail the properties 
of the type $\ttwo$ gravitational algebras.  
Section \ref{sec:constraints} is devoted to describing
the constraints appearing in gravity and their relation
to diffeomorphism invariance.  Following this,  section
\ref{sec:modham}  describes the relation between
the boost diffeomorphism and modular flow, and gives
evidence for the geometric modular flow conjecture.  
Section \ref{sec:crossprod} gives details related to the
type $\ttwo$ gravitational algebras, leading to
a demonstration that the algebraic entropy agrees
with the subregion generalized entropy up to a state-independent constant.
Several appendices are included that provide
further details on gravitational constraints,
von Neumann algebras, modular theory, and practical
calculations within the crossed product algebra.

\textit{Note about related work:} Shortly after this paper was first posted on the arXiv, two other papers appeared \cite{AliAhmad:2023etg, Leigh2023} which have conceptual overlap with this one.
We are also aware of forthcoming work by Kudler-Flam, Leutheusser, and Satishchandran \cite{KFLSforthcoming} which realizes the crossed product explicitly for free fields on certain backgrounds, and of forthcoming work by Freidel and Gesteau \cite{Freidel2023} which discusses connections between crossed products and gravitational edge modes.

\section{Outline of the construction}
\label{sec:outline}

We begin with an overview of the general arguments leading
to type $\ttwo$ algebras for gravitational
subregions and an associated calculation of generalized entropy, in order to clarify the major assumptions 
needed to reach the conclusion.   
The arguments will be based on purely bulk quantum gravitational considerations,
in a low energy and weak gravitational coupling limit, $\gc\rightarrow 0$, with 
$\gc = \sqrt{32\pi G_N}$.

\subsubsection*{Free graviton theory}

The first step is to consider the theory of Einstein gravity minimally coupled to matter in the strict $\gc\rightarrow 0$ limit. 
This limit suppresses gravitational backreaction, and hence is described
in terms of quantum fields propagating on a background
with a fixed metric $g_{ab}^0$ which we will take to be globally hyperbolic. Treating gravity as an effective field theory, the gravitons
can be quantized in a similar manner to the ordinary matter fields.  This is
done by expanding the metric around the background 
according to 
\beq \label{eqn:gg0+kh}
g_{ab} = g_{ab}^0 + \gc h_{ab},
\eeq
and quantizing the metric perturbation $h_{ab}$ as 
a free, massless, spin-2 field.  The coefficient of $h_{ab}$ is chosen to 
give it a canonical normalization in the quadratic action,\footnote{
I.e., so that the prefactor of the graviton
kinetic term is $\frac12$.}
and this also suppresses graviton interactions in the  $\gc\rightarrow 0$
limit. 

\begin{figure}[t]
    \centering
    \savebox{\vertbox}{\includegraphics[width=0.4\textwidth]{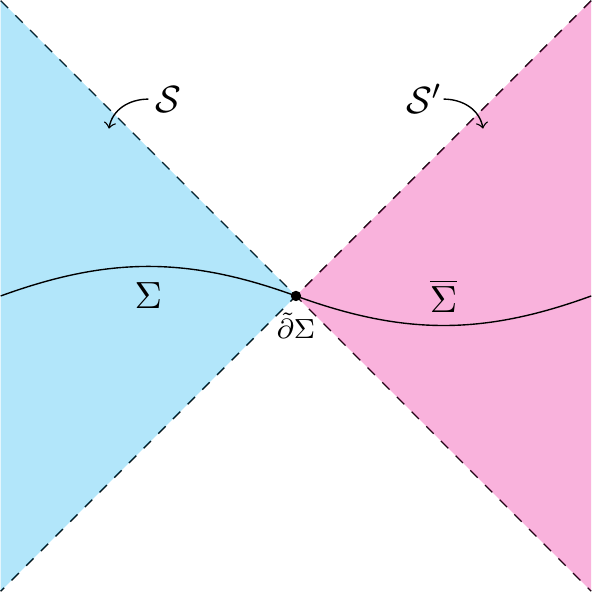}}
    \begin{subfigure}[b]{0.4\textwidth}
        \raisebox{\dimexpr.5\ht\vertbox-.5\height}{\includegraphics[width=\textwidth]{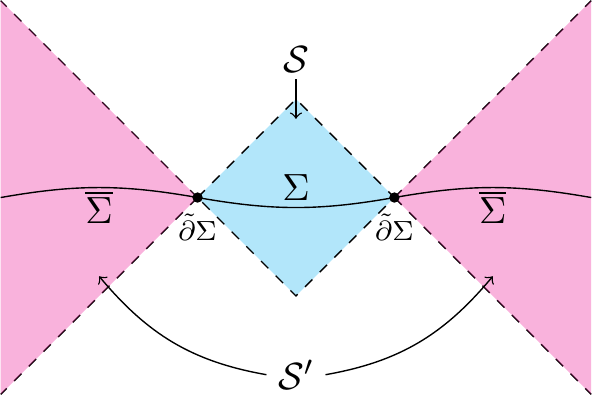}}
        \caption{}
        \label{fig:Sbd}
    \end{subfigure}
    \hspace{0.1\textwidth}
    \begin{subfigure}[b]{0.4\textwidth}
        \usebox{\vertbox}
        \caption{}
        \label{fig:Sunbd}
    \end{subfigure}
    \caption{Two examples of a partial Cauchy slice $\Sigma,$ its causal development $\sr$, the complementary region $\sr',$ and the entangling surface $\partial \Sigma.$ In example (\subref{fig:Sbd}), $\sr$ is bounded while $\sr'$ 
    is unbounded, while in example (\subref{fig:Sunbd}),
    both $\sr$ and $\sr'$ are unbounded.}
    \label{fig:S-and-Sprime}
\end{figure}

Diffeomorphisms that preserve the decomposition \eqref{eqn:gg0+kh} are generated by vector fields $\gc \xi^a$ proportional to $\gc,$ which act trivially on matter fields in the $\gc \to 0$ limit while generating an abelian algebra of linearized gauge transformations for the graviton, $\delta_{\gc \xi} h_{ab} = \lie_{\xi} g_{ab}^0.$
Because the action of diffeomorphisms is 
suppressed in $\gc$, there is no issue in defining arbitrary subregions in the background
geometry and analyzing the algebra of matter fields and free gravitons restricted
to these subregions.  
We will fix a subregion $\sr$ 
by choosing it to coincide with the domain of 
dependence $D(\Sigma)$ of a  partial Cauchy slice
$\Sigma$ with boundary $\fbdy\Sigma$,
where the $\fbdy$ notation refers to the finite-distance boundary
of $\Sigma$, and excludes any asymptotic boundaries (see figure \ref{fig:S-and-Sprime}).
We will use the symbol $\bar{\Sigma}$ to denote a complementary partial Cauchy slice, also with boundary $\fbdy \bar{\Sigma} = \fbdy \Sigma,$
so that $\csfull = \Sigma \cup \bar{\Sigma}$ is a Cauchy slice for the spacetime.
According to general arguments from algebraic quantum field theory \cite{fredenhagen1985modular, buchholz1995scaling}, the algebra $\aqft$ associated with $\sr$ is a von Neumann factor of type $\tthr_1$ for any quantum field theory with a UV fixed point.\footnote{Practically, the type $\tthr_1$ characterization means that the algebra contains no renormalizable density matrices, and each of its modular operators has spectrum equal to the full positive reals $[0, \infty)$. For a recent review of the formal definition of a type III$_1$ von Neumann factor, see \cite{sorce2023notes}.}
This algebra is realized as a collection of bounded operators acting on a Hilbert
space $\hs_\text{QFT}$.
By assuming Haag duality \cite{Haag1962, Haag1992}, 
the algebra of quantum fields for the complementary 
domain of dependence $\sr' = D(\bar\Sigma)$ 
can be taken to coincide with the 
commutant algebra $\aqft'$ consisting of all bounded operators acting on
$\hs_\text{QFT}$ that commute with $\aqft$. This commutant algebra
is also type $\tthr_1$.

The subregion $\sr$ can either be 
bounded, by which we mean that it 
has a bounded Cauchy surface $\Sigma$, as in 
figure
 \ref{fig:S-and-Sprime}(\subref{fig:Sbd}), 
or unbounded, meaning it  contains a complete
asymptotic boundary, as in figure \ref{fig:S-and-Sprime}(\subref{fig:Sunbd}).  
The constructions we are about to describe for
of the gravitational 
algebras in each case are similar, with the main
qualitative difference being that  bounded regions
will result in  type $\ttwo_1$ algebras while  unbounded regions
will result in  type $\ttwo_\infty$ algebras. Since the algebra of the 
causal complement $\sr'$ naturally arises as the 
commutant of the subregion algebra, both cases can be handled
at once if $\sr$ is chosen to be a bounded subregion in an
open universe, so that $\sr'$ is unbounded.  We therefore
restrict attention to this case for the remainder of this section.
Since $\Sigma$ then has no asymptotic boundaries, we will simply
write $\partial\Sigma$ for $\fbdy\Sigma$.

\subsubsection*{Gravitational interactions and constraints}

The next step is to consider corrections coming from the 
$\gc$ expansion.  A significant change  is that 
at first interacting order in $\gc$, all matter fields $\phi$
transform under rescaled diffeomorphisms, $\delta_\xi \phi = \gc
\lie_\xi\phi$.  The transformation of the graviton is similar,
$\delta_\xi h_{ab} = \gc \lie_\xi h_{ab}+ \lie_\xi g^0_{ab}$,
with the first term representing the diffeomorphism transformation
of the spin-2 field $h_{ab}$, and the second term still interpreted
as a linearized gauge transformation.  Because of these 
nontrivial transformations, care has to be taken in order to 
ensure that the algebra we construct is diffeomorphism-invariant. 
It is useful to break this problem into two parts: first, ensuring
that the algebra is invariant under diffeomorphisms that are 
supported locally
within the subregion $\sr$, and then  ensuring invariance under the wider
class of diffeomorphisms that act simultaneously on $\sr$ and
$\sr'$.

Diffeomorphism invariance within $\sr$ 
can be addressed either by gravitationally
dressing operators within the subregion, or by 
partially fixing the gauge to set up a well-defined coordinate
system within $\sr$.  
The local gravitational dressing can be constructed perturbatively
in the $\gc$ expansion \cite{Donnelly:2015hta, Giddings2022}, and it is generally expected that 
the  algebra $\aqft$ remains type $\tthr_1$ upon
including these perturbative corrections 
\cite{Leutheusser:2021qhd, Leutheusser:2021frk,
Witten2021}.  Operators in $\aqft'$ must similarly
be gravitationally dressed, and it is important to choose
this dressing to ensure that $\aqft$ and $\aqft'$ remain
commutants of each other.  A straightforward way to 
enforce this requirement is to dress both sets of operators
to the entangling surface $\partial\Sigma$ which is held fixed
(see, e.g.\ \cite{Almheiri2017});
doing so should prevent  the gravitational dressings  for the different
subregions from overlapping, 
thus preserving microcausality.

\begin{figure}[t]
    \centering
    \includegraphics[scale=1.5]{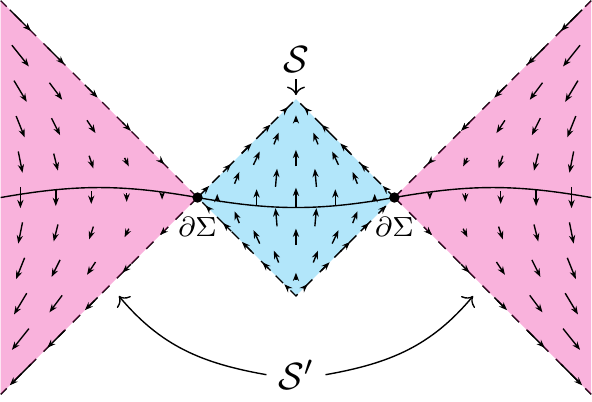}
    \caption{A vector field $\xi^a$ that is future-directed in $\sr,$ past-directed in $\sr',$ tangent to the null boundaries, and approximates a boost near the entangling surface $\partial \Sigma.$}
    \label{fig:vector-field}
\end{figure}

Because such dressings are necessarily quasilocal, there remain
additional conditions from requiring invariance under
diffeomorphisms that act in both $\sr$ and $\sr'$.  Of particular
importance are diffeomorphisms that generate boosts
around the entangling surface, as in figure \ref{fig:vector-field}.
These diffeomorphisms are generated by vector fields $\xi^a$ that are future-directed in $\sr$, past-directed in $\sr'$, and tangent to the null boundaries of the subregions so that they map $\sr$ and $\sr'$ into themselves.  Furthermore, $\xi^a$ must vanish at the entangling surface $\partial\Sigma$ and have constant surface gravity $\kappa$ on $\partial\Sigma$, defined by the relation
\beq \label{eqn:surfgrav}
\nabla_a \xi_b \overset{\partial\Sigma}{=} \kappa n_{ab},
\eeq
where $n_{ab}$ is the unit binormal to $\partial\Sigma$, 
i.e., the unique antisymmetric tensor that is normal to $\partial \Sigma$, co-oriented with the normal bundle of $\partial \Sigma$, and satisfies $n^{ab} n_{ab} = -2$.

We will choose one such diffeomorphism and study the effects of imposing it as a constraint on the algebras $\aqft$ and $\aqft'.$
This can be done by writing an expression for the associated constraint functional in the full nonlinear theory of gravity, and imposing that constraint on $\aqft$ and $\aqft'$ order by order in $\gc$.
However, just as in the CLPW construction \cite{Chandrasekaran2022a}, it is problematic
to impose this constraint
solely on the quantum field degrees of freedom comprising $\aqft$ and $\aqft'$.
Instead, we introduce an observer degree of freedom into the subregion
$\sr$ by tensoring in an additional 
Hilbert space $\hs_\text{obs} = L^2(\mathbb{R})$.  This 
observer is used both to define location
of the subregion beyond leading order in $\gc$ and
to serve as a
clock providing a physical notion of time
evolution for quantum fields within $\sr$.
It is not necessary to view the observer as a literal particle
following a worldline within the subregion, and,
as discussed in section \ref{sec:asymp}, the construction of the 
gravitational subregion algebra is largely agnostic
to the details of the observer model.
The main requirement is that the observer couple universally
to gravity via its energy-momentum, which implies that the 
observer Hamiltonian must appear in the gravitational constraints.
Following CLPW \cite{Chandrasekaran2022a}, 
we take the observer's Hamiltonian to be the position operator $H_\text{obs} = \hat{q}$, in which case the conjugate momentum $\hat{p} = -i\frac{d}{dq}$ has the interpretation of the time measured by the observer.
The full observer algebra is taken to be the set of all 
bounded operators acting on $\hs_\text{obs}$, $\Aobs = \mathcal{B}(\hs_\text{obs})$.  The complementary
region $\sr'$ must also have an observer degree of 
freedom, but because $\sr'$ contains an asymptotic
boundary, the role of the observer is played 
by the ADM Hamiltonian $H_\text{ADM}$.
It acts on a separate Hilbert space $\hs_\text{ADM}$,
and the full asymptotic observer algebra includes all
bounded operators on this Hilbert space, $\alg_\text{ADM}
 = \mathcal{B}(\hs_{\text{ADM}})$.
 
 Together, the full kinematical algebra is 
$(\aqft\vee\aqft')\otimes \alg_\text{obs} \otimes
\alg_\text{ADM}$, which acts on the Hilbert 
space $\hs_\text{kin} = \hs_\text{QFT}\otimes\hs_\text{obs}
\otimes \hs_{\text{ADM}}$.  
The tensor product structure reflects the fact that 
$\alg_\text{obs}$ and $\alg_{\text{ADM}}$ commute with the 
quantum field degrees of freedom before 
imposing the gravitational constraint.  As explained 
in section \ref{sec:constraints}, the constraint
is given by
\beq \label{eqn:Cxi-outline}
\mathcal{C}[\xi] = \Hxig + H_\text{obs} + H_\text{ADM},
\eeq
where $\Hxig$ is the operator generating the flow of $\xi^a$
on the quantum field algebras $\aqft$ and $\aqft'$ instantaneously
on $\csfull$. It takes the form of a local integral
of the matter and graviton stress tensors, as explained 
in section \ref{subsec:perturbative-constraints}.

\subsubsection*{Crossed product algebra}

To implement the constraint at the level of the 
subregion algebra, we need to
determine the  operators in 
$\aqft\otimes\alg_\text{obs}$ that commute with 
$\mathcal{C}[\xi]$.  Because both algebras already
commute with $H_\text{ADM}$, the desired subalgebra
consists of all operators commuting with the 
flow of $\mathcal{C} = \Hxig + H_\text{obs} = \Hxig + \hat{q}$.
This can alternatively be characterized as the set 
of operators on $\hs_\text{QFT}\otimes \hs_\text{obs}$
commuting with $\mathcal{C}$ as well as $\aqft'$.
As explained in appendix \ref{app:algebraic-background}, the resulting von Neumann algebra is the \textit{crossed product} of $\alg_{\text{QFT}}$ by the flow generated by $H_{\xi}^g.$
It is generated by elements of the form $e^{i\Hxig\hat{p}} \msf{a}
e^{-i\Hxig \hat{p}}$ with $\msf{a}\in\aqft$,
along with $e^{i\hat{q}t}$ for $t\in\mathbb{R}$;
in other words, the gauge-invariant algebra for the 
subregion $\sr$ is given by
\begin{equation} \label{eq:crossed-product-main-text}
   \ac
        = \{ e^{i H_{\xi}^g \hat{p}} \msf{a} e^{- i H_{\xi}^g \hat{p}}, e^{i \hat{q} t}\, | \, \msf{a} \in \aqft, t \in \mathbb{R}\}'',
\end{equation}
where $S''$ denotes the smallest von Neumann algebra containing the set $S$.
We can think of the operators $e^{i \hat{q} t}$ as generating the algebra of operators that are diagonal in the observer energy basis, and the operators $e^{i H_{\xi}^g \hat{p}} \msf{a} e^{- i H_{\xi}^g \hat{p}}$ as being dressed versions of operators in $\aqft$ where the observer clock has been synchronized with the time experienced by field-theoretic degrees of freedom.

Properly implementing the constraints at the level 
of the Hilbert space effectively eliminates the factor
of $\hs_\text{ADM}$ from $\hs_\text{kin}$ (see section
\ref{sec:crossconstr} 
and \cite{Chandrasekaran2022a}), leading to the 
representation of $\ac$ acting on $\hs_\text{QFT}\otimes
\hs_\text{obs}$ described above.  In this description,
the ADM Hamiltonian is represented by $H_\text{ADM} = -\mathcal{C}
=-\Hxig-\hat{q}$.  By construction, this operator, 
along with $\aqft'$,
generates the commutant algebra,
\beq
(\ac)' = \{\msf{b}', e^{i(\Hxig +\hat{q})s}|\msf{b}'\in
\aqft', s\in\mathbb{R}\}'',
\eeq
and this is naturally identified as the algebra 
associated with the complementary subregion $\sr'$.  
This algebra is an equivalent representation
of the crossed product of $\aqft'$ by the flow generated
by $\Hxig$.  

\subsubsection*{Geometric modular flow}

Having obtained the subregion algebra $\ac$ as a crossed product
with respect to the flow generated by $\Hxig$, the next task is to 
determine the type of the resulting von Neumann algebra.  
Our claim is that this algebra is type $\ttwo_\infty$, and 
coincides with the crossed product of $\aqft$ with respect to a modular
automorphism group, in direct analogy with previous
examples for subregions with boost symmetry \cite{Witten2021, Chandrasekaran2022a,Chandrasekaran2022b}.

This claim relies on a conjecture that $\Hxig$ is 
in fact proportional to the modular Hamiltonian for some
state on the algebra $\aqft$.
The intuitive argument for this conjecture is that any flow that agrees with 
the vacuum modular flow in the UV (i.e., on degrees of 
freedom localized close to the entangling surface) 
should define a valid modular
flow for some state on the algebra.   Since any 
entangling surface looks locally like Rindler space at short enough
distances, we need only require that the flow generated by $\Hxig$ agree
near the entangling surface 
with the vacuum modular flow for this local Rindler space.  
As is well known from the work of  Bisognano and Wichmann
\cite{Bisognano1975}, 
the vacuum modular flow of Rindler space is simply
the geometric flow of a boost that fixes the entangling surface.
Hence, by choosing $\xi^a$ to  look like 
a boost with constant surface gravity
near $\partial\Sigma$, we conjecture 
this ensures that $\Hxig$ generates a modular
flow of some state $|\Psi\rangle\in\hs_{\text{QFT}}$. 
The surface gravity determines the constant 
of proportionality between $\Hxig$ and $h_\Psi$, 
\beq
h_\Psi = \beta \Hxig = \frac{2\pi}{\kappa} \Hxig,
\eeq
as follows from the Unruh effect \cite{Unruh:1976db} associated with the local 
Rindler space near the entangling surface.  
Equivalently, 
this relation implies that 
$|\Psi\rangle$ satisfies
the KMS condition at inverse temperature 
$\beta = \frac{2\pi}{\kappa}$ for the flow
generated by $\Hxig$.
Additional arguments in 
favor of this geometric
modular flow conjecture are presented in section \ref{sec:modham}.

Note that when $\xi^a$ does not generate a symmetry of the 
background metric, the Hamiltonian generating the flow 
of $\xi^a$ will be time-dependent.  This means that the
time-independent
operator $\Hxig$ only generates this flow instantaneously 
on the initial Cauchy surface $\csfull$, and hence the 
modular flow looks local only in the vicinity of $\csfull$.
Fortunately, this is all that is needed to identify the 
modular crossed product algebra with the gauge-invariant
gravitational algebra.  
Due to time dependence, the Hamiltonian constructed 
on a different Cauchy slice will differ from $\Hxig$,
and therefore define a different KMS state.  This will
result in an isomorphic crossed-product algebra whose
states and operators are simply related to those 
of $\ac$.

\subsubsection*{Modular operators and density matrices}

In addition to supporting the conclusion that $\alg^{\mathcal{C}}$ 
is a type $\ttwo_\infty$ von Neumann algebra, the assumption that $\beta\Hxig$ 
is a modular Hamiltonian of some state on $\aqft$ 
allows one to leverage the full
machinery of modular theory (reviewed in 
appendix \ref{app:modth}) in order to compute 
density matrices and entropies 
in $\ac$.  
The existence of well-defined density matrices, 
along with the related existence of a renormalized trace,
are key features of type $\ttwo$ von Neumann factors, and 
both are uniquely determined up to a state-independent 
multiplicative constant.  
These properties follow from the fact that 
the modular operator $\Delta_{\wh{\Phi}}$ associated to $\ac$ for any state $|\wh{\Phi}\rangle \in \hs_{\text{QFT}} \otimes \hs_{\text{obs}}$ factorizes into separate operators respectively
affiliated with $\ac$ and $(\ac)'$ according to
\beq
\Delta_{\wh{\Phi}} = \rho_{\wh{\Phi}}(\rho_{\wh{\Phi}}')^{-1}.
\eeq
The factors in this relation determine the density matrix 
$\rho_{\wh{\Phi}}$ for $\ac$ and the density matrix $\rho_{\wh{\Phi}}'$
for $(\ac)'$.

As a concrete demonstration of this factorization, we consider 
a class of states of the form $|\wh{\Phi}\rangle = |\Phi\rangle
\otimes|f\rangle$ where $|\Phi\rangle \in\hs_\text{QFT}$
and $|f\rangle  = f(q)$ is a wavefunction in $\hs_\text{obs}$.
The factors of the modular operator can be determined 
exactly (see section \ref{sec:moddensity} and appendix 
\ref{app:modcomp}), resulting
in the density matrices
\begin{align}
\rho_{\wh{\Phi}} 
&= 
\frac1\beta e^{i\hat p\hb}f\Big(\hat{q}-\hb\Big)e^{\frac{\beta\hat q}{2}}
\Delta_{\Phi|\Psi} e^{\frac{\beta\hat q}{2}}
f^*\Big(\hat{q}-\hb\Big) e^{-i\hat{p}\hb} 
\label{eqn:rhoPhioutline}
\\
\rho_{\wh{\Phi}}' 
&=
\frac1\beta 
\Delta_{\Psi|\Phi}^{-\frac12}
J_{\Phi|\Psi} J_\Psi
e^{\frac{\beta\hat q}{2}}
\,\Big|f\Big(\hat{q}+\hb\Big)\Big|^2 e^{\frac{\beta\hat{q}}{2}}
J_\Psi J_{\Psi|\Phi}
\Delta_{\Psi|\Phi}^{-\frac12},
\label{eqn:rhoPhi'outline}
\end{align}
where the relative modular operators $\Delta_{\Phi|\Psi}$,
$\Delta_{\Psi|\Phi}$ and modular conjugations $J_\Psi$, $J_{\Psi|\Phi}$,
$J_{\Phi|\Psi}$ are defined in appendix \ref{app:modth}.

\subsubsection*{Generalized entropy}

With the expression for $\rho_{\wh{\Phi}}$ in hand, the entropy can be computed 
as the expectation value of $-\log\rho_{\wh{\Phi}}$,
\beq
S(\rho_{\wh{\Phi}}) = \langle\wh{\Phi}|-\log\rho_{\wh{\Phi}}
|\wh{\Phi}\rangle.
\eeq
In order to simplify the computation of the logarithm, we impose the same 
semiclassical assumption on the observer wavefunction $f(q)$ as employed in 
\cite{Chandrasekaran2022a, Chandrasekaran2022b}, namely that it is slowly
varying.  This amounts to assuming that the entanglement between the observer
and the quantum field degrees of freedom is negligible, and allows us to
ignore commutators of the form $[f\left(\hat{q}-\hb\right),\Delta_{\Phi|\Psi}]$
appearing in $\log\rho_{\wh{\Phi}}$.  Under this assumption, the entropy
can be expressed as (see section \ref{sec:crossprod})
\beq\label{eqn:SrhoSrel}
S(\rho_{\wh{\Phi}}) = -S_\text{rel}(\Phi||\Psi) - \beta\langle H_\text{obs}\rangle_f
+S_f^\text{obs} + \log\beta,
\eeq
where $S_\text{rel}(\Phi||\Psi)$ is the relative entropy
between the states $|\Phi\rangle$ and $|\Psi\rangle$ in the algebra $\aqft$, and 
$S_f^\text{obs}$ is the entropy associated with the probability distribution
derived from the observer's wavefunction.  
This expression for the entropy is manifestly UV finite
for a wide class of states
 and hence defines a good notion of renormalized
entropy for the subregion.  This formula for the 
entropy is closely related to expressions from
\cite{Chandrasekaran2022a, Chandrasekaran2022b} applicable
to subregions possessing boost symmetry.
Note that the multiplicative ambiguity in the definition of 
$\rho_{\wh{\Phi}}$ translates to a state-independent additive ambiguity
in $S(\rho_{\wh{\Phi}})$, reflected in the constant $\log\beta$ 
term in (\ref{eqn:SrhoSrel}).  This ambiguity is discussed 
in more detail in sections \ref{sec:moddensity} and \ref{sec:genent}.

We can also relate $S(\rho_{\wh\Phi})$ to the generalized entropy for the 
subregion.  Because $|\Psi\rangle$ is a KMS state
for $\Hxig$,  the relative entropy  in (\ref{eqn:SrhoSrel})
can be expressed as a free
energy 
with respect to the one-sided Hamiltonian  for the subregion $H^\Sigma_\xi$, 
\beq
S_\text{rel}(\Phi||\Psi) = \beta\langle H_\xi^\Sigma\rangle_\Phi -S_\Phi^\text{mat}
+\text{const.},
\eeq
with the constant state-independent.
Each term in this expression is separately UV divergent, but the 
combination is finite for states with finite relative entropy with respect to $|\Psi\rangle$.
To convert this to a generalized entropy, we note that when the local gravitational
constraints are satisfied on the subregion Cauchy surface $\Sigma$, the total 
energy in the subregion is related to the bounding area according to (see
section \ref{sec:nonlinear-constraints})
\beq
H_\xi^\Sigma+ H_\text{obs} = -\frac{\kappa}{2\pi} \frac{A}{4G_N}.
\eeq
This relation is the integrated form of the first law of local subregions, 
an analog of the first law of black hole mechanics that is applicable
to generic subregions in gravitational theories. Applying these relations to (\ref{eqn:SrhoSrel}),
we arrive at the result
\beq
S(\rho_{\wh{\Phi}}) = \left\langle\frac{A}{4G_N}\right\rangle_{\wh{\Phi}}
+ S^\text{mat}_\Phi+S_f^\text{obs} 
+\text{const.}= S_\text{gen} + \text{const.},
\eeq
demonstrating that the algebraic entropy $S(\rho_{\wh\Phi})$ computed in
$\ac$ agrees with the subregion generalized entropy up to 
a state-independent constant.
Note that invoking the local first law of subregions 
simplifies the derivation of the generalized entropy
from (\ref{eqn:SrhoSrel}) relative to the original
arguments appearing in
\cite{Chandrasekaran2022a, Chandrasekaran2022b}.

\subsubsection*{Type $\ttwo_1$/Type $\ttwo_\infty$ algebras 
from energy conditions}
We next turn to the question of energy conditions satisfied by the observer.  
Although we do not at present have a detailed model for the observer, 
a reasonable requirement to avoid instabilities is  that the observer
energy be bounded below, as was assumed by CLPW in \cite{Chandrasekaran2022a}.
This can be implemented by acting with a step function 
projection $\Pio = \Theta(H_\text{obs})= \Theta(\hat{q})$,
on all elements of $\ac$.  The resulting algebra $\aproj = \Pio \ac \Pio $
consists of all operators of the form $\Pio \, \wh{\msf{a}}\,\Pio$, 
with $\wh{\msf{a} }
\in\ac$.  
As explained in section \ref{sec:encond}
and appendix \ref{app:algebraic-background}, 
the effect of this projection on the algebra can be diagnosed by 
evaluating the trace of $\Pio$ in $\ac$.  
This trace is defined on $\ho a \in \ac$ by
\beq \label{eqn:outlinetrace}
\atr\;\ho a = 2\pi \beta \langle\Psi|\langle 0|_p
e^{-\frac{\beta\hat{q}}{2}} \,\ho a\,e^{-\frac{\beta\hat{q}}{2}}
|0\rangle_p|\Psi\rangle,
\eeq
where $|0\rangle_p$ is the zero momentum eigenstate.
$\atr$ can be viewed as a renormalized version of the standard
Hilbert space trace that preserves cyclicity $\atr(\ho a\,\ho b) = \atr(\ho b\, \ho a)$
and satisfies good physical properties including
faithfulness, semifiniteness, and normality.  
See section \ref{sec:moddensity} and 
appendix \ref{app:algebraic-background} for further discussion
of this trace.  

From this definition, the trace of $\Pio$ is readily evaluated,
\beq
\atr(\Pio) = \beta\int_{0}^\infty dy e^{-\beta y} = 1.
\eeq
Because $\atr(\Pio)$ is finite, the projected algebra $\aproj$ is 
a factor of type
$\ttwo_1$.  
This matches the algebra type obtained by CLPW for the static patch of de Sitter
space.  Type $\ttwo_1$ algebras have the property of possessing a maximum
entropy state whose density matrix coincides with the identity operator.  
This state is given by
\beq \label{eqn:Psimax}
|\Psi_\text{max}\rangle = |\Psi, \sqrt{\beta} e^{-\frac{\beta q}{2}} \Theta(q)\rangle.
\eeq
The existence of such a maximal entropy state immediately implies a version
of Jacobson's entanglement equilibrium hypothesis \cite{Jacobson2015},
which conjectured that the entropy of the vacuum for small causal diamonds
is maximal in quantum gravity theories.  Given the form of the maximal
entropy state (\ref{eqn:Psimax}), we see that it is the KMS state $|\Psi\rangle$
that defines the maximal entropy state, which reduces to the vacuum
state only for special choices of subregions and matter content.  

The energy conditions for the complementary region $\sr'$ 
can also be analyzed from the perspective of the commutant
algebra $(\ac)'$.  
As discussed in section \ref{sec:asymp}, the ADM Hamiltonian
is represented on $\hs_{\text{QFT}}\otimes \hs_\text{obs}$
by the operator $-\Hxig-\hat{q}$, and hence the projection
to positive ADM energy is implemented by $\Pi_\text{ADM}
= \Theta\left(-\hb - \hat{q} \right) \in (\ac)'$.  
Equation (\ref{eqn:outlinetrace}) 
also defines a trace on $(\ac)'$, and on $\Pi_\text{ADM}$ this 
trace is infinite,
\beq
\atr(\Pi_\text{ADM}) = \infty.
\eeq
Accordingly, the projected algebra $\Pi_\text{ADM} (\ac)'\Pi_\text{ADM}$
remains type $\ttwo_\infty$.  This reflects a generic feature of 
unbounded subregion algebras: the projection to positive ADM energy
is always infinite for such gravitational algebras due to the way
the ADM Hamiltonian appears in the gravitational constraint.  
The resulting picture is that bounded subregions
are associated with type $\ttwo_1$ algebras possessing maximal
entropy states, while unbounded subregions produce type $\ttwo_\infty$
algebras and correspondingly have no maximal entropy state.  

\subsection{List of assumptions}\label{sec:assumptions}

The construction outlined above provides evidence that local subregions in gravity should be associated with type $\ttwo$ von Neumann algebras, and that doing so leads to an algebraic interpretation of the generalized entropy. This conclusion relies on a number of assumptions, which we list here in order to clarify the logic of the argument. In much of the remainder of the paper, we discuss these assumptions in greater detail and give partial evidence for them.

\paragraph{Assumptions:} 
\begin{enumerate}
[label=A\arabic*., ref=A\arabic*]
    \item \label{assm:aqft}
        There exist algebras $\aqft$, $\aqft'$ (which we call ``kinematical'') describing the quantum field degrees of freedom (including linearized gravitons) associated with the causally complementary subregions $\sr$ and $\sr'$. These algebras are perturbatively definable order by order in the $\gc$ expansion, and remain type $\tthr_1$ and commutants of each other to all orders in $\gc$. 
    \item \label{assm:obs}
        There exist auxiliary observer degrees of freedom associated with the subregions $\sr$ and $\sr'$ described by type $\tone_\infty$ algebras $\alg_\text{obs}, \alg_{\text{obs}}'$ that commute with $\aqft$ and $\aqft'$ to all orders in $\gc$.
    \item \label{assm:constr}
        The physical gravitational algebra arises from imposing the constraint $\mathcal{C}[\xi] = \Hxig + H_\text{observers}$, where $\Hxig$ is the generator of a specific boost-like flow on $\aqft$ and $\aqft'$, and $H_\text{observers}$ refers to the Hamiltonian of the auxiliary observer degrees of freedom associated with $\sr$ and $\sr'$.
    \item \label{assm:mod} 
        The flow generated by $\Hxig$ coincides with the modular flow for some state on the algebras $\aqft$, $\aqft'$.
    \item \label{assm:1stlaw} 
        The local gravitational constraints hold on a Cauchy surface $\Sigma$ for the subregion $\sr$, allowing the application of the first law of local subregions in the computation of the entropy.
    \item \label{assm:bdenergy} 
        The energies of the auxiliary observer degrees of freedom in~\ref{assm:obs} with respect to a future-directed vector field are bounded below.
\end{enumerate}

We use assumptions \ref{assm:aqft}-\ref{assm:mod} to obtain a type $\ttwo$ algebra for the subregion as a crossed product with respect to a modular flow. We use~\ref{assm:1stlaw} to rewrite the entropy associated with this crossed product algebra as the generalized entropy up to a state-independent constant. Finally, we only use assumption~\ref{assm:bdenergy} to argue that the algebra for a bounded subregion is actually type $\ttwo_1$ and therefore possesses a maximal entropy  state; the preceding arguments connecting the algebraic entropy to generalized entropy are independent of assumption
\ref{assm:bdenergy}.

Assumption~\ref{assm:aqft} is the starting point for finding the crossed product algebra in section~\ref{sec:crossconstr} and implicitly has many working parts. For example, when we say that the kinematical algebras are definable order by order in $\gc$, we have in mind that the operators in $\aqft$ and $\aqft'$ commute with the constraints generated by diffeomorphisms compactly supported on $\sr$ and $\sr'$ respectively, order by order in $\gc$. That is, these operators are gravitationally dressed within $\sr$ and $\sr'$. 
It also implicitly assumes a prescription for 
specifying the boundary of the subregion
in a diffeomorphism-invariant manner, a topic on which
we briefly comment in section \ref{sec:fixregion}.
Furthermore we are assuming that all of the thorny questions related to Einstein gravity as a nonrenormalizable low-energy effective theory can be answered to produce renormalized and dressed operators (whose endpoints are presumably local up to a resolution scale $\gc$). These assumptions go beyond classic results \cite{fredenhagen1985modular, buchholz1995scaling}  proving that UV-complete, non-gravitational, Lagrangian field theories have type $\tthr_1$ factors associated with subregions. Even so, Assumption~\ref{assm:aqft} is not really new; it is in line with recent works \cite{Leutheusser:2021qhd,
Witten2021, Leutheusser:2021frk, Chandrasekaran2022a,
Chandrasekaran2022b, Leutheusser:2022bgi,
Bahiru:2022oas,
Bahiru:2023zlc} (including CLPW) concerning operator algebras in large $N$ theories. 

In introducing observers or using ADM energy as an effective observer in Assumption~\ref{assm:obs}, we are following the lead of CLPW~\cite{Chandrasekaran2022a} for a bounded subregion and~\cite{Witten2021, Chandrasekaran2022b} for one that includes an asymptotic boundary. In the first case this introduction is, in a sense, phenomenological, and it proves quite useful. We would however like to arrive at it from more fundamental considerations, perhaps as a consequence of specifying a subregion in a theory of gravity. Assumption~\ref{assm:constr} is analogous to the constraint considered by CLPW in the static patch of de Sitter space, although our constraint does not in general generate an isometry. This assumption is on solid ground and is discussed in detail in section~\ref{sec:constraints}. Assumption~\ref{assm:mod} really has two parts, since the state at this order in $\gc$ is a sum of two terms, one being a Gibbs-like distribution for the matter degrees of freedom and the other a state for linearized gravitons. When the subregion is a ball in flat space and the matter is a CFT, this Gibbs-like distribution 
coincides with the CFT vacuum, as follows 
from the Hislop-Longo theorem \cite{Hislop1982} as well
as the classic argument by Casini, Huerta, and Myers
\cite{Casini:2011kv}, 
 while in the static patch of de Sitter the vacuum is such a state (with $\xi$ generating time translations). We argue in section~\ref{sec:modham} that an analogous (generically excited) state exists more generally for subregions of matter QFT.
Note that for the specific case where $\sr$ admits a stationary null slice, a state with local modular flow on that slice can be realized using ideas from \cite{Wall2011, faulkner2016modular, casini2017modular}.

Assumption~\ref{assm:1stlaw} 
is well motivated from the perspective of gravitational constraints; however,  it is also somewhat schematic since it involves  sums of terms that are separately UV divergent.  The  resulting first law arising from this constraint  can nevertheless  be viewed as a Lorentzian argument in favor of finiteness of the generalized entropy, since it is used to convert the generalized entropy into an expression involving a relative  entropy. It and Assumption~\ref{assm:bdenergy} appear chiefly  in sections~\ref{sec:genent} and~\ref{sec:encond}.

\section{Gravitational constraints} \label{sec:constraints}

One of the main points of the present work is that type $\ttwo$ von Neumann algebras arise in the treatment of gravitational subsystems as a consequence of diffeomorphism invariance.
In any quantum theory with gauge symmetries, there are constraints that must be imposed on the Hilbert space and the algebra of observables.
At the classical level, the constraints generate gauge transformations via Poisson brackets, so at the quantum level, gauge-invariant operators are ones that commute with the quantized constraints.
Thus, to understand the consequences of diffeomorphism invariance for algebras of observables in gravity, we must begin by studying the structure of the corresponding classical constraints.

In this section, we explain how diffeomorphism constraints appear in the theory of perturbative gravitons coupled to matter quantized around a fixed background.
In subsection \ref{sec:nonlinear-constraints}, we explain the structure of diffeomorphism constraints in nonlinear general relativity minimally coupled to matter.
In subsection \ref{subsec:perturbative-constraints}, we study perturbative gravitons by taking the small-$G_N$ limit of the nonlinear constraints, and explain certain subtleties in the structure of the constraints via an analogy to $U(1)$ gauge theory.
In subsection \ref{sec:fixregion}, we discuss issues related to gauge-fixing the regions $\sr$ and $\sr'$; while we do not completely resolve the issue of gauge-fixing, we explain some features that a good gauge-fixing prescription should have.

\subsection{Constraints in nonlinear gravity}
\label{sec:nonlinear-constraints}

One key feature of a classical theory with gauge symmetries, as explained e.g.\ in \cite{Henneaux:1992ig}, is a redundancy of the configuration space variables for describing solutions to the equations of motion; even if initial data is specified for all configuration space variables, their values under dynamical evolution are not completely determined.
In a phase space formulation of the theory, this leads to a too-large ``kinematical'' phase space in which physical configurations live on a constraint submanifold.
In classical field theories, as explained e.g.\ in \cite{Lee:1990nz}, the kinematical phase space should be thought of as (a particular quotient of) the space of field configurations, with the constraint submanifold containing field configurations satisfying the equations of motion.
The kinematical phase space is equipped with a symplectic form whose restriction to the constraint submanifold develops degeneracies corresponding to gauge symmetries.
A gauge symmetry of the configuration space variables, written e.g.\ as $\phi \mapsto \phi + \epsilon \delta \phi,$ induces a flow on the constraint submanifold that is a degenerate direction for the induced symplectic form.
One can show, as in \cite{Lee:1990nz}, that for any such flow there exists a functional $\mathcal{C}$ on phase space that (i) vanishes on the constraint submanifold, and (ii) generates the flow via Poisson brackets, in the sense that for any function $f$ on phase space we have
\begin{equation}
    \{f, \mathcal{C}\}|_{\text{constraint submanifold}} = \delta f.
\end{equation}
Consequently, $\mathcal{C}$ commutes with gauge-invariant functions on the constrained phase space.

The story is similar in quantum theory.
Under canonical quantization, the phase-space functional $\mathcal{C}$ must turn into an operator $\hat{\mathcal{C}}$ that commutes with all gauge-invariant operators.
In place of the kinematical phase space of the classical theory, we consider a kinematical algebra of operators in the quantum theory.
The physical operators are the ones that commute with the constraints; these are called ``dressed operators.''
These dressed operators can be identified by studying the commutation relations between constraints and kinematical operators.

We now apply the above considerations to gravitational theories.
In any gravity theory, diffeomorphisms with compact support on a Cauchy slice are gauge symmetries.
For any such diffeomorphism, there is an associated constraint that must vanish in the physical theory.
More generally, diffeomorphisms with non-compact support are generated by a Hamiltonian that consists of a constraint term, which vanishes on physical configurations, and a boundary term that remains nonzero even after the constraints are imposed.
Taking $\zeta^a$ to be a vector field generating a diffeomorphism 
and $\csfull$ to be a complete Cauchy surface for the spacetime region where the diffeomorphism acts,
the expression for the gravitational Hamiltonian $H_\zeta^g$ is given by
\beq\label{eqn:Hzeta}
H_\zeta^g = \int_{\csfull} C_\zeta \;+ H_\zeta^\text{bdy}.
\eeq
Precise expressions for the constraint and boundary terms can be derived
from any canonical formulation of the classical gravitational theory; see
appendix \ref{app:CPS} for a review of the derivation using covariant
phase space techniques.
For general relativity minimally coupled to
matter, the constraint current $C_\zeta$ takes the form
\beq \label{eqn:Czeta}
C_\zeta = \left(\frac{1}{8\pi G_N}(G\indices{^a_b} +\Lambda \delta\indices{^a_b})
-T\indices{^a_b} \right)\zeta^b \epsilon_{a\ldots}
\eeq
where $G\indices{^a_b}$ is the Einstein tensor, $\Lambda$ is the cosmological
constant, $T\indices{^a_b}$ is the matter stress tensor, and $\epsilon_{a\ldots}$
is the spacetime volume form.\footnote{For some theories with 
tensor matter, there are additional contributions to the 
constraint involving the matter equations of motion,
see appendix \ref{app:CPS}.}

To obtain the crossed product in section \ref{sec:outline}, we imposed a constraint associated with a vector field $\xi^a$ that generates a boost around an entangling surface (see again figure \ref{fig:vector-field}).
More specifically, we considered splitting a spacetime Cauchy surface $\Sigma_c$ into two pieces $\csfull = \Sigma\cup\bar\Sigma.$
The domain of dependence of $\Sigma$ was called $\sr$ and the domain of dependence of $\bar{\Sigma}$ was called $\sr'.$
We required that $\xi^a$ be future-directed in the interior of $\sr$, past directed in the interior of $\sr'$, vanishing at the entangling surface $\partial\Sigma$, and tangent to the null boundaries 
of $\sr$ and $\sr'$ (see again figure \ref{fig:vector-field}).
We also required that $\xi^a$ approach a global time translation at any asymptotic boundaries, and that on $\partial \Sigma$ there is a constant $\kappa$ satisfying
\beq \label{eqn:nabxi}
\nabla_a \xi_b \overset{\partial\Sigma}{=} \kappa n_{ab},
\eeq
where $n_{ab}$ is the unit binormal to $\partial\Sigma$.
The constancy of $\kappa$ is a quasilocal version of the zeroth law of black hole mechanics applicable to general subregions, and we show in 
some examples in section \ref{sec:modham} that it is tied to the existence of a KMS state associated with the flow of $\xi^a$.
Note that for reasons discussed in footnote \ref{ftn:rindler}, we also required that the entangling surface $\partial \Sigma$ be compact.

The boundary term in equation \eqref{eqn:Hzeta} for the vector field $\xi^a$ is determined by the topology of the Cauchy surface $\csfull.$
For every asymptotic boundary in $\csfull,$ the boundary term $H_{\xi}^{\text{bdy}}$ picks up a corresponding ADM Hamiltonian.
Due to the time orientation of $\xi^a$, the ADM Hamiltonian comes with a positive sign for an asymptotic boundary of $\Sigma,$ and a negative sign for an asymptotic boundary of $\bar{\Sigma}.$
As explained in \cite{Chandrasekaran2022a}, imposing the identity $H_{\xi}^g = 0$ directly on the kinematical algebras $\aqft$ or $\aqft'$ completely trivializes the algebra.
For regions with asymptotic boundaries this is not an issue, because the boundary term in equation \eqref{eqn:Hzeta} is nonzero, so imposing the constraint does not set $\Hxig$ to zero, but rather relates it to the ADM Hamiltonian.
If either $\Sigma$ or $\bar{\Sigma}$ does not have an asymptotic boundary, then it is necessary to introduce an auxiliary ``observer'' degree of freedom in that region to take the place of the boundary term.
We assume that the observers are weakly coupled to the matter degrees of freedom, but couple to gravity via their energy-momenta.  
We will remain agnostic about the details of the observers --- see section \ref{sec:observer} for further discussion --- but will assume that the observers act as clocks that measure time along the flow $\xi^a$, in that we have
\begin{equation}
    H_{\text{obs}}
        = - \int_{\Sigma} (T_\text{obs})^{a}{}_{b} \xi^b \epsilon_{a\ldots}
\end{equation}
in the region $\sr,$ or
\begin{equation}
    H_{\text{obs}}'
        = \int_{\bar{\Sigma}} (T_\text{obs}')^{a}{}_{b} \xi^b \epsilon_{a\ldots}
\end{equation}
in the region $\sr'.$
The sign difference between these two equations is due to the fact that $\xi^a$ is past-directed on $\bar{\Sigma}.$

When an observer is coupled to gravity, its stress-energy must be included as a contribution to the stress-energy tensor appearing in the constraint current \eqref{eqn:Czeta}.
The total Hamiltonian for $\xi^a,$ computed via equation \eqref{eqn:Hzeta}, can then be written in an explicit form.
For convenience, as in section \ref{sec:outline}, we now restrict to the case where $\Sigma$ is bounded and $\bar{\Sigma}$ is unbounded.
In this case, the full gravitational Hamiltonian, including the observer contribution, is given by
\begin{equation} \label{eqn:main-nonlinear-constraint}
    H_{\xi}^{\text{total}} = \int_{\csfull} C_{\xi}^{\text{mat}} + H_{\text{obs}} - H_{\text{ADM}}^{\bar{\Sigma}},
\end{equation}
where $C_{\xi}^{\text{mat}}$ denotes the constraint current \eqref{eqn:Czeta} \textit{without} the observer-stress energy included.
Going forward, we will reserve the symbol $H_{\xi}^g$ for the Hamiltonian that generates the flow of $\xi^a$ purely on the gravitational and matter degrees of freedom, without acting on the observer.
With this choice of notation, equation \eqref{eqn:Hzeta} can be expressed in convenient form as
\begin{equation} \label{eqn:ham-constraint-equation}
    H_{\xi}^g + H_{\text{obs}} + H_{\text{ADM}}^{\bar{\Sigma}} = \int_{\csfull} C_{\xi}^{\text{mat+obs}} \equiv \mathcal{C}[\xi].
\end{equation}
After quantization, $\mathcal{C}[\xi]$ becomes an operator $\hat{\mathcal{C}}[\xi]$ that must commute with physical observables.
If $\Sigma$ were unbounded, we would replace $H_{\text{obs}}$ by $- H_{\text{ADM}}^{\Sigma}$; if $\bar{\Sigma}$ were bounded, we would replace $H_{\text{ADM}}^{\bar{\Sigma}}$ by $- H_{\text{obs}}'.$
Going forward, we will remain in the $\Sigma$-bounded, $\bar{\Sigma}$-unbounded scenario, and therefore will drop the superscript ``$\bar{\Sigma}$'' from $H_{\text{ADM}}^{\bar{\Sigma}}$; analogous results for other scenarios can be obtained by appropriate substitution of observer Hamiltonians for ADM Hamiltonians:
\begin{align}
    \begin{split}
        \mathcal{C}_{\text{$\sr$ bounded, $\sr'$ unbounded}}[\xi]
            & = H_{\xi}^g + H_{\text{obs}} + H_{\text{ADM}}^{\bar{\Sigma}}. \\
        \mathcal{C}_{\text{$\sr$ unbounded, $\sr'$ bounded}}[\xi]
            & = H_{\xi}^g - H_{\text{ADM}}^{\Sigma} - H_{\text{obs}}'. \\
        \mathcal{C}_{\text{$\sr$ bounded, $\sr'$ bounded}}[\xi]
            & = H_{\xi}^g + H_{\text{obs}} - H_{\text{obs}}'. \\
        \mathcal{C}_{\text{$\sr$ unbounded, $\sr'$ unbounded}}[\xi]
            & = H_{\xi}^g - H_{\text{ADM}}^{\Sigma} + H_{\text{ADM}}^{\bar{\Sigma}}.
    \end{split}
\end{align}
As explained in section~\ref{sec:asymp}, the sign difference between observer and ADM Hamiltonians as they appear in these equations is responsible for producing a type $\ttwo_\infty$
algebra for unbounded regions after imposing a positive energy condition, instead of a type $\ttwo_1$ algebra in the bounded case.

In addition to the global constraints discussed above, 
it is also important to consider
the individual contributions
to the constraint coming from $\Sigma$ and $\bar\Sigma$.
Formally, since the global constraint is expressible as an integral
over the complete Cauchy surface $\csfull$, it can 
 be expressed as a sum of two quasilocal
contributions
\beq \label{eqn:Csplit}
\mathcal{C}[\xi] = \int_\Sigma C_\xi 
+ \int_{\bar\Sigma}C_\xi.
\eeq
These quasilocal constraints lead to important relations
that are used to interpret the entropies of the type $\ttwo$
gravitational algebras in terms of generalized entropies.  
Since the partial Cauchy surface $\Sigma$ has a non-asymptotic boundary, we may apply equation \eqref{eqn:Hzeta} to obtain an expression for $\int_{\Sigma} C_{\xi}$ in terms of a gravitational Hamiltonian $H_{\xi}^{\Sigma}$ and a boundary term coming from the entangling surface $\partial \Sigma.$
In general relativity, the boundary term is proportional to the area of $\partial \Sigma,$ which can be derived by relating it to the Noether charge of \cite{wald1993black, Iyer1994} and using the 
constancy of the surface gravity $\kappa$ (see appendix \ref{app:CPS}).
The expression is
\beq \label{eqn:HSigmareln}
\int_\Sigma C_\xi^{\text{mat}+\text{obs}} 
= H_\xi^\Sigma + H_\text{obs} + \frac{\kappa}{2\pi} \frac{A}{4G_N}.
\eeq
An analogous relation can be derived for the complementary region $\bar{\Sigma}.$
If we write $H_{\xi}^{g} = H_{\xi}^{\Sigma} - H_{\xi}^{\bar{\Sigma}}$ to emphasize that $\xi^a$ is past-directed on $\bar{\Sigma},$ the identity is
\beq\label{eqn:HbarSigmareln}
    \int_{\bar\Sigma}C^{\text{mat}}_\xi = -H_\xi^{\bar\Sigma} + H_\text{ADM}  - \frac{\kappa}{2\pi}\frac{A}{4G_N}.
\eeq
Note that adding these two equations together gives
\begin{align}
\mathcal{C}[\xi]
    & = 
\left(H_\xi^\Sigma +H_\text{obs}+\frac{\kappa}{2\pi}
\frac{A}{4G_N}\right) +\left(-H_\xi^{\bar\Sigma} +
H_\text{ADM} - \frac{\kappa}{2\pi}\frac{A}{4 G_N}\right) \nonumber \\
& = H_{\xi}^{\Sigma} - H_{\xi}^{\bar{\Sigma}} + H_{\text{obs}} + H_{\text{ADM}}
\end{align}
in agreement with equation \eqref{eqn:ham-constraint-equation}.

From equation \eqref{eqn:HSigmareln}, 
we see that if the constraints $C_\xi^{\text{mat}+\text{obs}} = 0$
are satisfied locally on the partial Cauchy slice $\Sigma$, then the total $\xi$-energy within the subregion is related to the area of the boundary.
We may assume this for the present purposes, as it is part of our assumption \ref{assm:aqft} from section \ref{sec:assumptions}.
In section \ref{sec:crossprod}, we will use equation \eqref{eqn:HSigmareln} to relate the entropy computed in a crossed product algebra to the generalized entropy of Bekenstein.

To connect with familiar concepts from gravitational thermodynamics, it is useful to take a variation of equation \eqref{eqn:HSigmareln} at fixed $\kappa$, which leads to an infinitesimal relation
\beq
\delta H_\xi^\Sigma + \delta H_\text{obs} = -\frac{\kappa}{2\pi}
\delta\frac{A}{4G_N}.
\eeq
We call this the {\it first law of local subregions}.  
It is a generalization to arbitrary subregions of various other thermodynamic relations that have appeared previously in gravity such as the first law of black hole mechanics \cite{Bardeen1973}, the first law of event horizons \cite{gibbons1977cosmological}, and the first law of causal diamonds \cite{Jacobson2015, Bueno2016, Jacobson2018}.
The integrated form of the first law (\ref{eqn:HSigmareln}) could therefore
be referred to as a quasilocal equation of state or Smarr relation.
Note that quasilocal Smarr relations and first laws have recently been explored in  \cite{Banihashemi:2022htw}.
\subsection{Perturbative constraints for nonlinear gravitons}
\label{subsec:perturbative-constraints}

In the previous subsection, we described the structure of diffeomorphism constraints in general relativity coupled to matter.
The setting of section \ref{sec:outline} is the $G_N \to 0$ limit of this theory, where general relativity is treated as an effective field theory of gravitons.
The constraints of this theory can be studied by expanding the exact nonlinear constraints of the previous section order by order in the graviton coupling.\footnote{See \cite{Giddings2022} for a recent discussion of this perturbative expansion about generic backgrounds.
}

Perturbative gravitons around a fixed background are field configurations of the form
\begin{equation} \label{eqn:linearized-metric}
    g_{ab} = g^0_{ab} + \gc h_{ab},
\end{equation}
with $\gc = \sqrt{32 \pi G_N}$, and where $G_N$ is treated as a vanishingly small formal parameter.
The tensor $g^0_{ab}$ is a metric solving Einstein's equations (possibly with a classical source or cosmological constant), and $h_{ab}$ is a generic symmetric tensor that we call a graviton field.
The gauge symmetries of the perturbative graviton theory are inherited from the full nonlinear theory of gravity.
Every compactly supported diffeomorphism is a gauge symmetry of the nonlinear theory; however, when studying perturbative gravitons, we have already done a partial gauge-fixing by restricting the background metric to be exactly $g^0_{ab},$ and the residual gauge symmetries correspond to compactly supported diffeomorphisms that do not alter this choice.
In practice, this means that the gauge symmetries of perturbative gravitons are compactly supported diffeomorphisms that are formally proportional to $\gc,$ i.e., $\delta_{\gc \zeta} g_{ab} = \gc \lie_{\zeta} g_{ab}.$
Because $g_{ab}^0$ is held fixed under these transformations,
their effect is to alter the metric fluctuation $h$ by 
\begin{equation}
    \delta_{\gc \zeta} h_{ab} = \lie_{\zeta} g^0_{ab} + \gc \lie_{\zeta} h_{ab} = \overset{0}{\nabla}_a\zeta_b + \overset{0}{\nabla}_b \zeta_a + \gc \lie_{\zeta} h_{ab}.
    \end{equation}
If a matter field $\phi$ is present, then these diffeomorphisms act on the matter field by
\begin{equation}
    \delta_{\gc \zeta} \phi = \gc \lie_{\zeta} \phi.
\end{equation}
In the limit $\gc \to 0,$ the transformation of matter fields is neglected, and the graviton field is transformed by the addition of the pure-gauge term $\overset{0}{\nabla}_a \zeta_b + \overset{0}{\nabla}_b \zeta_a.$
This the usual abelian gauge symmetry of the free graviton theory; it is abelian because the commutator $[\gc \zeta_1, \gc \zeta_2] = O(\gc^2)$ is neglected in the $\gc \to 0$ limit.

If the background metric $g^0_{ab}$ admits a compactly supported Killing vector field $X^a$, then the associated diffeomorphism is a symmetry of the background metric.
This has an important effect on the theory of gravitons, which can be thought of in two different ways.
The traditional perspective, which is called the study of ``linearization instabilities'' \cite{fischer1973linearization, brill1973instability, moncrief1975spacetime}, notes that the vector field $\gc X^a$ produces no change in the fields at leading order.
Consequently, the constraints of the full theory cannot be treated by considering only perturbative corrections to the leading-order constraints; to fix this issue, a constraint corresponding to $X^a$ must be imposed at leading order.
An alternative perspective notes that $X^a$ generates a transformation that maps field configurations of the form \eqref{eqn:linearized-metric} into other field configurations of that form, so it must be imposed as a gauge symmetry at leading order in any consistent truncation of the full nonlinear theory.
In either perspective, the linear theory of gravitons can only be consistently embedded into a nonlinear theory of gravity if one takes into account the gauge transformation $\delta_X g_{ab} = \lie_{X} g_{ab},$ which acts on the metric fluctuation $h_{ab}$ by
\begin{equation}
    \delta_{X} h_{ab} = \lie_{X} h_{ab},
\end{equation}
and on matter fields by
\begin{equation}
    \delta_{X} \phi = \lie_{X} \phi.
\end{equation}
Crucially, this transformation acts on all fields at leading order.
In \cite{Chandrasekaran2022a}, a crossed product algebra was obtained for the static patch of de Sitter space by imposing a constraint corresponding to the static patch's boost isometry. 
In the same work, a crossed product algebra was obtained for the exterior of a static black hole by requiring the (non-compactly supported) Schwarzschild time translation to generate the same physical flow as the ADM Hamiltonian.

The main point of this paper is to argue that crossed product algebras and generalized entropies can be associated to subregions in general backgrounds in the $\gc \to 0$ limit of quantum gravity, even in the absence of isometries.
We contend that the linearization instability is a red herring --- \textit{every} constraint has a contribution at subleading order that can affect the linearized theory, whether or not the leading-order contribution of that constraint vanishes. 
Without taking these effects into account, it is possible to miss important aspects of the theory. For example, the gravitational Gauss law that expresses
the Hamiltonian in gravity as a boundary term only becomes 
nontrivial at first interacting order in $\gc$ beyond
the linearized theory, which Marolf has argued is a crucial point behind the holographic nature of gravity \cite{Marolf2008}. We will argue here that 
a similar effect is responsible for producing the crossed-product subregion algebras and finite renormalized
entropies. 
To understand this claim, we will expand the quantities from subsection \ref{sec:nonlinear-constraints} as power series in the gravitational coupling $\gc.$

The constraint current $C_{\gc \zeta},$ computed via equation \eqref{eqn:Czeta}, admits an expansion in $\gc$ as
\begin{equation} \label{eqn:constraint-expansion} 
    C_{\gc \zeta}
        = 4 (G^{(1)})^{a}{}_{b} \zeta^b \epsilon_{a \ldots}
        +  \gc \left( 4 (G^{(2)})^{a}{}_{b} - (T^{(0)})^{a}{}_{b} + 2 h (G^{(1)})^{a}{}_{b} \right) \zeta^b \epsilon_{a \ldots} + O(\gc^2),
\end{equation}
where we have expanded the Einstein tensor as
\begin{equation}
    G^{a}{}_{b} = (G^{(0)})^{a}{}_{b} + \gc (G^{(1)})^{a}{}_{b} + \gc^2 (G^{(2)})^{a}{}_{b} + \dots,
\end{equation}
introduced the notation $h \equiv (g^0)^{ab} h_{ab},$ and have assumed that the background metric $g^0$ solves Einstein's equations with cosmological constant $\Lambda$.\footnote{More generally, it could solve Einstein's equations with a semiclassical matter source, which would appear as a background contribution to the matter stress tensor proportional to $1/\gc^2.$}
As in \cite{Giddings2022}, we can construct physical observables in the theory of a nonlinear graviton coupled to matter by imposing the integral of equation \eqref{eqn:constraint-expansion} as a constraint order by order in $\gc.$
Once the linearized constraints have been imposed, we may neglect terms proportional to $G^{(1)},$ as terms proportional to $G^{(1)}$ generate linearized diffeomorphisms on the graviton field.
The residual constraint current is
\begin{equation} \label{eqn:linearized-constraint-current}
    C_{\gc \zeta}
        = \gc \left( 4 (G^{(2)})^{a}{}_{b} - (T^{(0)})^{a}{}_{b} \right) \zeta^b \epsilon_{a \dots} + O(\gc^2).
\end{equation}
In the language of section 2.1, restricting our attention to this expression is part of assumption \ref{assm:aqft}, which implies that the kinematical algebras consist of dressed operators that already satisfy all of the linearized constraints.
In practice, this means that the kinematical operators are gauge-invariant under the abelian gauge symmetry of the free graviton; constructing them is analogous to constructing gauge-invariant operators in pure Maxwell theory.
Note that imposing the linearized constraints also entails fixing the location of the subregion boundary in a diffeomorphism-invariant way at lowest perturbative order, which we discuss further in subsection \ref{sec:fixregion}.

The next step in studying the quantum theory is to restrict to the subalgebra of operators that commute with an operator version of
\begin{equation}
    \mathcal{C}^{(1)}[\gc \zeta]
        \equiv \int_{\csfull} \left( 4 (G^{(2)})^{a}{}_{b} - (T^{(0)}_{\text{mat}+\text{obs}})^{a}{}_{b} \right)\zeta^b\epsilon_{a\ldots}.
\end{equation}
We will restrict our attention to the vector field $\xi^a$ defined in the previous subsection.
The constraint $\mathcal{C}^{(1)}[\gc \xi]$ can be written in terms of fundamental physical quantities using equation \eqref{eqn:ham-constraint-equation}.
Using the fact that the observer and ADM Hamiltonians rescale linearly under the substitution $\xi^a \to \gc \xi^a,$
we have
\begin{equation}
    \mathcal{C}^{(1)}[\gc \xi]
        = (H_{\gc \xi}^{g})^{(1)} + H_{\text{obs}}^{(0)} + (H_{\text{ADM}}^{\bar{\Sigma}})^{(0)},
\end{equation}
where we have expanded $H_{\gc \zeta}^g$ as
\begin{equation} \label{eqn:H-gc-expansion}
    H_{\gc \xi}^{g} = \frac{1}{\gc} (H_{\gc \xi}^{g})^{(-1)} + (H_{\gc \xi}^{g})^{(0)} + \gc (H_{\gc \xi}^{g})^{(1)} + O(\gc^2).
\end{equation}
Note that while $(H_{\gc \xi}^g)^{(1)}$ appears multiplying $\gc$ in equation \eqref{eqn:H-gc-expansion}, it is quadratic in the graviton field.

The commutators of $\mathcal{C}^{(1)}[\gc \xi]$ in the quantum theory can be studied at leading order in $\gc$ by studying the Poisson brackets of the corresponding classical quantity.
As explained in the previous subsection, the functional $H^{g}_{\gc \xi}$ generates diffeomorphisms with respect to $\gc \xi$ on the kinematical algebra of gravity and matter fields, i.e.,
\begin{align} \label{eqn:h-suppressed-flow}
    \{h_{ab}, H_{\gc \xi}^g\}
        & = \lie_\xi g^0_{ab}+\gc \lie_{\xi} h_{ab}, \\
    \{\phi, H_{\gc \xi}^g \} \label{eqn:phi-suppressed-flow}
        & = \gc \lie_{\xi} \phi.
\end{align}
By matching linear-in-$\gc$ terms in equation \eqref{eqn:h-suppressed-flow}, we see that $(H_{\gc \xi}^g)^{(1)}$ must generate diffeomorphisms on kinematical fields with no $\gc$ suppression.\footnote{There is a small subtlety here: if the graviton fluctuation $h_{ab}$ is regarded as a formal power series in $\gc,$ then the linear-in-$\gc$ contribution to $H_{\gc \xi}^g$ generates diffeomorphisms only on the lowest term in that power series; higher-order terms in $H_{\gc \xi}^g$ generate diffeomorphisms acting on higher-order terms in $h_{ab}$.}
I.e., we have
\begin{align} \label{eqn:H-1-h-poisson}
    \{h, (H_{\gc \xi}^g)^{(1)}\}
        & = \lie_{\xi} h, \\
    \{\phi, (H_{\gc \xi}^g)^{(1)}\}
        & = \lie_{\xi} \phi. \label{eqn:H-1-phi-poisson}
\end{align}
$(H_{\gc \xi}^{g})^{(-1)}$ is a constant and has vanishing Poisson brackets with all fields, and $(H_{\gc \xi}^{g})^{(0)}$ generates the linearized diffeomorphism $\delta h_{ab} = \lie_{\xi} g^0_{ab}.$
Note that while $(H_{\gc \xi}^g)^{(1)}$ appears multiplying $\gc$ in equation \eqref{eqn:H-gc-expansion}, it is quadratic in the graviton field.

Our conclusion is that there \textit{is} a constraint operator corresponding to $\mathcal{C}^{(1)}[\gc \xi]$ that must commute with physical operators, and that it consists of pieces corresponding to observer and ADM Hamiltonians, together with a piece that generates diffeomorphisms on the kinematical algebra of observables.
There is a small subtlety associated to the fact that the constraint we are really told to impose is $\gc \mathcal{C}^{(1)}[\gc \xi],$ not $\mathcal{C}^{(1)}[\gc \xi].$
From the perspective of the perturbative graviton theory, where $\gc$ is a formal parameter, imposing one of these constraints is not the same as imposing the other.
But in order for the perturbative graviton theory to embed consistently within the full nonlinear theory, where $\gc$ really is just a number, both constraints must be imposed.

\begin{figure}[t]
    \centering
    \includegraphics{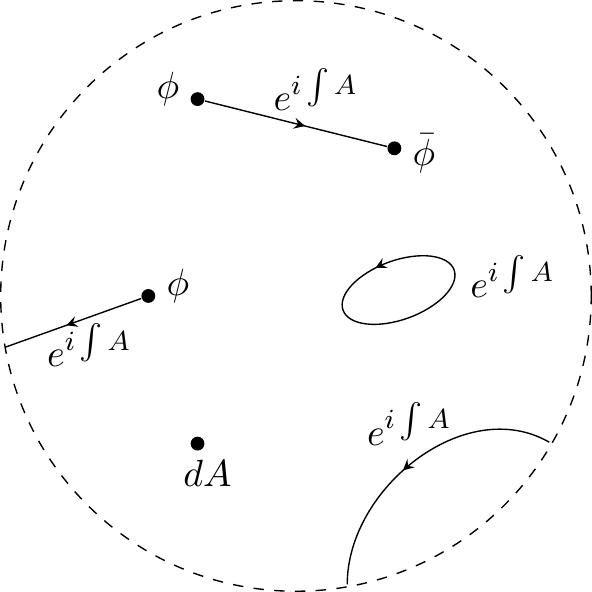}
    \caption{Several types of ``kinematical'' operators in scalar-Maxwell theory that are invariant under compact gauge transformations supported within a region. There are charged scalars dressed to each other, charged scalars dressed to the boundary of the region, local field strength operators, and Wilson lines with no endpoints in the interior of the region. Any kinematical operator including a Wilson line with an endpoint on the boundary is not invariant under the global constraints of the theory.}
    \label{fig:quasilocal-operators}
\end{figure}

To understand this last point, it may be helpful to consider an analogy to $U(1)$ Maxwell theory coupled to a charged scalar.
The fundamental fields are a gauge field $A_{\mu}$ and a complex scalar $\phi$, which transform under gauge transformations as $A_{\mu} \mapsto A_{\mu} + \partial_{\mu} \lambda,$ $\phi \mapsto e^{i \lambda} \phi.$
The quasilocal constraints in a region $\sr$ correspond to gauge transformations for functions $\lambda(x)$ that are compactly supported within $\sr.$
The algebra of operators commuting with these constraints is generated by (i) local field strength operators $dA$, (ii) Wilson lines with no endpoints in the interior of $\sr$, (iii) scalar fields $\phi$ dressed to conjugate fields $\bar{\phi}$ by Wilson lines,  (iv) scalar fields $\phi$ dressed by Wilson lines that end on the boundary of $\sr$, and (v) other extended operators.
See figure \ref{fig:quasilocal-operators}.
In our language, the algebra generated by these operators is the kinematical algebra $\aqft.$

This algebra only satisfies some of the constraints associated with the full theory.
One additional constraint that one could consider comes from a gauge transformation for a function $\lambda(x)$ that is constant in a neighborhood of $\sr,$ and that vanishes at infinity.
This constraint does not commute with any operators in $\aqft$ that involve Wilson lines ending on the boundary of $\sr,$ so restricting to the subalgebra commuting with this constraint would remove all operators of this kind from $\aqft.$
If one wants to keep these operators in the theory in order to have quasilocal charged scalars localized to the region $\sr,$ it is necessary to augment the theory by an auxiliary Hilbert space $\hs_{\text{charge}}$ whose algebra is generated by a single operator $Q$ that transforms, under gauge transformations constant in a neighborhood of $\sr$, as $Q \mapsto e^{- i \lambda} Q.$
This operator plays the role of the observer Hamiltonian in the gravity construction described above.
The kinematical operators $\phi e^{i \int A}$ and $Q$ are not invariant under the constant gauge transformation, but combinations like $\phi e^{i \int A} Q$ are.
Properly accounting for the constant gauge transformation therefore requires restricting to a subalgebra of $\aqft \otimes \mathcal{B}(\hs_{\text{charge})},$ just as accounting for the boost transformation in gravity required restricting to a subalgebra of $\aqft \otimes \alg_{\text{obs}}.$

To make a more precise analogy between the Maxwell theory example and the gravity example, we may treat the Maxwell field $A_{\mu}$ perturbatively around the vacuum as $A_{\mu} = 0 + \alpha a_{\mu},$ where $\alpha$ is the Maxwell coupling.
The constant gauge transformation should be suppressed by a factor of $\alpha,$ so it acts as
\begin{equation}
    \delta_{\alpha} a_{\mu} = 0, \qquad \delta_{\alpha} \phi = i \alpha \phi, \qquad \delta_{\alpha} Q = - i \alpha Q.
\end{equation}
The leading order piece of the constraint corresponding to this transformation generates the $\alpha \to 0$ part of this transformation, so it is trivial on the kinematical algebra $\aqft \otimes \mathcal{B}(\hs_{\text{charge}}).$
In analogy with equation \eqref{eqn:H-1-phi-poisson}, the subleading piece $\mathcal{C}^{(1)}[\alpha]$ satisfies the Poisson brackets
\begin{equation}
    \{\phi, \mathcal{C}^{(1)}[\alpha]\} = i \phi, \qquad \{Q, \mathcal{C}^{(1)}[\alpha]\} = - i Q.
\end{equation}
The operators $\phi e^{i \int A}$ and $Q$, which must be removed from $\aqft \otimes \mathcal{B}(\hs_{\text{charge}})$ in the full theory, fail to commute with $\mathcal{C}^{(1)}[\alpha].$
The operator $\phi e^{i \int A} Q,$ which remains in the full theory, does commute with $\mathcal{C}^{(1)}[\alpha].$
So we conclude, as claimed above, that treating a gauge theory perturbatively around a fixed configuration requires applying subleading constraints to the leading order kinematical algebras, at least if one wants to retain quasilocal operators associated to subregions.

From this point of view, our approach in gravity is incomplete, but it has the virtue of being in the right direction. In assumption~\ref{assm:aqft} we assume that the kinematical algebras $\aqft$ and $\aqft'$ built from metric fluctuations and matter commute with the leading $O(\gc^0)$ part of the constraints, at least for diffeomorphisms that have compact support inside $\sr$ and $\sr'$. This is in complete analogy with the quasilocal algebras constructed in the Maxwell theory example described above. While imposing a single constraint at subleading order is clearly not the end of the story, we expect the other constraints at $O(\gc)$ will not significantly change the structure of the algebras. Our expectation is that dressing within the subregions $\sr$ and $\sr'$ can account for $O(\gc)$ terms in the constraints that generate diffeomorphisms compactly supported in $\sr$ and $\sr'.$ As for diffeomorphisms that ``straddle'' $\sr$ and $\sr'$, like the boost constraint we impose, our expectation is that these will only lead to a richer crossed product. We discuss this along with a potential relation to gravitational edge modes in section~\ref{sec:edgemodes}.

\subsection{Fixing a region}
\label{sec:fixregion}

An important issue to address when working
with generic subregions in gravity 
is the problem of specifying the 
entangling surface in a diffeomorphism-invariant manner.  
Even in the linearized theory at $\gc = 0$, if $\partial\Sigma$ 
is not extremal in the background spacetime, linearized 
diffeomorphisms of the graviton field, $\delta_{\gc \zeta} h_{ab}
 = \lie_\zeta g_{ab}^0$, can result in $\mathcal{O}(\gc)$ changes in the 
area, which translate to large $\mathcal{O}(\gc^{-1})$ changes
in the generalized entropy.  Hence, appropriate
gauge-fixing conditions are needed to specify the surface
location at linear order.
Although we do not give a complete treatment of this issue, we outline
a set of gauge-fixing conditions that lead to sensible results 
for the entropy calculations in section \ref{sec:crossprod}.  

A convenient condition to impose at leading order in $\gc$ 
is that the quasilocal gravitational Hamiltonian $H_\xi^\Sigma$
appearing in the subregion constraint (\ref{eqn:HSigmareln})
have no contribution at first order in perturbations around
its background value.  In terms of the $\gc$ expansion, 
\beq
H_\xi^\Sigma = \frac{1}{\gc^2} (H_\xi^\Sigma)^{(-2)}
+ \frac{1}{\gc} (H_\xi^\Sigma)^{(-1)}
+ (H_\xi^\Sigma)^{(0)} + \ldots,
\eeq
$(H_{\xi}^\Sigma)^{(-2)}$ denotes the constant background value,
and $(H_\xi^\Sigma)^{(-1)}$
is the quantity we propose to set 
to zero as a gauge-fixing condition. Note that $(H_\xi^\Sigma)^{(-1)}$
is linear in $h_{ab}$ and receives no contribution from the matter fields.

One reason for choosing this condition is that 
it holds automatically when $\xi^a$ is a Killing vector of 
the background metric, such as in the de Sitter static patch
or a black
hole exterior.  The fact that $(H_\xi^\Sigma)^{(-1)}$ vanishes
identically in these cases leads to various first law
relations, as is  apparent in the Iyer-Wald formalism
\cite{wald1993black, Iyer1994}.  Applied, for example,
to the exterior region of a static black hole, 
the vanishing of $(H_\xi^\Sigma)^{(-1)}$ results in 
the first law of black hole mechanics, relating
 the first-order change in the black hole
area to the first-order change in the ADM Hamiltonian, assuming
the constraints hold.  
Since $(H_\xi^\Sigma)^{(-1)}$ is identically zero when $\xi^a$
is a Killing vector of $g_{ab}^0$, 
it does not define a gauge-fixing condition in this case; however,
the entangling surface is also extremal when $\xi^a$ is Killing, 
which suppresses the effect of linearized diffeomorphisms
in calculations of the entropy.

A context in which $(H_\xi^\Sigma)^{(-1)}=0$ does define a gauge-fixing
condition is for a causal diamond in a maximally symmetric space
\cite{Jacobson2015}.
In that case, $(H_\xi^\Sigma)^{(-1)}$ is proportional to the first 
order change in the volume of $\Sigma$.
The gauge-fixing condition thus requires that the 
radius of the ball be adjusted to compensate for metric
perturbations that change the volume.  This is representative
of the generic case, where small transverse deformations of the 
entangling surface can be used to enforce the gauge-fixing condition
$(H_\xi^\Sigma)^{(-1)} = 0$.

When enforcing this gauge condition and imposing 
the constraints, the local 
first law relation (\ref{eqn:HSigmareln}) expanded to first
order in perturbations gives
\beq
H_\text{obs}^{(-1)} = -4\kappa A^{(1)},
\eeq
where $H_\text{obs}^{(-1)}$ denotes the coefficient 
of the $\mathcal{O}(\gc^{-1})$ contribution
to $H_\text{obs}$.  
The analogous relation derived 
from (\ref{eqn:HbarSigmareln}) 
when working on $\bar{\Sigma}$ with
an asymptotic boundary reads $H_\text{ADM}^{(-1)} = 4\kappa
A^{(1)}$.  Here  there is a choice of 
whether to allow $\mathcal{O}(\gc^{-1})$ changes
in the observer energy and ADM Hamiltonian.  
If taking the perspective that $H_\text{obs}$ should 
enter at the same order as ordinary matter, the natural
condition is to set $H_\text{obs}^{(-1)} = 
H_\text{ADM}^{(-1)} = 0$, which
then fixes $A^{(1)} = 0$.  This is the perspective 
we will take in this work. However, it appears consistent to 
formally allow the observer and ADM Hamiltonian to have 
$\mathcal{O}(\gc^{-1})$ contributions, which 
appear in the entropy formulas derived in
section \ref{sec:genent} as $\mathcal{O}(\gc^{-1})$ 
contributions to the generalized entropy.
This latter
choice appears to be related to the canonical ensemble
discussed in \cite{Chandrasekaran2022b} for the AdS black
hole, since in that case the fluctuations in the area
can appear at order $\gc$.  Imposing instead that 
$H_\text{obs}$ and $H_\text{ADM}$ have no $\gc^{-1}$
contribution is then analogous to the microcanonical ensemble
of \cite{Chandrasekaran2022b}, whose corresponding 
area fluctuations are order $\gc^2$.

As a final comment, note that the condition
$(H_\xi^\Sigma)^{(-1)} = 0$
does not fully fix the location of the entangling surface;
rather, it should be viewed as a single condition determining
the overall size of the region.  
Linearized diffeomorphisms of the graviton affect the 
area at order $\gc^2$ and hence the entropy at order 
$\gc^0$, which highlights the importance of fixing the 
entangling surface location at this order. 
Although we do not treat this problem in detail in the 
present work, we offer a proposal for how this 
gauge-fixing might work.  
We first note that the entangling surface of a causal
diamond in a maximally symmetric space can be 
viewed as an extremum of the functional $A - k V$,
where $V$ is the spatial volume and $k$ is a parameter
determining the radius.  Since in this case $V$ is related
to the subregion Hamiltonian $H_\xi^\Sigma$ at first order,
this suggests that for more generic subregions,
the surface could be fixed by demanding that it 
extremize the functional $\frac{\kappa A}{8\pi G_N} +
\mathcal{V}[\xi]$, where $\mathcal{V}[\xi]$ 
is a geometric functional whose first order variation
agrees with $(H_\xi^\Sigma)^{(-1)}$.\footnote{See
\cite{Jacobson:2017hks} for an exploration of this
extremization procedure in the case of causal diamonds.}  
This extremization prescription generalizes the 
Ryu-Takayanagi procedure \cite{Ryu:2006ef, Ryu:2006bv, Hubeny:2007xt}, 
which corresponds to
$\mathcal{V}[\xi] = 0$, resulting in an extremal area
entangling surface.  
Although the details of the gauge-fixing prescription for the region do not appear to 
affect the relation between algebraic entropy and 
generalized entropy derived in section 
\ref{sec:genent}, a more careful treatment 
of this issue is an important goal for future work.

\section{Local modular Hamiltonian} \label{sec:modham}

A  key assumption that underlies many  of the results 
in this work is assumption \ref{assm:mod} from
section \ref{sec:assumptions}, asserting that 
$\Hxig$ is proportional to a modular Hamiltonian for some state
on the type $\tthr_1$ algebras $\aqft$, $\aqft'$.
As discussed in 
section \ref{subsec:perturbative-constraints}, $\Hxig$ can be constructed as a local integral
over the complete Cauchy surface $\csfull$
of the matter and graviton stress tensors 
weighted by the vector $\xi^a$.
Thus the assumption that $\Hxig$ is proportional to a modular Hamiltonian may seem at first surprising.  
Except for 
special symmetric configurations such as regions bounded 
by a Killing horizon or a conformal Killing horizon for a CFT,
vacuum modular Hamiltonians of subregions are generically
given by complicated, nonlocal expressions.  The  path
integral construction of the density matrix  for a subregion \cite{Holzhey:1994we,Calabrese:2004eu}
gives some indications as to why this is the case.  
The density matrix can be expressed as a Euclidean time-ordered
exponential of the integral of the stress tensor
\cite{Wong2013}, but 
unless this Euclidean time evolution is a symmetry so that 
the generator is conserved, the time-ordered exponential
does not reduce to a simple exponential of a local
Hamiltonian.  
This clearly precludes $\Hxig$ from being proportional to the modular Hamiltonian
for the vacuum state for most choices of subregions; however, 
the possibility remains that $\Hxig$ corresponds to the modular
Hamiltonian of some other excited state,  provided that $\xi$ approximates a boost near $\partial\Sigma$. 

In this section we discuss this assumption and collect evidence in its favor from a number of viewpoints. We begin by pointing out that the requisite
KMS states exist 
in regulated quantum field theories whose associated von Neumann algebras
are type $\tone$, a canonical example of which is lattice field theory.  
This argument extends to any algebra possessing a faithful semifinite normal trace,
and hence applies to type $\ttwo$ algebras as well.  
Going to the continuum in which the quantum field theory
algebra becomes type $\tthr$, we note that the converse of Connes's cocycle
derivative theorem provides a characterization of the set of operators
that can serve as modular Hamiltonians for this algebra.  The theorem 
suggests that given a modular flow of an arbitrary state, one can 
subtract off the nonlocal terms from its modular Hamiltonian to 
arrive at the generator of the local diffeomorphism flow, as asserted in assumption
\ref{assm:mod}.  
As a final piece of evidence, we adapt the arguments of Casini, Huerta, and Myers \cite{Casini:2011kv}, to demonstrate the existence of the proposed states for causal diamonds in flat-space conformal field theory weakly deformed by relevant operators.

\subsection{Regulated vs.\ continuum KMS states}
\label{sec:regulated-KMS}

Suppose we consider non-gravitational field theory in a lattice approximation, 
so that the algebra of operators in a subregion is type $\tone$ and carries well-defined density matrices (any other regulator producing a type $\tone$ algebra suffices 
for this argument).  Then there is a  
procedure for constructing the desired state.
We can 
split the Hamiltonian into separate local contributions from the 
subregion Cauchy surface $\Sigma$ and its complement $\bar{\Sigma}$, 
\beq \label{eqn:Hxisplit}
\Hxig = H_\xi^\Sigma - H_\xi^{\bar\Sigma}.
\eeq
We then form a  density matrix for the algebra $\aqft$
that is thermal with respect to the subregion Hamiltonian,
$\rho = e^{-\beta H_\xi^\Sigma}/Z_\xi$.  This density matrix can be used to
compute expectation values of operators  $\msf{a}\in\aqft$ by taking 
traces, $\vev{\msf{a}}_\rho = \tr(\rho\msf{a})$, and therefore defines 
a state on the algebra.  Furthermore, we can verify that the flow generated
by $\Hxig$ satisfies the KMS condition for the state defined by $\rho$.  
Defining the flowed operator $\msf{a}_z \equiv e^{iz\Hxig} \msf{a} e^{-iz\Hxig}$
where $z$ is a complex parameter, the KMS condition is the statement that
\beq\label{eqn:KMS}
\vev{ \msf{a}_s \msf{b} }_\rho = \vev{\msf{b}\msf{a}_{s+i\beta} }_\rho.
\eeq
This statement follows by noting that $\Hxig$ and $H_\xi^\Sigma$ generate
the same flow on $\msf{a}$, since $H_\xi^{\bar\Sigma}$ commutes with $\msf{a}$.  
Hence, $\msf{a}_z = \rho^{-i\frac{z}{\beta}}\msf{a}\rho^{i\frac{z}{\beta}}$,
and by expressing the expectation values in (\ref{eqn:KMS}) in terms of 
the trace, the equality of the two expressions follows by the cyclicity
of the trace.

This argument seems to imply that for any reasonable choice of Hamiltonian $H$, the associated thermal density matrix defines a state for which the flow generated by $H$ satisfies the KMS condition. The main restriction on the form of $H$ is that its spectrum be bounded below, or, equivalently,
that the correlation function $\vev{\msf{b} \msf{a}_z}_\rho$ is analytic in the strip $0\leq \Im(z) \leq \beta$.  There can also be additional physical
restrictions on $H$ when taking a continuum limit,
such as requiring the energy of the KMS state to remain
finite as the lattice spacing is taken to zero.  
Since the KMS condition for the type $\tone$ algebra
follows from the existence of a density matrix and cyclicity of the trace,
it is clear that similar arguments apply for type $\ttwo$ von Neumann algebras
as well.

However, in the continuum limit, the
algebra $\aqft$ is type $\tthr_1$, which means that the trace employed in the above argument does not actually exist. Hence it is a nontrivial task to determine if 
there is a state which satisfies the KMS condition with respect to a given flow.
Another subtlety is that the modular flow for any state $|\Phi\rangle$ on a type $\tthr_1$ algebra is an outer automorphism, which implies that its modular Hamiltonian $h_\Phi$ cannot be split into local, one-sided contributions.  Although it is common practice to formally make the split $h_\Phi = h_\Phi^\Sigma-h_\Phi^{\bar\Sigma}$, the objects $h_\Phi^\Sigma$, $h_\Phi^{\bar\Sigma}$ have UV divergent fluctuations, and hence do not even define unbounded operators.  Such divergences also occur when splitting $\Hxig$ into local contributions $H_\xi^\Sigma$, $H_\xi^{\bar\Sigma}$ as in~\eqref{eqn:Hxisplit}. The fact that the splittings of both $\Hxig$ and $h_\Phi$ exhibit similar divergences offers a clue for how one would argue that $\Hxig$ defines a valid modular Hamiltonian.  

The point is that the inability to split either operator is a UV issue,
related to the infinite entanglement between degrees of freedom 
localized close to the entangling surface.  
Modular flow is strongly constrained by the requirement that it preserve
this entanglement structure close to $\partial\Sigma$, but is largely 
unconstrained on how it acts on degrees of freedom well-separated from the
boundary.  Zooming in close to the entangling surface, the subregion
locally resembles Rindler space.  As is well-known from the results
of Bisognano and Wichmann  \cite{Bisognano1975}, the 
vacuum modular flow of a Lorentz-invariant quantum field theory 
in Rindler space coincides with the flow generated by the boost Hamiltonian.  
Hence, the natural expectation is that the modular flow of any state
will approximate a geometric flow that looks like a boost close to
the entangling surface.  

Here, we would like to employ a stronger conjecture, 
namely that {\it any} one-parameter group of automorphisms 
of $\aqft$ that looks 
like a boost near $\partial\Sigma$ (and possibly subject 
to additional restrictions close to $\partial\Sigma$)
coincides with the modular flow
of some state on $\aqft$.
Since $\Hxig$ generates such a flow,
this conjecture implies that it is proportional to the modular 
Hamiltonian of some state.
Note that the modular flow $U_{\text{mod}}(s) = \exp[i s H_{\xi}^g]$ will generally not be the same as the flow generated by the vector field $\xi^a.$
This is because when $\xi^a$ is not an isometry, the Hamiltonian 
generating the flow along $\xi^a$ is time-dependent, i.e., it depends on
the choice of Cauchy slice.  Denoting the time-dependent generator
$H(\lambda)$, the flow along the vector $\xi^a$ is given by a time-ordered
exponential $U_\xi(s) = \mathcal{T}\exp[i\int_0^s d\lambda H(\lambda)]$. 
The important point is that 
$\Hxig$ agrees with $H(\lambda)$ at $\lambda = 0$, 
and hence it generates the action
of the diffeomorphism instantaneously on the initial Cauchy surface.
This action approaches a boost near the entangling surface,
where $\xi^a$ approximates a Killing vector for the local
Rindler space.  Because of this, the effects of 
time-dependence should be suppressed near the entangling surface,
suggesting that in that region the modular 
flow approximates the local diffeomorphism flow.  

\subsection{Converse of the cocycle derivative theorem}

More evidence for this conjecture comes from the following characterization
of the space of modular Hamiltonians on a von Neumann algebra.  Suppose
$h_0$ is the modular Hamiltonian of some state $|\Phi_0\rangle$.  
Then choosing any two
Hermitian operators $\msf{a}\in \aqft$, $\msf{b}'\in\aqft'$,
the operator
\beq \label{eqn:innerrelated}
h_{\msf{a}\msf{b}'} \equiv \msf{a} + h_0 + \msf{b}'
\eeq
is a modular Hamiltonian for some other state $|\Phi_{\msf{a}\msf{b}'}\rangle$.
This fact follows from the converse of the cocycle derivative theorem
in the theory of modular automorphism groups \cite{Connes1973}, \cite[Theorem 3.8]{TakesakiII},
and is explained in more detail in appendix \ref{sec:convcoc}.  
This is 
the converse statement of the fact that any two modular Hamiltonians are 
related by operators from $\aqft$ and $\aqft'$ as in equation 
(\ref{eqn:innerrelated}), with $\msf{a}$ and $\msf{b}'$ constructed 
from Connes cocycles \cite{Connes1973, TakesakiII, Haag1992}.

The relevance of the relation (\ref{eqn:innerrelated}) to the present problem is that $\msf{a}$ and $\msf{b}'$ should be viewed as operators localized away from the entangling surface, so that adding them to the modular Hamiltonian does not  affect the flow close to $\partial\Sigma$.  Any two flows that agree near
$\partial\Sigma$ should be generated by Hamiltonians that are related as in (\ref{eqn:innerrelated}).  Since we expect all modular flows to look like a boost near $\partial\Sigma$, this then suggests that any flow that looks like a boost near $\partial\Sigma$ is the modular flow for some state. It further suggests that although a generic modular Hamiltonian will be a sum of a local integral of the stress tensor and additional multilocal contributions, the multilocal contributions should be given by operators from within the algebras $\aqft$, $\aqft'$. If this is the case, the nonlocal pieces of the modular Hamiltonian can be canceled by adding operators as in (\ref{eqn:innerrelated}), resulting in a modular Hamiltonian consisting of only a local piece that generates the boost about the entangling surface $\partial\Sigma$. 

It is possible that these ideas arguing for $\Hxig$ to be 
proportional to a modular
Hamiltonian for some state could be upgraded into a proof of the 
conjecture.  Such an investigation would be fruitful also in determining
what the precise conditions on the vector field $\xi^a$ must be.  
In section \ref{sec:outline}, we described some conditions on its behavior
within the subregions $\sr$ and $\sr'$.  Some of these just ensure 
that the flow generated by $\Hxig$ will preserve the algebras $\aqft$ and 
$\aqft'$.  The condition that $\xi^a$ have constant surface gravity $\kappa$
goes beyond this, and is likely an important point in proving that
there is a KMS state for the flow.  Constancy of $\kappa$ has an
interpretation of a zeroth law of thermodynamics, which is an equilibrium
statement about the ability to define a constant temperature for the system.
Since the KMS condition is also an equilibrium statement, it seems likely
that the requirement of constant surface gravity is an important 
characterization of generic modular flows.  It remains to be seen what
further conditions can be derived for modular flows.

\subsection{CFT and weakly deformed CFT}
\label{sec:CHM}
As a final justification for assumption 
\ref{assm:mod}, we consider a nontrivial example
in which the necessary KMS state is not the 
vacuum.  Note that
assumption~\ref{assm:mod} implicitly contains two parts. At leading order in the $\gc\to 0$ limit the generator $\Hxig$ is a sum of two terms. One is constructed from the matter stress tensor, and the other is quadratic in metric fluctuations. To show that $\Hxig$ generates a modular flow one must then show that there is a KMS state for the flow generated by $\Hxig$ both of the matter fields and of metric fluctuations. 

Finding the metric fluctuation part of such a KMS state is in principle a matter of direct computation. After all, the graviton part of $\Hxig$ is a known quadratic ``Hamiltonian'' for a free spin-2 field. We relegate it to future work and do not consider it further. 

In the rest of this section we focus on the matter part of the problem where we can draw upon existing results. There are two notable examples where  KMS states of the sort we want are known to exist. The first is in flat-space conformal field theories, where $\sr$ is the causal development of a ball $\Sigma$ of radius $R$ on the surface $t=0$. In that case Hislop and Longo \cite{Hislop1982} and Casini, Huerta, and Myers~\cite{Casini:2011kv} have shown that the CFT vacuum is in fact a KMS state of the sort we wish, with a modular Hamiltonian given by the matter part of $\Hxig$. In that case $\xi^a$ is a conformal killing vector that fixes $\sr$, with $\xi^\alpha\partial_\alpha = \frac{(R^2-(t^2+r^2))}{2R^2}\partial_t - \frac{tr}{R^2}\partial_r $
where $r$ is the distance from the center of the ball. Notably this vector field approaches a boost near the entangling surface $r=R$, and has constant surface gravity at $\partial\Sigma$, consistent with our expectations for $\xi^a$ mentioned above.
In this instance the flow generated by (the matter part of) $\Hxig$ is geometric not only at the instant in time $t=0$ where we define the KMS state, but throughout the causal development $\sr$. Shortly we will show that 
a small perturbation of this state continues to serve as a KMS state for relevant deformations that are small relative to the size of the ball.

The second instance where there are known examples of the desired KMS states is in the theory of a free 1+1-dimensional Dirac fermion where they have been constructed numerically in~\cite{Swingle:2022vie} when $\sr$ is the causal development of an interval.\footnote{Those authors also considered states of a fermion on an interval times $\mathbb{R}^{d-1}$ in $d$ spacetime dimensions, but factorized according to the momentum along the $\mathbb{R}^{d-1}$, so that the states were effectively those of a tower of fermions in 1+1-dimensions.} By~\cite{Hislop1982, Casini:2011kv} the vacuum is a KMS state, but many other states were considered where the modular flow was generated by $H_{\xi}$, characterized by an effective local temperature $\xi^t = \beta(x)$. In particular those authors found a finite energy density in the interval when $\xi$ approached a boost near $\partial\Sigma$~\cite{MvRprivate}, i.e.\ when $\beta$ vanished linearly at the edge of the interval. Moreover, for general $\beta(x)$ these states have a time-dependent stress tensor one-point function, as one would expect for a superposition of excited states. 

Now let us go beyond the case of a flat space CFT in the causal development of a ball by turning on relevant deformations. First let us recall the methods of Casini, Huerta, and Myers for an undeformed CFT. By mapping the causal development $\sr$ of a ball $\Sigma$ of radius $R$ to a Rindler wedge, they showed that the CFT vacuum, restricted to the ball, has a modular Hamiltonian
\beq
    H_{\xi} = \int d\Sigma_{\alpha}\xi_{\beta}T^{\alpha\beta} = \frac{\beta}{2R^2}\int_{r<R} d^{d-1}x (R^2-r^2)T^{tt}(x)\,, \qquad \beta = 2\pi R\,.
\eeq
Here $\xi$ is the conformal Killing vector that fixes $\sr$ and in the second equality we have written the integral over the constant time slice at $t=0$. The flat space stress tensor $T^{\mu\nu}$ can be mapped through a Weyl transformation to one $\widetilde{T}^{\mu\nu}$ on hyperbolic space  through the combination of the coordinate transformation
\begin{align}
\begin{split}
    t &= R\frac{\sinh\left( \frac{\tau}{R}\right)}{\cosh(u)+\cosh\left( \frac{\tau}{R}\right)}     \,,
    \\
    r & = R \frac{\sinh(u)}{\cosh(u) + \cosh\left( \frac{\tau}{R}\right)}\,,
\end{split}
\end{align}
followed by the Weyl rescaling $\eta_{\mu\nu} \to g_{\mu\nu}=\Omega^2\eta_{\mu\nu}$ with 
\beq
    \Omega = \cosh(u) + \cosh\left( \frac{\tau}{R}\right)\,.
\eeq
Up to a state-independent piece coming from the Weyl anomaly we have
\beq
    H_{\xi} = \beta \int d\Sigma \,\widetilde{T}_{\tau\tau}\,,
\eeq
with $d\Sigma$ the volume form on hyperbolic space. Here we have used that, at $t=\tau=0$, $T_{\tau\tau} = \Omega^d \widetilde{T}_{\tau\tau}$. The CFT vacuum restricted to $\sr$ can be mapped to the thermal state on hyperbolic space at temperature $T = \frac{1}{\beta} = \frac{1}{2\pi R}$. In particular, vacuum correlation functions in $\sr$ can be obtained from hyperbolic space where modular flow is equivalent to time translation. Since the state is thermal in hyperbolic space, this guarantees that these correlation functions obey the KMS condition with respect to the flow generated by $H_{\xi}$.

As alluded to above, these expressions for $H_{\xi}$ should not be taken too literally since we cannot split the modular Hamiltonian into a piece living entirely inside the ball and a piece outside. What is true is that vacuum correlation functions of operators inside $\sr$ can be computed on hyperbolic space at finite temperature, and so these expressions for $H_{\xi}$ should be understood as a recipe that defines an algebraic state.

Now, starting in Minkowski space, let us turn on a very small relevant deformation $\lambda$ for an operator $O$ of dimension $\Delta$. The flat space stress tensor is deformed as $T_{\mu\nu} = T^{(0)}_{\mu\nu} - \lambda O \eta_{\mu\nu}$, where $T^{(0)}_{\mu\nu}$ is the stress tensor of the undeformed CFT. By a small deformation, what we really mean is that we work in conformal perturbation theory in the coupling $\lambda$, which we expect to be valid when $|\lambda|R^{d-\Delta} \ll 1.$ We then postulate the existence of an algebraic state described by a ``modular Hamiltonian''
\beq
    H_{\xi} =\int d\Sigma_{\alpha}\xi_{\beta}T^{\alpha\beta} =  \frac{1}{2R^2} \int_{r<R} d^{d-1}x (R^2-r^2)\beta(r)T^{tt}(x)\,,
\eeq
where in addition to deforming the stress tensor we have allowed ourselves the freedom to adjust the vector field $\xi^{\alpha}$ perturbatively in $\lambda$ away from the expression above, i.e. at $t=0$,
\beq
    \xi^t = \frac{R^2-r^2}{2R^2}\beta(r)\,,
\eeq
where $\beta(r)  =2\pi R(1+\delta \beta(r))$ where $\delta\beta$ is a correction that is suppressed by powers of $\lambda$.

Now the algebraic state that corresponds to $H_{\xi}$, roughly speaking the density matrix $e^{-H_{\xi}}/\tr(e^{-H_{\xi}})$, is our candidate KMS state. We would now like to see if it satisfies two conditions. First, do correlation functions in this state respect the KMS condition? Second, is the energy in the ball finite in this state? To answer both of these questions it is convenient to map this modular Hamiltonian into one in hyperbolic space.\footnote{Weyl rescalings are not a symmetry of the deformed theory. However, up to anomalies, they equate the deformed CFT in the presence of position-dependent sources $(g_{\mu\nu},\lambda)$ to the deformed CFT in the presence of new sources $(\Omega^2 g_{\mu\nu}, \Omega^{\Delta}\lambda)$.} The map above tells us that 
\beq
\label{E:hyperbolicH}
    H_{\xi} =  \int d\Sigma \,\beta(u)\left( \widetilde{T}^{(0)}_{\tau\tau} + \lambda(u) \widetilde{O}\right)\,.
\eeq
The freedom to adjust the vector field $\xi^0$ has turned into a position-dependent temperature $\beta(u)$, while the relevant deformation has turned into an effective position-dependent coupling in the Hamiltonian 
\beq
\label{E:bdyLambda}
    \lambda(u) = \lambda \Omega^{\Delta-d} = \lambda \left( 2 \cosh^2\left( \frac{u}{2}\right)\right)^{\Delta - d}\,.
\eeq
Also $\widetilde{O}$ is the transformed version of $O$, $\widetilde{O} = \Omega^{-\Delta} O$. Note that this coupling dies off near the boundary of hyperbolic space, which corresponds to the entangling surface $\partial\Sigma$ back in Minkowski space. With the position-dependent coupling, the hyperbolic space stress tensor is $\widetilde{T}_{\tau\tau} = \widetilde{T}_{\tau\tau}^{(0)} + \lambda(u)\widetilde{O}$. Because $\beta(u) = 2\pi R (1+ \delta \beta(u))$, $\delta\beta$ then multiplies the $\tau\tau$ component of the stress tensor evaluated at $\delta \beta =0$, and so the correction proportional to $\delta\beta$ can be interpreted as a perturbation in the $\tau\tau$ component of the metric, this thermal state corresponds to a deformed CFT in the deformed geometry
\beq
\label{E:bdyMetric}
    ds^2 \approx -(1+\delta \beta(u))d\tau^2 + R^2 \left( du^2 + \sinh^2(u)d\Omega_{d-2}^2\right)\,,
\eeq
where $\tau \sim \tau - 2\pi i R$. Note that the sources, the metric and coupling $\lambda(u)$, have a time translation symmetry, and consequently there are no issues associated with time-dependence of the modular Hamiltonian. In this example, modular flow is just translation in $\tau.$\footnote{Note that $\tau$ is not the physical time inside the causal diamond. Operators off of the $t=0$ slice are related to those on the $t=0$ slice by evolution under the Minkowski space Hamiltonian, which, from the point of view of $H_{\xi}$, is generated by a $\tau$-dependent Hamiltonian. This is consistent with the observations we made at the end of subsection~\ref{sec:regulated-KMS}.}

From this last form~\eqref{E:hyperbolicH} of $H_{\xi}$ we see that $t=0$ correlation functions in this state, upon being mapped to thermal correlators in hyperbolic space, respect the KMS condition. The reason is the same one mentioned above for an undeformed CFT in the vacuum. Namely, modular flow simply acts as time translation on operators in hyperbolic space, combined with the fact that the state in hyperbolic space is thermal (albeit in a way that depends locally in space, a form of hydrostatic equilibrium). As for finite energy, we evaluate the energy density in conformal perturbation theory, and then use a Weyl rescaling to map it back to the energy density in Minkowski space.

The conformal integrals in this problem involve the vacuum CFT correlators $\langle OO\rangle$ and $\langle TOO\rangle$ and are in general intractable. For this reason we content ourselves with the asymptotic behavior of the flat space energy density near the entangling surface, or, in hyperbolic space, near the conformal boundary of hyperbolic space. To compute that behavior we consider CFTs with a holographic dual and perform the computation using the AdS/CFT dictionary. We can do that here since the correlators $\langle OO\rangle$ and $\langle TOO\rangle$ are universal in any CFT and are fixed by the dimension of $O$. 

The gravitational problem we wish to solve is that of Einstein gravity with negative cosmological constant minimally coupled to a massive scalar field $\phi$ dual to $O$, where our boundary conditions are that the metric on the conformal boundary is~\eqref{E:bdyMetric} and the source encoded in the near-boundary behavior of $\phi$ is the position-dependent coupling~\eqref{E:bdyLambda}. Setting the AdS radius to unity and introducing a formal expansion parameter $\varepsilon$ that counts powers of the original source $\lambda$, the scalar and metric profiles read
\begin{align}
    \phi & = \varepsilon \phi^{(1)} + O(\varepsilon^3)\,,
    \\
    ds^2& = -\left( \frac{r^2}{R^2}-1\right)d\tau^2 + r^2 (du^2 + \sinh^2(u)d\Omega_{d-2}^2) + \frac{dr^2}{\frac{r^2}{R^2}-1} + \varepsilon^2 h_{\mu\nu}^{(2)}dx^{\mu}dx^{\nu} + O(\varepsilon^4)\,.
    \nonumber
\end{align}
The leading order configuration is just the topological hyperbolic black hole with temperature $\frac{1}{2\pi R}$. The Klein-Gordon equation for $\phi$ and the Einstein's equations for $h_{\mu\nu}$ can be solved order by order in $\varepsilon$, with the result that $\phi^{(1)}$ solves the Klein-Gordon equation in the leading order metric, while $h_{\mu\nu}^{(2)}$ solves the linearized Einstein's equations with a stress tensor source generated by $\phi^{(1)}$, of the form
\begin{align}
    \begin{split}
        \Box^{(0)}\phi^{(1)} & = \Delta (\Delta - d)\phi^{(1)}\,,
        \\
        \mathcal{D}^{(0)}_{\mu\nu}{}^{\rho\sigma} h_{\rho\sigma}^{(2)} &= T_{\mu\nu}^{(2)}\,,
    \end{split}
\end{align}
where $\mathcal{D}^{(0)}$ is the differential operator that generates the left-hand-side of the linearized Einstein's equations and $T_{\mu\nu}^{(2)} = \partial_{\mu}\phi^{(1)}\partial_{\nu}\phi^{(1)} - \frac{1}{2}\left( g^{(0)\mu\nu}\partial_{\mu}\phi^{(1)}\partial_{\nu}\phi^{(1)} + \Delta (\Delta - d)(\phi^{(1)})^2\right)$ is the stress tensor sourced by $\phi^{(1)}$. The metric fluctuation $h^{(2)}$ encodes the boundary stress tensor of order $\lambda^2$, which is what we want to find.

To proceed we solve these equations by separation of variables. We decompose the position-dependent source for $O$ into normalizable eigenfunctions $\mathcal{U}_k(u)$ of the scalar Laplacian on hyperbolic space with eigenvalues $k(k+2-d)$. At large $u$ these eigenfunctions behave as $e^{-ku}$. That is,
\beq
    \phi^{(1)} = \sum_k \mathcal{R}_k(r) \mathcal{U}_k(u)\,,
\eeq
where the sum runs over those $k$ appearing in the decomposition of the source~\eqref{E:bdyLambda}. The
radial functions $\mathcal{R}_k(r)$ then decouple by symmetry and satisfy a radial equation. The lowest value of $k$ dominates the behavior of $\phi$ at large $u$ near the boundary of hyperbolic space and is given by $k = d-\Delta$. For that value the radial equation can be simply solved to give $\mathcal{R}_{d-\Delta}(r) \propto \lambda r^{\Delta-d}$, and it encodes the large-$u$ expectation value of $\widetilde{O}$, $\langle \widetilde{O}\rangle \propto \lambda e^{(\Delta - d)u}$. Going back to flat space we have $\langle O\rangle \propto \lambda (R-r)^{-2\left( \Delta - \frac{d}{2}\right)}$.

Now this scalar profile generates a stress tensor which in turn backreacts to create a metric fluctuation $h_{\mu\nu}^{(2)}$, which we can gauge-fix to be of the form
\beq
    h^{(2)}_{\mu\nu}dx^{\mu}dx^{\nu} = h^{(2)}_{\tau\tau}(u,r)d\tau^2 + h^{(2)}_{uu}(u,r)du^2 + h^{(2)}(u,r)\sinh^2(u)d\Omega_{d-2}^2\,.
\eeq
It is in general quite complicated to solve these linearized Einstein's equations in the presence of this source. For the moment let us dial $\delta \beta(u)=0$. As in solving the Klein-Gordon equation, it is convenient to exploit the symmetries of hyperbolic space. One can expand the metric fluctuations and stress tensor into a basis of appropriate eigenfunctions. For example the $\tau\tau$ components are scalars from the point of view of hyperbolic space and can be expanded into a basis of eigenfunctions of the scalar Laplacian, while the other components can be expanded in a basis of tensor eigenfunctions. The smallest eigenvalue appearing in $\phi^{(1)}$ fixes the smallest eigenvalues appearing in the stress tensor $T^{(2)}$. In particular, the large-$u$ behavior of the $\tau\tau$ and $uu$ components of the stress tensor is $\sim \lambda^2 e^{2(\Delta - d)u}$, while the large-$u$ behavior of the angular components is $\sim \lambda^2 e^{2(\Delta-d+1)u}$. These source the metric fluctuations $(h^{(2)}_{\tau\tau}, h^{(2)}_{uu},h^{(2)})$ to have the form $\sim \lambda^2 e^{2(\Delta - d)u}$. These in turn, generate a boundary stress tensor $\langle \widetilde{T}_{\mu\nu}\rangle$ whose $\tau\tau$ and $uu$ components scale as $\sim \lambda^2e^{2(\Delta - d)u}$, and whose angular components scale as $\sim \lambda^2 e^{2(\Delta -d + 1)u}$. Mapping back to flat space we see that this state has an energy density
\beq
\label{E:flatEnergy}
    \langle T_{tt}\rangle \sim \lambda^2(R-r)^{2\left( \frac{d}{2} - \Delta\right)}\,.
\eeq
There are corrections that, at large $u$, are suppressed relative to this leading contribution by $e^{-2nu}\sim (R-r)^{2n}$ for $n=1,2,\hdots$. This energy density remains finite for dimensions in the range $\Delta \in \left( \frac{d-2}{2},\frac{d}{2}\right)$, diverging for $\Delta > \frac{d}{2}$. However these states have finite energy in the larger range $\Delta \in \left( \frac{d-2}{2},\frac{d+2}{2}\right)$.

This is almost, but not quite what we wanted. On the one hand we have good KMS states for sufficiently relevant deformations, but on the other these states have diverging energy for large enough conformal dimension $\Delta > \frac{d+2}{2}$. This is where the freedom to tune the vector field in $H_{\xi}$ becomes useful. Let us now turn on a perturbation $\delta \beta \sim \lambda^2 e^{2(\Delta - d)u}$. (Really we mean $\lambda^2$ times the normalizable hyperbolic harmonic with $k=2(d - \Delta)$.) By the same arguments above, this gives another contribution to the energy density $\langle \widetilde{T}_{\tau\tau}\rangle \sim \lambda^2 e^{2(\Delta - d)u}$. By tuning the strength of this perturbation in $\delta \beta$ we can eliminate the contribution above, so that now the asymptotic growth of the flat space energy density is given by the first correction to~\eqref{E:flatEnergy}, going as $\langle T_{tt}\rangle \sim \lambda^2 (R-r)^{2\left( \frac{d+2}{2}-\Delta\right)}$. After this tuning the energy density is finite in the range $\Delta \in \left( \frac{d-2}{2},\frac{d+2}{2}\right)$, and the energy is finite in the range $\Delta \in \left( \frac{d-2}{2},\frac{d+4}{2}\right)$. We can keep going and tune the coefficient of a perturbation $\delta \beta \sim \lambda^2 e^{2(\Delta - d -1)u}$ to further extend the range in which the energy is finite. 

Taking stock, we see that, provided our region is the causal development of a ball at $t=0$ in flat space, we can construct a KMS state to quadratic order in conformal perturbation theory for a relevant coupling. We map this state to a generalized thermal state in hyperbolic space, implying that equal-time correlation functions on the $t=0$ slice obey the KMS condition. And, with some tuning of the vector field $\xi$ near the entangling surface, we can arrange for the state to have finite energy.

\section{Crossed product algebra}
\label{sec:crossprod}

We now have all the ingredients needed to construct the 
type $\ttwo$ von Neumann algebra associated with the 
subregion $\sr$, utilizing the 
assumptions \ref{assm:aqft}-\ref{assm:mod} described in 
section \ref{sec:assumptions}.  The point of these four 
assumptions is to reduce the construction of the subregion algebra
to the same sequence of steps as were employed by CLPW in their
construction of the algebra for the de Sitter static patch
\cite{Chandrasekaran2022a}.  
Assumption \ref{assm:aqft} asserting the existence of the 
type $\tthr_1$ algebras $\aqft$ and $\aqft'$ is no different
from the analogous statement employed by CLPW, and has been justified
by a number of recent works on large $N$ limits in holography 
\cite{Leutheusser:2021qhd, Leutheusser:2021frk,Leutheusser:2022bgi, 
Bahiru:2022oas,
Bahiru:2023zlc}.  
Assumption \ref{assm:obs} asserting the existence 
of an observer and an associated type $\tone_\infty$ algebra is 
also directly analogous to the assumption made by CLPW.  
Additional evidence in favor of this assumption will be 
described in section \ref{sec:asymp} when we consider asymptotic
dressing, but for the moment we will simply take it as a postulate
and see that it produces sensible results for the subregion entropy. 

Assumptions \ref{assm:constr} and \ref{assm:mod} 
are novel ideas employed in the present work, 
and both are necessary to lift the boost symmetry requirement
in the CLPW construction.  
Sections \ref{sec:constraints} and \ref{sec:modham}
have been devoted to justifying these assumptions, and together
they imply that a constraint must be imposed to obtain the gravitational
subregion algebra, and that the matter and graviton contribution to this constraint
generates a modular flow on $\aqft$ and $\aqft'$.

Once these assumptions have been made, the type $\ttwo$ algebra for the 
subregion follows directly from the CLPW construction \cite{Chandrasekaran2022a}.
We will repeat the analysis below, emphasizing two improvements we make to
the original work.  First, we provide an exact computation 
of the modular operators and density matrices for a natural set of states on the type $\ttwo$
algebra without assuming any semiclassical conditions on the observer's 
wavefunction.  
The semiclassical conditions are only employed in the computation
of the entropy from the density matrix, where they can be interpreted
as the statement that the observer is weakly entangled with the quantum
fields. This ensures that the observer and matter
contributions to the total entropy of the state
takes the form of a simple sum $S^\text{obs} + S^\text{QFT}$.
Second, we show that the resulting expression for the entropy can be directly
converted to a generalized entropy by employing the integrated first law of 
local subregions, equation 
(\ref{eqn:HSigmareln}), assuming the local gravitational
constraints are satisfied.  The applicability of this local
constraint is the content of assumption \ref{assm:1stlaw}
of section \ref{sec:assumptions}.  
Applying this first law
avoids having to consider the dynamics of the 
subregion horizon as was done
in \cite{Chandrasekaran2022a, Chandrasekaran2022b}, 
and allows the entropy to be expressed in terms of 
instantaneous quantities on the subregion Cauchy surface $\Sigma$.  

The analysis in this section will be done for a bounded subregion
within a spacetime with an asymptotic boundary, so that $\sr$ 
is bounded and $\sr'$ is unbounded.  The other cases described in
 section \ref{sec:nonlinear-constraints} can be handled analogously.  
The distinction between bounded and unbounded only appears in
deriving the consequences of assumption \ref{assm:bdenergy},
which demands
that the observer degree of freedom have energy that 
is bounded below.  For a bounded subregion, 
this results in
an algebra of type $\ttwo_1$, while for an unbounded subregion
it yields a type $\ttwo_\infty$ algebra.  The implications
of these different algebra types for the bounded and unbounded 
cases are explored in sections \ref{sec:encond} and \ref{sec:asymp}.

Throughout this section, we  extensively employ results from
modular theory for von Neumann algebras.  
We encourage the reader to refer to appendix \ref{app:modth} 
for a brief overview of this topic and an explanation of the notation
employed for the various modular operators appearing in this section.
For more general background on von Neumann algebras and 
crossed products, see appendix \ref{app:algebraic-background}.

\subsection{Crossed product from constraints}
\label{sec:crossconstr}
As outlined in section \ref{sec:outline}, the starting point is the type $\tthr_1$
algebra $\aqft$ describing the quantum field theory degrees of freedom
of matter fields and gravitons in the bounded 
subregion $\sr$.  These operators
act on a Hilbert space $\hs_\text{QFT}$, and,
assuming Haag duality, the commutant 
algebra $\aqft'$ acting on this Hilbert space describes
the quantum field theory degrees of freedom in the causal
complement $\sr'$, which is assumed to include
asymptotic boundaries.  
In addition, the auxiliary observer degree of freedom
acts on a Hilbert space $\hs_\text{obs} = L^2(\mathbb{R})$, and its
algebra is taken to be the algebra of all bounded 
operators on this Hilbert space
$\alg_\text{obs} = \mathcal{B}(\hs_\text{obs})$.  
Hence, the total Hilbert space for the 
subregion algebra and its commutant is $\hs_{\sr}=\hs_\text{QFT}\otimes \hs_\text{obs}$,
and the kinematical algebra before imposing the gravitational
constraint is $\aqft\otimes\alg_\text{obs}$.  In particular, $\alg_\text{obs}$
is assumed to commute with $\aqft$; it is only after imposing the 
constraint that these degrees of freedom become noncommuting.

The gravitational constraint to impose is given by \eqref{eqn:ham-constraint-equation},
which we reproduce here,
\beq \label{eqn:Cfull}
\mathcal{C}[\xi] = \Hxig + H_\text{obs} + H_\text{ADM}.
\eeq
$\Hxig$ is an operator
acting on $\hs_\text{QFT}$ that generates 
the action of the boost 
vector $\xi^a$ on the algebras $\aqft$, $\aqft'$ infinitesimally
near the subregion Cauchy surface $\Sigma$.  
The observer Hamiltonian acts on a single-particle
Hilbert space $\hs_\text{obs}$, 
and hence the simplest choice for this Hamiltonian
is 
 $H_\text{obs} = \hat{q}$. 
The final component of the constraint is the ADM Hamiltonian.  
Since $H_\text{ADM}$ is assumed to commute with the kinematical algebras
$\aqft$, $\aqft'$, and $\alg_\text{obs}$, it should be represented
as an operator acting on a separate Hilbert space $\hs_{\text{ADM}}$,
which is tensored with $\hs_{\sr}$. This is directly
analogous to the setup considered by CLPW when including an observer
in the complementary patch of de Sitter \cite[section 4.2]{Chandrasekaran2022a},
which was argued to be necessary when imposing the constraint
at the level of the Hilbert space.  The result of the CLPW analysis
is that the constrained Hilbert space can be mapped to 
the subregion Hilbert  space $\hs_\sr$, and the algebra of 
operators commuting with $\mathcal{C}[\xi]$ can be taken to act on  
$\hs_\sr$.  

Since $H_\text{ADM}$  commutes with $\aqft$ and $\alg_\text{obs}$,
its presence in the constraint $\mathcal{C}[\xi]$ does not affect the construction
of the subregion algebra when representing it on $\hs_\sr$.  Hence,
in terms of operators acting on $\hs_\sr$, the subregion algebra 
can be characterized as the algebra 
$\ac$ of operators that commute with 
$\mathcal{C} = \Hxig + H_\text{obs}=\Hxig + \hat{q}$, 
as well as with $\aqft'$.
The simplest such operator is $\hat{q} = H_\text{obs}$.  The remaining
operators are constructed as dressed versions of the operators 
in $\aqft$.  Defining the momentum operator $\hat{p} = -i\frac{d}{dq}$,
it is straightforward to see that $\mathcal{C}$ 
commutes with operators of the form $e^{i\hat{p}\Hxig} \msf{a}
e^{-i\hat{p}\Hxig}$ with $\msf{a}\in\aqft$.
As explained in appendix \ref{app:algebraic-background}, 
these operators generate 
the full algebra $\ac$, in that we have 
\beq \label{eqn:AC}
\ac = \{e^{i\hat{p}\Hxig} \msf{a}e^{-i\hat{p}\Hxig}, 
e^{i\hat{q}t}|\msf{a}\in\aqft, t\in\mathbb{R} \}'',
\eeq
where we recall that  $''$ denotes taking the double 
commutant of the operators appearing in this set.
Note that this algebra contains arbitrary bounded functions of $\hat{q}$,
although since $\hat{q}$ itself is unbounded, it is only an operator affiliated with 
$\ac$.\footnote{An unbounded operator is said to be 
{\it affiliated} with the algebra $\ac$ 
if every bounded function of that operator
is in $\ac$, equivalently, if it commutes with every operator in $(\ac)'$.
See for example \cite[remark 5.3.10]{pedersen1979c}.  For brevity, we will occasionally
say an unbounded operator affiliated with $\ac$ is in $\ac$.}
Appendix \ref{app:algebraic-background} 
also explains that this algebra is 
identified as
the crossed product of 
$\aqft$ by the flow generated by $\Hxig$.  
Since $\ac$ was realized as a commutant of the set of operators
$\{\cloc, \aqft'\}$, these operators generate the commutant 
algebra $(\ac)'$.
Hence the commutant algebra is immediately identified as
\beq\label{eqn:AC'}
(\ac)' = \{\msf{b}', e^{i(\Hxig + \hat{q})s}|\msf{b}'\in\aqft',
s\in\mathbb{R}\}''.
\eeq
Here we note that $\mathcal{C} = \Hxig + \hat{q}$ is an operator
affiliated with  $(\ac)'$.
This operator should be identified with $-H_\text{ADM}$ in
order to realize the full constraint (\ref{eqn:Cfull}) as
the trivial operator $0$ when acting on $\hs_\sr$.  This 
identification is derived in \cite{Chandrasekaran2022a} (with
$-H_\text{ADM}$ replaced with $H_\text{obs}'$, the observer Hamiltonian
in the complementary patch of de Sitter space)
through a proper treatment of the constraint in terms of a simple BRST complex.

We now apply assumption \ref{assm:mod} to identify the 
flow generated by $\Hxig$ with the modular flow of some state 
$|\Psi\rangle\in\hs_{\text{QFT}}$.
This implies that the modular Hamiltonian for $|\Psi\rangle$
is given by
\beq
h_\Psi = \beta \Hxig, \qquad \beta = \frac{2\pi}{\kappa}.
\eeq
The inverse temperature $\beta$ is determined in this relation
by matching to the Unruh temperature upon zooming in close to 
the entangling surface \cite{Unruh:1976db}, where the flow approaches 
a local Rindler boost.  Since this then implies that $\ac$
is the crossed product of a type $\tthr_1$ algebra
by its modular automorphism group, 
we conclude that 
it is a type $\ttwo_\infty$ von Neumann factor, as explained in 
appendix \ref{app:algebraic-background} and \cite{Takesaki1973, 
Witten2021}.

\subsection{Modular operators and density matrices}
\label{sec:moddensity}

One of the most useful features of a type $\ttwo_\infty$ algebra
is that any modular operator factorizes into a product of an operator
affiliated with
the algebra and an operator affiliated with the commutant.
This coincides
with the fact that modular flow is an inner automorphism
for type $\ttwo$ algebras, and hence is generated by an
element in the algebra (see e.g. \cite[chapter V.2.4]{Haag1992}).  
This is in stark contrast 
with type $\tthr_1$ algebras, where modular flow is 
an outer automorphism, and, as discussed in
section \ref{sec:modham}, any attempt to split a modular
Hamiltonian into an element of the algebra and an element
of the commutant leads to UV divergences.

The factorization of the modular operator gives rise to two 
interconnected properties of the von Neumann algebra: the existence
of a trace and the existence of density matrices. These 
properties are discussed in detail in \cite{Witten2021}; here, we 
briefly review how they arise.  
Consider a cyclic-separating 
state $|\wh{\Phi}\rangle\in \hs_{\sr}$ whose modular 
operator $\Delta_{\wh{\Phi}}$ for $\ac$ factorizes
according to 
\beq \label{eqn:Delfactor}
\Delta_{\wh{\Phi}} =
\rho_{\wh{\Phi}}(\rho_{\wh{\Phi}}')^{-1},
\eeq
with $\rho_{\wh{\Phi}} \in
\ac$ and $\rho_{\wh{\Phi}}'\in(\ac)'$.  The operator $\rho_{\wh\Phi}$ 
can be used to construct a trace on the algebra, given by
\beq\label{eqn:trdefn}
\atr(\ho a) = \langle \wh{\Phi}|\, \rho_{\wh{\Phi}}^{-1}\ho a\,|\wh\Phi\rangle,
\eeq
for which the cyclicity property $\atr (\ho a\,\ho b) = \atr(\ho b\, \ho a)$
follows straightforwardly from standard identities for the modular
operator.
This definition of the trace also immediately
implies that $\rho_{\wh{\Phi}}$ functions as a density matrix,
since it is a Hermitian, positive operator in $\ac$ satisfying 
\beq
\atr(\rho_{\wh{\Phi}} \ho a) = \langle\wh\Phi|\,\ho a\,|\wh{\Phi}\rangle,
\qquad
\atr(\rho_{\wh{\Phi}}) = 1.
\eeq

A subtlety associated with the above definitions of $\rho_{\wh{\Phi}}$ and $\atr$ is that the factorization property 
(\ref{eqn:Delfactor}) 
defines the density matrices only up to rescaling
$\rho_{\wh{\Phi}}\rightarrow
\lambda\rho_{\wh{\Phi}}$, $\rho_{\wh{\Phi}}' \rightarrow
\lambda\rho_{\wh{\Phi}}'$, 
where $\lambda$ is an element of the center of the von Neumann algebra.
Since $\ac$ is a factor, this coincides with a constant
rescaling ambiguity for $\rho_{\wh{\Phi}}$, and an associated rescaling ambiguity for the trace defined by (\ref{eqn:trdefn}).
Importantly, the rescaling ambiguity is state-dependent, since the density matrices $\rho_{\wh{\Phi}}$ for different states $|\wh{\Phi}\rangle$ can be rescaled independently.
Since this would lead to a state-dependent additive ambiguity in the entropy $S(\rho_{\wh{\Phi}}) = \langle \wh{\Phi}| - \log \rho_{\wh{\Phi}} |\wh{\Phi}\rangle$, it is important to resolve.
An easy way to fix the relative normalization of density matrices for different states is to use the fact, reviewed in appendix \ref{app:algebraic-background}, that the trace defined by equation \eqref{eqn:trdefn} for any state $|\wh{\Phi}\rangle$ is faithful, normal, and semifinite, and that any two traces with these properties are related by a constant factor.
So to resolve the state-dependent normalization ambiguity in $\rho_{\wh{\Phi}}$ into a single state-independent ambiguity, we simply choose the normalization of $\rho_{\wh{\Phi}}$ so that equation \eqref{eqn:trdefn} is independent of $|\wh{\Phi}\rangle$.
In practice, a straightforward way of doing this
is to choose a fixed operator $\ho e\in\ac$ with finite trace,
and normalize the density matrices $\rho_{\wh{\Phi}}$ so that 
the trace of $\ho e$, as defined by (\ref{eqn:trdefn}), is unity:
\beq
\atr (\ho e) = \langle\wh{\Phi}|\, \rho_{\wh{\Phi}}^{-1}\ho e\, |\wh{\Phi}
\rangle = 1.
\eeq
This allows density matrices for different states to be compared in a meaningful way by reducing the normalization ambiguity to a single choice, which is the normalization of $\wh{\msf{e}}.$
Equivalently, one can fix a choice of trace $\wh{\tr},$ then normalize all the density matrices $\rho_{\wh{\Phi}}$ to satisfy $\wh{\tr}(\rho_{\wh{\Phi}}).$

To demonstrate the factorization of the modular operator for the algebra
$\ac$ explicitly, we  consider a 
class of classical-quantum states  $|\wh\Phi\rangle\in
\hs_\sr$, defined 
as  tensor product states of the form
\beq
|\wh\Phi\rangle =|\Phi,f\rangle \equiv |\Phi\rangle \otimes f(q)
\eeq
where $|\Phi\rangle$ is a state in $\hqft$ and $f(q)$ is a normalized
wavefunction for a state in $\hs_\text{obs} = L^2(\mathbb{R})$.
The modular operator $\Delta_{\wh\Phi}$ for this state 
is determined by the relation 
\beq
\langle \wh{\Phi}|\,\ho a\,\ho b\, |\wh\Phi\rangle
=
\langle \wh{\Phi}|\,\ho b\, \Delta_{\wh{\Phi}}\, \ho a \,|\wh{\Phi}\rangle
\eeq
for any two 
operators $\ho{a}, \ho b \in \ac$.  We can determine 
$\Delta_{\wh\Phi}$ by solving this relation on an
additive basis of algebra elements of the form
$\ho a = e^{i\hat{p}\hb} \msf{a} e^{-i\hat{p}\hb} e^{iu\hat{q}}$,
$\ho b = e^{i\hat{p}\hb} \msf{b} e^{-i\hat{p}\hb} e^{iv\hat{q}}$.
The calculation is somewhat involved, so the details 
are presented in appendix \ref{app:modcomp}.  
A technical tool
that facilitates the computation is to assume
that the state $|\Phi\rangle$ for the quantum
field degrees of freedom is canonically purified with 
respect to the KMS state $|\Psi\rangle$.  This implies 
that it is fixed by the modular conjugation operation
$J_\Psi |\Phi\rangle = |\Phi\rangle$, and furthermore that 
$\Delta_{\Phi|\Psi}^{\frac12}|\Psi\rangle = |\Phi\rangle$,
where $\Delta_{\Phi|\Psi}$ is the relative modular operator
of the states $|\Phi\rangle$, $|\Psi\rangle$ for the 
algebra $\aqft$.
However, the assumption of canonical purification is not strictly necessary, and we present expressions for general states $\ket{\Phi}$ at the end of this subsection.
See appendix \ref{app:modth} for additional
details on modular theory and canonical purifications.

With the assumption that $|\Phi\rangle$ is canonically purified,
the factors of the modular operator $\Delta_{\wh{\Phi}}$ are given by 
\begin{align}
\rho_{\wh{\Phi}} &= \frac1\beta e^{i\hat p\hb}f\Big(\hat{q}-\hb\Big)e^{\frac{\beta\hat q}{2}}
\Delta_{\Phi|\Psi}e^{\frac{\beta\hat q}{2}}f^*\Big(\hat{q}-\hb\Big) e^{-i\hat{p}\hb} 
\label{eqn:rhoPhi}
\\
\rho_{\wh{\Phi}}' &=\frac1\beta \Delta_{\Psi|\Phi}^{-\frac12}
e^{\frac{\beta\hat q}{2}}
\,\Big|f\Big(\hat{q}+\hb\Big)\Big|^2e^{\frac{\beta\hat q}{2}}
\Delta_{\Psi|\Phi}^{-\frac12}.
\label{eqn:rhoPhi'}
\end{align}
To see  that $\rho_{\wh{\Phi}}$ is in $\ac$,
 we  move the factors of $e^{i\hat{p}\hb}$ 
inward in the expression (\ref{eqn:rhoPhi}) using the 
relation $e^{i\hat{p}\hb} g(\hat q)e^{-i\hat{p}\hb} = 
g(\hat{q}+\hb)$, 
allowing $\rho_{\wh\Phi}$ to equivalently be expressed as
\beq \label{eqn:equivrhophi}
\rho_{\wh\Phi} = \frac1\beta f(\hat{q}) e^{\frac{\beta\hat{q}}{2}}
e^{i\hat{p}\hb}\Delta_\Psi^{-\frac12}\Delta_{\Phi|\Psi}
\Delta_\Psi^{-\frac12}
e^{-i\hat{p}\hb}
e^{\frac{\beta\hat{q}}{2}}f^*(\hat{q}),
\eeq
where $\Delta_\Psi = e^{-h_\Psi}$ is the modular operator 
for the KMS state $|\Psi\rangle$.
Although neither $\Delta_\Psi$ nor $\Delta_{\Phi|\Psi}$ is an element
of $\aqft$, the product 
$\Delta_\Psi^{-\frac12}\Delta_{\Phi|\Psi}\Delta_{\Psi}^{-\frac12}$
is in $\aqft.$
This can be checked by formally expressing the modular operators in terms of factorized density matrices $\Delta_{\Phi|\Psi} = \rho_\Phi\otimes (\rho_\Psi')^{-1}$, $\Delta_{\Psi} = \rho_\Psi\otimes
(\rho_\Psi')^{-1}$, with a more rigorous argument given in
the discussion surrounding equation \eqref{eqn:affiliated-operators-cocycles} of appendix \ref{app:modth}.
Hence, the expression (\ref{eqn:equivrhophi}) is explicitly a 
product of elements of $\ac$.  Similarly, we can express
$\rho_{\wh{\Phi}}'$ as
\beq\label{eqn:equivrhophi'}
\rho_{\wh{\Phi}}' = \frac1\beta
\Delta_{\Psi|\Phi}^{-\frac12} \Delta_\Psi^{\frac12}e^{\beta \hat q + h_\Psi}
\,\Big|f\Big(\hat{q}+\hb\Big)\Big|^2
\Delta_\Psi^{\frac12}\Delta_{\Psi|\Phi}^{-\frac12}.
\eeq
Again applying the density matrix expression for the modular
operators or the arguments leading to equation (\ref{eqn:affiliated-operators-cocycles}),
one sees that  $\Delta_{\Psi|\Phi}^{-\frac12} \Delta_\Psi^{\frac12}$
and $\Delta_{\Psi}^\frac12\Delta_{\Psi|\Phi}^{-\frac12}$
are in $\aqft'$.
Thus, (\ref{eqn:equivrhophi'}) expresses
$\rho_{\wh{\Phi}}'$ manifestly as a product of 
elements of $(\ac)'$.

The normalization ambiguity in both $\rho_{\wh{\Phi}}$ and 
its associated trace $\atr$ defined by
(\ref{eqn:trdefn}) can be resolved by requiring
$\atr( \Pio) =1$, where $\Pio = \Theta(\hat{q})$,
with $\Theta$ the Heaviside step function.  This choice will
prove convenient later in section \ref{sec:encond} when projecting 
to the  type $\ttwo_1$ subalgebra associated with positive
observer energy.  The calculation of $\atr(\Pio)$ directly
from (\ref{eqn:trdefn}) is messy, but because any (faithful, semifinite,
normal) trace is equivalent up to rescaling, we can instead
use the definition of the trace defined  in appendix \ref{app:algebraic-background}, 
given by
\beq\label{eqn:atrdefn}
\atr(\ho a) = 2\pi \beta \langle \Psi|\langle0|_p
e^{-\frac{\beta\hat{q}}{2}}\, \ho a\, e^{-\frac{\beta\hat{q}}{2}}
|0\rangle_p|\Psi\rangle,
\eeq
where $|0\rangle_p$ is the zero momentum eigenstate.  It is straightforward
to check that this trace is normalized to satisfy $\atr(\Pio) = 1$.
It will therefore define the same trace as (\ref{eqn:trdefn})
as long as $\atr(\rho_{\wh{\Phi}}) = 1$.  
Using the identity $h_\Psi|\Psi\rangle = 0$, one verifies that
\begin{align}
\atr(\rho_{\wh{\Phi}}) 
& = 
2\pi \langle \Psi|\langle 0|_p f(\hat{q})
\Delta_{\Phi|\Psi}  f^*(\hat{q})|0\rangle_p|\Psi\rangle
=
\int dy |f(y)|^2\langle \Psi|\Delta_{\Phi|\Psi}|\Psi\rangle = 
\langle \Phi|\Phi\rangle =1.
\end{align}
Hence, the constant prefactors in (\ref{eqn:rhoPhi}) and 
(\ref{eqn:rhoPhi'}) have been chosen
consistently to yield density matrices that are normalized 
in a state-independent way.  

Finally, the density matrices for the most general classical quantum
state $|\wh{\Phi}\rangle = |\Phi,f\rangle$ where $|\Phi\rangle$
is not assumed to be canonically purified can be obtained 
from (\ref{eqn:rhoPhi}) and (\ref{eqn:rhoPhi'}) from the observation
that any such state can always be expressed as $|\Phi\rangle
 = \msf{u}'|\Phi_c\rangle$, where $|\Phi_c\rangle$ is a canonically
purified, and $\msf{u}'$ is a unitary in $\aqft'$
\cite{Araki1974}.  Since
$\msf{u}'\in(\ac)'$, we see that $|\wh{\Phi}\rangle$
is obtained from a canonically purified classical-quantum
state $|\wh{\Phi}_c\rangle = |\Phi_c,f\rangle$ by the action of a unitary from
the commutant algebra
$(\ac)'$.  The modular operators for the two states are then 
related by a simple conjugation,
\beq
\Delta_{\wh{\Phi}} = \msf{u'}\Delta_{\wh{\Phi}_c}(\msf{u}')^\dagger
\eeq
implying the relation for the density matrices
\beq\label{eqn:conjrho'}
\rho_{\wh{\Phi}} = \rho_{\wh{\Phi}_c}, \qquad \rho_{\wh{\Phi}}'
 = \msf{u}' \rho_{\wh{\Phi}_c}' (\msf{u}')^\dagger.
\eeq
The unitary $\msf{u}'$ is given explicitly in terms of modular
conjugations (see appendix \ref{app:modth}),
\beq
\msf{u}' = J_{\Phi|\Psi} J_\Psi.
\eeq
In addition, note that the relative modular operators
are related to the canonically purified versions according
to
\beq \label{eqn:DelPsiPhiconj}
\Delta_{\Phi|\Psi} = \Delta_{\Phi_c|\Psi},\qquad
\Delta_{\Psi|\Phi} = \msf{u}'
\Delta_{\Psi|\Phi_c}(\msf{u}')^\dagger.
\eeq
Because of this, the density matrix on the algebra $\ac$
is unchanged, and the assumption that $|\Phi\rangle$
is canonically purified has no effect on
computations related to $\ac$, such as expectation
values of operators, entropies, or other quantum
information quantities constructed from the 
density matrix.  

For the commutant algebra $(\ac)'$, the full expression
for the density matrix when $|\Phi\rangle$ is not a canonical
purification follows from  (\ref{eqn:rhoPhi'}), 
(\ref{eqn:conjrho'}), and (\ref{eqn:DelPsiPhiconj}),
which lead to
\beq \label{eqn:rho'text}
\rho_{\wh{\Phi}}' = \frac1\beta \Delta_{\Psi|\Phi}^{-\frac12}
J_{\Phi|\Psi} J_\Psi e^{\frac{\beta\hat{q}}{2}}
\Big|f\Big(\hat{q}+\hb\Big)\Big|^2e^{\frac{\beta\hat{q}}{2}}
J_\Psi J_{\Psi|\Phi} \Delta_{\Psi|\Phi}^{-\frac12}.
\eeq
The factors of $J_{\Phi|\Psi}J_\Psi$
and $J_\Psi J_{\Psi|\Phi}$ can have an effect on the entropy
computation,
but we will argue in section \ref{sec:asymp}
that these terms drop when imposing a semiclassical
assumption on the observer wavefunction $f$.

\subsection{Generalized entropy} \label{sec:genent}

Having obtained the density matrix (\ref{eqn:rhoPhi}) for the 
classical-quantum state $|\wh{\Phi}\rangle$, the next task is 
to compute the entropy for this state and relate
it to the generalized entropy for the subregion $\sr$.
This involves computing the logarithm of $\rho_{\wh{\Phi}}$,
which is a nontrivial task since $\Delta_{\Phi|\Psi}$ 
does not commute with the other operators $f(\hat{q}-\hb)$, 
$f^*(\hat{q}-\hb)$ appearing in the expression
for the density matrix.  However, following 
\cite{Chandrasekaran2022a, Chandrasekaran2022b}, 
we can simplify this computation by making a 
semiclassical assumption on the
observer wavefunction $f(q)$ 
by requiring that it be slowly varying,
so that commutators of the form $[\Delta_{\Phi|\Psi},f(\hat{q}-\hb)]$
are suppressed, being proportional to the derivative of $f(q)$.
This assumption amounts to requiring that
the observer is not strongly entangled
with the quantum field degrees of freedom.  To see this,
note that although the classical-quantum state
$|\Phi\rangle\otimes f(q)$ naively looks unentangled, 
the quantum field operators that act on this state
involve a conjugation by $e^{i\hat{p}\hb}$,
as seen in equation (\ref{eqn:AC}).  
This conjugation generates entanglement between the observer
degree of freedom and the quantum field algebra whenever $|\Phi\rangle$
has nonzero modular energy.
Since the operator $h_{\Psi} \otimes \hat{p}$ appears in the conjugation, a state is only unentangled if it is a zero eigenfunction of either $\hat{p}$ or $h_{\Psi}.$
Zero eigenfunctions of $\hat{p}$ are not normalizable, so any normalizable state  in $\hs_\sr$ either involves entanglement between the quantum field operators and the observer degree of freedom, or has zero modular energy.
For example, any state of the form $\ket{\Psi} \otimes \ket{f}$ has zero modular energy due to the condition $h_{\Psi} \ket{\Psi} = 0.$
For states with nonzero modular energy, the point of the semiclassical assumption is to reduce the entanglement
as much as possible in order to be able to treat the observer and the quantum
fields independently in their contributions to the entropy of the state.

Having made this restriction on $f(q)$, we can compute $\log \rho_{\wh{\Phi}}$ 
by retaining only the leading terms in the Baker-Campbell-Hausdorff 
expansion of the logarithm.  
Noting that the factors of $e^{\pm i\hat{p}\hb}$ in (\ref{eqn:rhoPhi}) can be moved outside
the logarithm, and recalling that the 
relative modular Hamiltonian is defined by $h_{\Phi|\Psi}
 = -\log\Delta_{\Phi|\Psi}$, 
 we obtain 
\begin{align}
-\log\rho_{\wh{\Phi}}
&=
e^{i\hat{p}\hb}\left(
h_{\Phi|\Psi}-\beta\hat{q} 
-\log\Big|f\Big(\hat{q}-\hb\Big)\Big|^2+\log\beta
\right) e^{-i\hat{p}\hb}
\label{eqn:logrholine1}
\\
&=
e^{i\hat{p}\hb}(h_\Phi-h_{\Psi|\Phi} + h_\Psi)e^{-i\hat{p}\hb}
-h_\Psi -\beta\hat{q} - \log|f(\hat{q})|^2 +\log\beta
\\
&=
e^{i\hat{p}\frac{h_{\Psi|\Phi}}{\beta}} h_\Phi e^{-i\hat{p} \frac{h_{\Psi|\Phi}}{\beta}}
-h_{\Psi|\Phi} -\beta\hat q - \log |f(\hat{q})|^2 +\log\beta
\label{eqn:logrho}
\end{align}
Arriving at the second line requires the modular Hamiltonian identity
$h_{\Phi|\Psi}-h_\Psi = h_\Phi - h_{\Psi|\Phi}$, which follows from the 
two equivalent definitions of the Connes cocycle $u_{\Phi|\Psi}(s)$
(see appendix \ref{app:modth}).  The third line uses the fact that 
$h_{\Phi} - h_{\Psi|\Phi}$ is an operator affiliated with
$\aqft$, being proportional to the derivative of $u_{\Phi|\Psi}(s)$ at $s=0$, 
and the fact that the flow generated by $h_{\Psi}$ agrees with the 
flow generated by $h_{\Psi|\Phi}$ on elements of $\aqft$.

The entropy is then defined by the expectation value 
\beq\label{eqn:Sdefn}
S(\rho_{\wh{\Phi}}) = \langle\wh\Phi|-\log\rho_{\wh\Phi} |\wh\Phi\rangle = -\atr\left(\rho_{\wh{\Phi}} \log\rho_{\wh\Phi}\right).
\eeq
When evaluating this, the only complicated term
in (\ref{eqn:logrho})
is the first one, due to the conjugation by $e^{i\hat{p}\frac{h_{\Psi|\Phi}}{\beta}}$.
However, when this operator acts on the state $|\Phi,f\rangle$, it has a negligible 
effect due to the assumption that $f(q)$ is slowly varying, which
equivalently means its Fourier transform
is sharply peaked around zero momentum.
Hence, this assumption implies the approximation
\beq \label{eqn:twirlhPhi}
\langle\Phi,f| e^{i\hat{p}\frac{h_{\Psi|\Phi}}{\beta}}
h_\Phi e^{i\hat{p}\frac{h_{\Psi|\Phi}}{\beta}}
|\Phi,f\rangle \approx \langle \Phi |h_\Phi |\Phi\rangle.
\eeq
Of course, this expression is actually zero since $h_\Phi|\Phi\rangle = 0$,
but keeping it in the expression of the entropy makes it clear that 
only operators in the subregion algebra $\ac$ appear.
Note that because $e^{i\hat{p}\hb}$ produces a small change
in the state $|\Phi,f\rangle$ under the semiclassical
assumption, it is tempting to instead use
equation (\ref{eqn:logrholine1}) directly when evaluating 
the entropy, and simply drop the factors of $e^{i\hat{p}\hb}$ 
when taking the expectation value in (\ref{eqn:Sdefn}).
The issue with this, as explained in
\cite[section 3.2]{Chandrasekaran2022b},
is that the operator $\hat{q}$ has 
large fluctuations when $f$ is slowly varying,
so that a small change in the state can produce a order $1$
change in the expectation value of $\hat{q}$.  This argument 
does not apply to the expectation value of $h_\Phi$, 
which is why it is justified to drop the twirling 
factors in (\ref{eqn:twirlhPhi}), but not when 
taking an expectation value of $\hat{q}$.

The  expectation values to evaluate in the computation of the entropy 
are therefore given by
\begin{align}
\langle\wh{\Phi}|h_\Phi-h_{\Psi|\Phi}|\wh{\Phi}\rangle
&=
\langle\Phi|h_\Phi-h_{\Psi|\Phi}|\Phi\rangle = -S_{\text{rel}}(\Phi||\Psi) 
\label{eqn:Srel}
\\
\langle\wh{\Phi}|-\beta\hat{q}|\wh{\Phi}\rangle 
&= 
-\int_{-\infty}^\infty
dq |f(q)|^2 \beta q
= -\beta\vev{H_\text{obs}}_f
\\
\langle\wh{\Phi}|-\log|f(\hat{q})|^2|\wh{\Phi}\rangle
&= -\int_{-\infty}^\infty dq |f(q)|^2 \log|f(q)|^2
=S_\text{obs}^f
\end{align}
where (\ref{eqn:Srel}) employs Araki's expression for the relative entropy, $S_\text{rel}(\Phi||\Psi) = \langle\Phi|h_{\Psi|\Phi}
|\Phi\rangle = \langle\Phi|h_{\Psi|\Phi} - h_\Phi|\Phi\rangle$
\cite{Araki1975}.  
This results in the following expression for the entropy of the classical-quantum
state $|\wh\Phi\rangle$,
\beq \label{eqn:SrhoSrelobs}
S(\rho_{\wh\Phi}) = -S_\text{rel}(\Phi||\Psi) -\beta\vev{H_\text{obs}}_f
+S_\text{obs}^f +\log\beta.
\eeq
Since both $|\Phi\rangle$ and $|\Psi\rangle$ are cyclic and separating, 
the relative entropy in this expression is finite, and hence we find that the 
entropy $S(\rho_{\wh\Phi})$ computed in the type $\ttwo$ algebra is finite.

The expression (\ref{eqn:SrhoSrelobs}) 
agrees with the analogous result in \cite{Chandrasekaran2022b}
for the AdS black hole,
equation (3.7), upon identifying our $H_\text{obs}$ with $-h_R$ in their equation.
This is the expected identification when working with an asymptotic boundary
instead of a local subregion, as discussed in section
\ref{sec:nonlinear-constraints}.  (\ref{eqn:SrhoSrelobs})
differs from the expression given by CLPW, equation (3.18) in \cite{Chandrasekaran2022a},
by a term involving $\langle \Phi|h_\Psi|\Phi\rangle$.
We believe this occurs due to an inaccuracy in \cite{Chandrasekaran2022a}
in computing the density matrix for the state under consideration,
as explained in appendix \ref{app:modcomp}.
However, we note that this term does not seem to contribute 
significantly to the final expression for the entropy as a generalized
entropy, possibly due to the semiclassical assumptions for the observer
wavefunction.

Having obtained the expression (\ref{eqn:SrhoSrelobs}) for the entropy of the 
state $|\wh{\Phi}\rangle$, we next would like to demonstrate that it agrees with 
the generalized entropy for the subregion up to a state-independent constant.
This involves rewriting the relative entropy in terms of  the entropy of the quantum
fields restricted to the subregion $\sr$.  This will obviously introduce UV divergent 
quantities into the expression for $S(\rho_{\wh{\Phi}})$, but given that the starting 
expression is UV finite, such divergent quantities will always appear in combinations
in which the  divergences cancel.  We start by introducing the one-sided density matrices $\rho_\Phi$ and $\rho_\Psi$ for the quantum field degrees of freedom
in the states $|\Phi\rangle$ and $|\Psi\rangle$. 
Using the identities $h_\Phi = -\log \rho_\Phi + \log\rho_\Phi'$ and 
$h_{\Psi|\Phi} = -\log\rho_\Psi + \log\rho_\Phi'$, the relative
entropy defined by (\ref{eqn:Srel}) can be converted to the 
standard expression\footnote{Retaining $h_\Phi$ in this 
expression guarantees that only $\rho_\Phi$ density
matrices appear, as opposed to $\rho_\Phi'$.
However, since $0=\langle\Phi|h_\Phi|\Phi\rangle = S(\rho_\Phi) - 
S(\rho_{\Phi}')$, it is equivalent 
to work with just $\langle \Phi|h_{\Psi|\Phi}|\Phi\rangle$.}
\beq
S_{\text{rel}}(\Phi||\Psi)
=
\tr\left(\rho_\Phi\log\rho_\Phi - \rho_\Phi \log\rho_\Psi\right)
=-S_\Phi^\text{QFT} - \vev{\log\rho_\Psi}_\Phi.
\eeq
Since $|\Psi\rangle$ is a KMS state for the flow generated 
by the Hamiltonian $H_\xi^g$, the one-sided density matrix for this state will be 
thermal with respected to the one-sided Hamiltonian $H^\Sigma_\xi$,
introduced in equation (\ref{eqn:Hxisplit}).  Hence it may be expressed as 
\beq
\rho_\Psi = \frac{e^{-\beta H_\xi^\Sigma}}{Z_\xi},
\eeq
where $Z_\xi$ is a  normalization factor.
This then allows the relative entropy to be expressed as a difference
in free energy, 
\beq\label{eqn:Srelfreeen}
S_\text{rel}(\Phi||\Psi) = \beta\vev{H_\xi^\Sigma}_\Phi-S_\Phi^\text{QFT} +\log Z_\xi
\eeq
where 
\beq\label{eqn:logZ}
\log Z_\xi = -\beta\langle H_\xi^\Sigma\rangle_\Psi +S^\text{QFT}_\Psi.
\eeq

The final step is to employ the integrated first law of local
subregions, obtained from (\ref{eqn:HSigmareln}) after assuming that the 
gravitational constraints $C_\xi=0$ hold locally on the Cauchy slice for the subregion. 
In the quantum theory, the vanishing of the constraints holds as an operator
equation, and hence we should view the area and Hamiltonians  appearing in 
(\ref{eqn:HSigmareln}) as operators as well.  Taking the expectation value
of this equation in the state $|\wh{\Phi}\rangle$
 implies
\beq
\beta \vev{H_\xi^\Sigma}_\Phi + \beta \vev{H_\text{obs}}_f = -\left\langle\frac{A}{4G_N}
\right\rangle_{\wh{\Phi}}.
\eeq 
Combining this equation with (\ref{eqn:Srelfreeen}) and (\ref{eqn:SrhoSrelobs})
results in the generalized entropy,
\beq \label{eqn:SrhoSgen}
S(\rho_{\wh\Phi}) = \left\langle\frac{A}{4G_N}\right\rangle_{\wh\Phi} +
S^\text{QFT}_\Phi + S^\text{obs}_f + c,
\eeq
where the state-independent constant 
$c$ is given by
\beq
c = \log \beta - \log Z_\xi.
\eeq  
Since the constant term involves $\log\beta$, which appeared
as a normalization constant ensuring $\atr \rho_{\wh{\Phi}} = 1$,
we see that the entropy is sensitive to the normalization
of the trace $\atr$.  
When working with a type $\ttwo_\infty$ algebra, there is not
any preferred normalization, so this additive ambiguity
is unavoidable.  However, when working with the projected 
type $\ttwo_1$ algebra in section \ref{sec:encond}, there 
is a preferred normalization defined by
imposing that $\atr \,\mathbbm{1}=1$
(this condition is not possible on a type $\ttwo_\infty$ algebra
since in that case the identity has infinite trace; see appendix \ref{app:algebraic-background}).  
We will use this preferred normalization to interpret the constant 
terms in the entropy in section \ref{sec:encond}.
We therefore see that the entropy in the type $\ttwo$ gravitational
algebra agrees with the semiclassical generalized entropy
$S_\text{gen}(\wh{\Phi})$, up to an
additive ambiguity.  

Note that although the area term in (\ref{eqn:SrhoSgen}) 
appears to be order $\gc^{-2}$, the constant term $c$ compensates
the divergent piece so that $S(\rho_{\wh{\Phi}})$ is finite.
This relies on the area $A$ having no contribution at order
$\gc$, which is the convention employed when fixing 
the subregion, as discussed in section \ref{sec:fixregion}.
There we noted that one could allow $A$ to be nonzero
at order $\gc$ by allowing $H_\text{obs}$ to have energies
at order $\gc^{-1}$.  This produces an order $\gc^{-1}$ contribution
to the entropy, as is also apparent in the expression 
for $S(\rho_{\wh\Phi})$ in (\ref{eqn:SrhoSrelobs}).  Although
such divergent observer energies complicate
the crossed product construction of the algebra, it 
is interesting that the final entropy
formula seems to allow for such formally large contributions.  

The derivation of the generalized entropy
presented here has the advantage over the analogous 
one in \cite{Chandrasekaran2022a, Chandrasekaran2022b} of 
only using fields and states defined on the subregion Cauchy surface.  
This is enabled by assumption \ref{assm:1stlaw}, and again ties the finiteness
of the generalized entropy (or, rather, generalized entropy differences)
to the imposition of the local gravitational constraints.
It also avoids having to consider dynamics of the Cauchy horizon for the 
subregion, as was done in \cite{Chandrasekaran2022a, Chandrasekaran2022b},
which likely would be more complicated in the present
context where the subregion has no time-translation symmetry.
Note the connection between relative entropy and generalized entropy 
employed here has appeared previously in studies of semiclassical
entropy for Killing horizons.  
It was an important observation in Casini's proof of the 
Bekenstein bound \cite{Casini:2008cr},
and also features in Wall's proof of the generalized 
second law \cite{Wall2011}.

\subsection{Type $\ttwo_1$ algebra for bounded subregions}
\label{sec:encond}

Up to this point, we have employed assumptions
\ref{assm:aqft}-\ref{assm:1stlaw}, which are sufficient
to obtain the local gravitational algebra as a crossed
product and to relate the entropy of a set of 
states on this algebra to the generalized entropy.  
It remains to explore the consequences of 
assumption \ref{assm:bdenergy}, which requires the 
energy of the observer be bounded below.  This 
assumption is implemented in the same way as in the CLPW
construction \cite{Chandrasekaran2022a}, and our discussion and results are closely
related to theirs.  

The energy condition on the observer is imposed 
by way of a projection 
to states of positive observer energy.\footnote{For
simplicity, we choose the lower bound of the observer
energy to be zero, although any finite lower bound
will result in the same overall type of the projected
von Neumann algebra.  There may nevertheless be interesting
situations where the lower bound is different from
zero. } 
This projection is given by $\Pio = 
\Theta(H_\text{obs}) = \Theta(\hat{q})$,
where $\Theta$ is the Heaviside step function.
Since $\Pio$ is a bounded function of $\hat{q}$, it is an
element of $\ac$, and hence acting with it 
on $\ac$ results in a subalgebra $\aproj$.  
This subalgebra consists of all operators of the form
$\Pio \,\ho a\, \Pio$, which we denote by
\beq
\aproj = \Pio \ac \Pio.
\eeq
This algebra acts on the projected Hilbert 
space $\widetilde{\hs} = \Pio \hs_\sr = \hqft\otimes
L^2(\mathbb{R}^+)$.\footnote{A possible concern one might have is that after projecting,
the operator $\hat{p}$ fails to be self-adjoint when acting 
on $\widetilde{\hs}$.  Although true, this does not cause any problems
when defining the projected algebra,
since $\hat{p}$ itself is not in $\ac$.   Even though $\ac$
involves operators that are twirled by factors of $e^{i\hat{p}\hb}$,
the projection $\Pio$ is Hermitian, and so any operator
that was Hermitian in $\ac$ will project to a Hermitian
operator in $\aproj$.  Hermitian 
operators that are fixed by the projection
(i.e.\ are already elements of $\aproj\subset \ac$)
therefore will remain Hermitian when acting on $\widetilde{\hs}$.}

In order to determine the type of the algebra $\aproj$,
we need to evaluate the trace of the projection 
$\Pio$.  
In section \ref{sec:moddensity}, we alluded to the fact that 
the trace on $\ac$ was chosen to assign $\atr\,\Pio = 1$, 
which we can verify directly using the definition
of the trace given in (\ref{eqn:atrdefn}),
\beq
\atr\,\Pio = 2\pi \beta\langle \Psi|\langle 0|_p e^{-\beta\hat{q}}\Theta(\hat{q})
|0\rangle_p|\Psi\rangle
=
\beta\int_0^\infty dy e^{-\beta y} = 1.
\eeq
The trace $\atr$ descends to a trace on the projected algebra 
$\aproj$, and, since $\Pio$ acts as the identity 
in $\aproj$, we see that in this algebra, the trace of the identity
is $1$.  As explained in appendix \ref{app:algebraic-background},
this implies that $\aproj$ is a type $\ttwo_1$ von Neumann
algebra, in direct analogy with the algebra of the static 
patch of de Sitter \cite{Chandrasekaran2022a}.  

A crucial feature of type $\ttwo_1$ algebras is that they
possess a maximal entropy state, whose density matrix 
is given by the identity $\mathbbm{1}$.
The purification of this state in  $\vphantom{\Big[}\widetilde{\hs}$
can be taken to be
\beq
|\wh{\Psi}_\text{max}\rangle 
= |\Psi, \sqrt{\beta}e^{-\frac{\beta q}{2}} \Theta(q)\rangle,
\eeq
so we see that the KMS state $|\Psi\rangle$ 
for the quantum fields along with the Boltzmann distribution
for the observer determines 
the maximal entropy state for
$\aproj$. 
Using this expression for the maximal entropy state, we can 
interpret the constant $c$ that appears in the 
entropy formula (\ref{eqn:SrhoSgen}).  First, applying the integrated 
first law to the expression for the partition function (\ref{eqn:logZ}),
we have 
\beq
\log Z_\xi = \left\langle\frac{A}{4G_N}\right\rangle_{\Psi_\text{max}}
+S_\Psi^\text{QFT}+ \beta\langle H_\text{obs}\rangle_{f_\text{max}},
\eeq
where we have defined the observer wavefunction $f_\text{max}(q) = \sqrt{\beta}e^{\frac{-\beta q}{2}}\Theta(q)$.
We also have that the observer
entropy in the Boltzmann state is given by
\begin{align}
S_{f_\text{max}}^\text{obs} &= -\int_0^\infty dq|f_\text{max}(q)|^2\log|f_\text{max}(q)|^2
=-\int_0^\infty dq \beta e^{-\beta q}(\log\beta - \beta q) 
\nonumber \\
&= 
-\log \beta +\beta\vev{H_\text{obs}}_{f_\text{max}}.
\label{eqn:Sobsmax}
\end{align}
The constant term that appears in the entropy
formula (\ref{eqn:SrhoSgen}) is then given by
\beq
c=\log \beta - \log Z_\xi = -\left\langle\frac{A}{ 4G_N}\right\rangle_{\Psi_\text{max}}
-S_\Psi^\text{QFT} - S_{f_\text{max}}^\text{obs}  = - S_\text{gen}(\Psi_\text{max}),
\eeq
and hence the entropy formula for states on the type $\ttwo_1$ algebra reduces
to a difference in generalized entropy from the maximal entropy state,
\beq\label{eqn:SrhoSgendiff}
S(\rho_{\wh{\Phi}}) = S_\text{gen}(\wh{\Phi}) - S_\text{gen}(\Psi_\text{max}).
\eeq
Finally, we can derive another convenient
expression for the entropy by 
noting that the relative
entropy of the observer wavefunction 
with respect to the maximal entropy
state is given by
\beq
S_\text{rel}(f||f_\text{max}) = 
\int dq |f(q)|^2(\log|f(q)|^2 
- \log |f_\text{max}(q)|^2)
=
\beta\vev{H_\text{obs}}_f -S_f^\text{obs}
-\log\beta.
\eeq
Plugging this into equation (\ref{eqn:SrhoSrelobs}), we find that 
the entropy on the algebra 
is given in terms of a sum of relative 
entropies,
\beq
S(\rho_{\wh{\Phi}}) =
-S_\text{rel}(\Phi||\Psi)
-S_\text{rel}(f||f_\text{max}).
\eeq
Since relative entropies are positive, this expression makes 
manifest that $S(\rho_{\wh{\Phi}})$ is always negative, as one expects
from the interpretation provided by (\ref{eqn:SrhoSgendiff}) as 
an entropy difference from the maximal entropy state.  
The connection between relative entropy and entropy
of a type $\ttwo_1$ algebra has recently been explored 
in \cite{Longo:2022lod}.

Intriguingly, the existence 
of a maximal entropy state connects to Jacobson's 
entanglement equilibrium conjecture \cite{Jacobson2015}.
This conjecture was formulated for small causal diamonds
in maximally symmetric background geometries,
and it was shown that 
Einstein's equation can be derived from the assumption that 
the vacuum state
of a CFT coupled to gravity  
has maximal generalized entropy when
restricted to 
the diamond.
Here, we have confirmed 
the existence of a maximal entropy state for a generic
subregion in semiclassical quantum gravity once an
energy condition is imposed on the observer.  Instead 
of the vacuum state defining the maximal entropy
state, we find that it is the KMS state
associated with the boost flow within the 
diamond that determines the maximal entropy state.
Further comments on the application of von Neumann
algebras to the entanglement equilibrium conjecture 
are given in section \ref{sec:semiclassical}.

\subsection{Type $\ttwo_\infty$ algebra for asymptotic subregions} \label{sec:asymp}

The discussion up to this point has focused on the gravitational algebras for
bounded subregions.  However, 
the crossed product construction applies
equally well for subregions that include complete asymptotic boundaries.  
These arise naturally as causal complements in open
universes of the bounded subregions considered
above, or otherwise occur when dividing a spacetime with multiple asymptotic
boundaries, such as a two-sided black hole.  The algebra for such an unbounded
region could be constructed from scratch following an identical procedure
as described in section \ref{sec:crossconstr}.  One simply replaces the observer
Hilbert space with a Hilbert space for the ADM Hamilton, $\mathcal{H}_\text{ADM}$,
and constructs a crossed-product algebra acting on $\mathcal{H}_\text{QFT}\otimes
\mathcal{H}_\text{ADM}$.  Alternatively, in the case that the asymptotic 
subregion $\sr'$ is the causal complement of a bounded subregion $\sr$,
the algebra can be realized as the commutant $(\ac)'$ of the bounded 
subregion algebra.  Just as in the discussion of the complementary static
patch of de Sitter space considered by CLPW \cite{Chandrasekaran2022a},
these two procedures will yield unitarily equivalent descriptions of the 
algebra for the subregion $\sr'$.  The identification of $(\ac)'$ as the 
algebra associated with the causal complement $\sr'$ realizes a version
of Haag duality for gravitational subregion algebras.  

Since the commutant algebra $(\ac)'$ has already been identified 
in (\ref{eqn:AC'}), we will use this description in the present
section to examine the entropy associated with the region $\sr'$.
$(\ac)'$ 
contains a type $\tthr_1$ subalgebra associated with 
quantum fields restricted to $\sr'$. 
In addition, it also
contains the operator $\mathcal{C} = \Hxig +\hat{q}$, which,
as discussed below (\ref{eqn:AC'}), is identified with 
$-H_\text{ADM}$, in order to be consistent with 
the global gravitational constraint (\ref{eqn:Cfull}).
In this sense, the ADM Hamiltonian plays the role 
of an asymptotic observer, allowing the outside
algebra to be interpreted as an algebra of operators 
dressed to the asymptotic boundary together with the 
global ADM Hamiltonian.  
 
The entropy of classical-quantum states $|\wh{\Phi}\rangle
=|\Phi,f\rangle$
for this 
algebra is again consistent with the generalized entropy
of the outside subregion.  Utilizing the density matrix
(\ref{eqn:rhoPhi'}) and employing the 
same semiclassical condition as in section \ref{sec:genent},
the logarithm of the density matrix can be written\footnote{Although this 
density matrix assumes that $|\Phi\rangle$ is 
a canonical purification with respect to $|\Psi\rangle$, 
it is also a valid expression  when utilizing 
the semiclassical approximation,
which implies
$[J_{\Phi|\Psi}J_\Psi , f(\hat{q}+\hb)]\approx 0$.
When applied to the exact density matrix
(\ref{eqn:rho'text}), the factors of $J_{\Phi|\Psi}J_\Psi$
can be commuted past $|f(\hat{q}+\hb)|^2$ and 
canceled against $J_{\Psi}J_{\Psi|\Phi}$, after which
the expression for the density matrix reduces to
(\ref{eqn:rhoPhi'}).} 
\begin{align}
-\log\rho_{\wh{\Phi}}' 
&= 
-h_{\Psi|\Phi}
-\beta\hat{q} - \log\Big|f\Big(\hat{q}+\hb\Big)\Big|^2 
+ \log\beta \nonumber \\
&= -h_{\Psi|\Phi} + h_\Psi + \beta H_{\text{ADM}} - 
\log|f(-H_\text{ADM})|^2 +\log\beta \label{eqn:logrhoPhi'}
\end{align}
Since $h_{\Psi} - h_{\Psi|\Phi} = h_{\Phi|\Psi} - h_\Phi = -h_{\Psi|\Phi}'+h_{\Phi}'$, the expectation value of  this term
can be expressed as minus a relative entropy for $\aqft'$.  
As in section \ref{sec:genent}, 
we can directly interpret this expectation
value as a free energy in $\sr'$ by making the formal
split
\beq
h_\Psi = -\log\rho_\Psi + \log\rho_{\Psi}',\qquad
h_{\Psi|\Phi} = -\log\rho_\Psi + \log\rho_{\Phi}',
\eeq
and expressing $\rho_\Psi'$ as a thermal state 
for the one-sided Hamiltonian $H_\xi^{\bar\Sigma}$,
\beq
\rho_\Psi' = \frac{e^{-\beta H_\xi^{\bar\Sigma}}}{Z_\xi'}.
\eeq
This then leads to
\beq
\langle \Phi|h_\Psi-h_{\Psi|\Phi}|\Phi\rangle
=
\Tr\big[\rho_\Phi'(\log\rho_{\Psi}' - \log\rho_{\Phi}')\big]
=
S_{\Phi'}^\text{QFT} -\beta\vev{H_\xi^{\bar\Sigma}}_\Phi-
\log Z_\xi'.
\eeq
Identifying $\langle\wh{\Phi}|-\log|f(-H_\text{ADM})|^2|
\wh{\Phi}\rangle$ with the entropy $S_{\text{ADM}}$ associated
with the uncertainty in the ADM Hamiltonian, the outside
entropy is given by
\beq
S(\rho_{\wh{\Phi}}') = \langle\wh{\Phi}|-\log\rho_{\wh{\Phi}}'
|\wh{\Phi}\rangle
= \beta\vev{H_\text{ADM}}_{\wh{\Phi}} -
\beta\vev{H_\xi^{\bar\Sigma}}_{\wh{\Phi}}
+S_{\Phi'}^{\text{QFT}} + S_\text{ADM} +\log\beta - \log Z_\xi'
\eeq
Finally, applying the first law relation for the outside 
region (\ref{eqn:HbarSigmareln}) converts this expression
to a generalized entropy, up to a state-independent 
constant,
\beq
S(\rho_{\wh{\Phi}}') 
= \left\langle\frac{A}{4G_N}\right\rangle_{\wh{\Phi}} 
+S_{\Phi'}^\text{QFT} +S_\text{ADM} + c'.
\eeq

This algebra can also be restricted to the positive
energy sector for $H_\text{ADM}$.  As in section
\ref{sec:encond}, this is implemented via the 
projection $\Piadm =\Theta(H_\text{ADM}) = 
\Theta(-\hat{q}-\hb)$.  The trace of this projection
can be evaluated by noting that the  formula
(\ref{eqn:atrdefn})  also defines a trace on $(\ac)'$.
Hence, using the identity $h_\Psi|\Psi\rangle =0$, the 
trace of $\Piadm$ evaluates to
\beq
\tr\Piadm=2\pi \beta
\big\langle\Psi|\big\langle0\big|_p
\,\Theta\Big(-\hat{q}-\hb\Big)
\frac{\beta e^{-\beta\hat{q}}}{|f(\hat{q})|^2} \,
\big|0\big\rangle_p\big|\Psi\big\rangle
=\beta\int_{-\infty}^\infty dq e^{-\beta q}\Theta(-q)
=\infty
\eeq
Because this trace is infinite, the projected algebra
$\widetilde{\alg'} = \Piadm (\ac)'\Piadm$ remains
type $\ttwo_\infty$.  Hence, the outside region that 
includes the asymptotic boundary behaves more 
like the AdS black hole with a type $\ttwo_\infty$
algebra \cite{Witten2021, Chandrasekaran2022b}, 
while the bounded subregion
is more analogous to the static patch of de Sitter,
possessing a type $\ttwo_1$ algebra~\cite{Chandrasekaran2022a}.

As mentioned above, a unitarily equivalent representation
of the outside algebra can given as a crossed product 
acting on $\hs_{\sr'} = \hs_{\text{QFT}}\otimes \hs_{\text{ADM}}$.
The resulting algebra $\alg_\text{out}$ is given by
\beq
\alg_{\text{out}}
= \{ e^{i\hat{p}'\Hxig} \msf{a}'e^{-i\hat{p}'\Hxig}, e^{is\hat{q}'}|
 \msf{a}'\in\aqft', s\in\mathbb{R}\}''
\eeq
where $\hat{q}' = -H_\text{ADM}$ and $\hat{p}' = -i\frac{d}{dq'}$
is the conjugate momentum. 
Superficially,  the operator
associated with the subregion observer $H_\text{obs}$ 
has  been eliminated in this 
description: all operators acting
on $\hs_{\sr'}$ are constructed from the ADM Hamiltonian
$-\hat{q}'$, its conjugate momentum $-\hat{p}'$, and 
the quantum field operators from $\aqft$ and $\aqft'$.
Of course, the observer is still present, with the observer
Hamiltonian given by the operator $H_\text{obs} = \hat{q}'
+\Hxig$ in the commutant algebra $\aout'$.

What is striking is that when the outside algebra $\aout$
is represented on $\hs_{\sr'}$, it can be viewed 
as an algebra constructed entirely from operators
dressed to the asymptotic boundary, supplemented by the 
ADM Hamiltonian.  In the context of holography, such operators
are expected to be constructible in the CFT using standard
techniques as in \cite{hamilton2006holographic, Faulkner:2013ana, Jafferis:2015del, Dong:2016eik, cotler2019entanglement, Faulkner:2017vdd}.
Assuming the validity of the arguments in the present
work, this algebra of boundary-dressed operators will
be type $\ttwo_\infty$ in $\gc\rightarrow 0$ limit,
corresponding to the large $N$ limit in the CFT.
Since any representation of a type $\ttwo$ algebra must 
have a nontrivial commutant, this implies the existence
of an algebra that is naturally associated with the complementary 
local
subregion $\sr$.  Since the CFT should furnish 
such a representation
in the large $N$ limit, there must be
operators  in the CFT corresponding
to localized observables in the bulk subregion $\sr$.\footnote{
Similar arguments have been made 
for the emergence of type $\tthr_1$ algebras corresponding
to local bulk subregions, both at infinite $N$ \cite{Leutheusser:2022bgi}
and to all
orders in the $1/N$ expansion \cite{Bahiru:2022oas,
Bahiru:2023zlc}. }
If, as suggested by the bulk geometry,
this CFT algebra contains a type $\tthr_1$ subalgebra
that is naturally isomorphic to the algebra of bulk quantum
fields restricted to $\sr$, we can furthermore conclude that 
there must be an additional degree of freedom
associated with the bulk observer.  This follows from
the simple fact that the commutant of a type $\ttwo$ algebra
is always type $\ttwo$, so there must be some mechanism
 to convert the type $\tthr_1$ quantum
field theory algebra to a type $\ttwo$ algebra that
can serve as the commutant.  A natural explanation for this 
mechanism is to conclude that there is some observer degree
of freedom in $\sr$ that implements a crossed product
on the algebra.

Note however that  $\aout$ is insensitive to 
the details of how the observer is modeled on the inside 
algebra.  These details are present only at the level of 
choice of representation of the outside algebra.  For example, 
one could instead construct the standard representation
of $\aout$,
in which the commutant algebra $\aout'$ is isomorphic
to $\aout$, and hence would also be type $\ttwo_\infty$.  
This would correspond to a  bulk wormhole geometry, 
with each subregion $\sr'$, $\sr$ containing
an asymptotic boundary.\footnote{A related construction
occurs in the canonical purification of an entanglement
wedge in holography \cite{engelhardt2019coarse, Dutta:2019gen}.}    
In this case, the observer Hamiltonian for
the complementary region is naturally interpreted
as the ADM Hamiltonian at the second asymptotic
boundary of the wormhole geometry.  
However, in representations where $\aout'$ is type
$\ttwo_1$, a natural expectation is that the 
commutant algebra describes a local subregion in the 
bulk with an associated local observer degree of freedom.

Finally, it is worth mentioning that since $\aout$ includes
the ADM Hamiltonian, the commutant algebra consists
of operators with zero ADM energy, since they commute
with $H_\text{ADM}$.  From the bulk perspective, ensuring this 
commutativity is the reason for introducing the observer 
degree of freedom, allowing  operators to be dressed to the 
observer instead of the asymptotic boundary.  
This is puzzling from the perspective of the dual
CFT, since the ADM Hamiltonian is dual to the CFT Hamiltonian,
which only commutes with a small number of topological
operators.  
However, the large $N$ limit required for the emergence 
of the local bulk algebras generally involves restricting to 
a code subspace in the CFT, and in some explicit 
examples operators can be constructed that commute with the 
CFT Hamiltonian within this code subspace to all orders in the 
$1/N$ expansion \cite{Bahiru:2022oas,
Bahiru:2023zlc}.  
Investigating such constructions in more detail may therefore
provide insight into the nature of the observer degree
of freedom.

%---------------------------------------
\section{Discussion}
%---------------------------------------
\label{sec:discussion}

In this work we have associated an algebra of operators to a generic subregion $\sr$ for theories of Einstein gravity coupled to matter in the $G_N\to 0$ limit. When $\sr$ is a bounded bulk region, the associated algebra is of type \text{II}$_1$. When $\sr$ includes an asymptotic boundary, the associated algebra is of type \text{II}$_{\infty}$. In both cases, physics in the region has some of the properties of ordinary quantum mechanics, in that there are well defined density matrices.
In the type II$_1$ case there is a maximum entropy state; in both cases, there is no minimum entropy state. The existence of density matrices lets us assign UV-finite entropies to regions that are almost well defined up to an ambiguous state-independent constant, which in ordinary quantum mechanics would be fixed through a state counting prescription that sets the entropy of a pure state to zero.
After regulating, we see that up to the ambiguous universal constant, our entropy is the generalized entropy of Bekenstein.

Our construction generalizes the work of~\cite{Chandrasekaran2022a}, in which the authors associated a type \text{II}${_1}$ algebra to the static patch of de Sitter space and a type II$_{\infty}$ algebra to the exterior of a static black hole (see also~\cite{Witten2021,Chandrasekaran2022b}). The main ingredients used in our work resemble theirs. We introduce a model of an observer weakly entangled with gravity; account for an integrated form of a Hamiltonian constraint in the subregion $\sr$; use a gravitational First Law associated with a generic subregion; and, crucially, assume the existence of a local state we refer to as a KMS state, whose modular Hamiltonian  is proportional to an integrated Hamiltonian constraint. In the absence of gravity there is good reason to think that such states exist in generic local quantum field theories. With gravity, we assume such a state exists and find that this is a consistent assumption leading to a type $\text{II}$ algebra and ultimately the generalized entropy.
The careful treatment of nonlinear constraints, the gravitational First Law, and the existence of the KMS state are the essential ingredients that allow us to generalize \cite{Witten2021, Chandrasekaran2022a, Chandrasekaran2022b} to subregions without boost isometries.

In the rest of this section we discuss a large number of open questions and applications in gravity and von Neumann algebras that are suggested by our work. More broadly, this paper and other recent works suggest a new perspective on observables in theories of quantum gravity that we begin to map out below.

%---------------------------------------
\subsection{Applications to semiclassical entropy}  \label{sec:semiclassical}
%---------------------------------------

One of the most significant outcomes of our analysis 
is the connection between the manifestly finite entropy
for the type $\ttwo$ local gravitational algebras 
and the semiclassical generalized entropy, as exhibited in 
equation (\ref{eqn:SrhoSgen}).  
This relationship was first identified in \cite{Chandrasekaran2022a, 
Chandrasekaran2022b} for algebras associated with the de Sitter
static patch and black hole exteriors; the present paper
shows that the relationship continues to hold 
for a much broader class of subregions in semiclassical 
 gravity. 
We additionally clarified the role that the local gravitational
constraints play in deriving the relationship, which also 
yielded a simplified derivation by invoking the first law of 
local subregions that follows from the constraints.  
Since generalized entropy is such an important quantity
in semiclassical quantum gravity,
it is worth considering what implications this finite type $\ttwo$
entropy has on the question of UV-finiteness of the generalized entropy.

On this point, it is actually the relationship between generalized 
entropy and relative entropy that is key.  
Once we have made the assumption \ref{assm:mod} on the existence 
of a KMS state $|\Psi\rangle$ for the flow generated 
by $\Hxig$, the relative entropy can be converted to the free energy
expression (\ref{eqn:Srelfreeen}).  
Imposing the local gravitational constraints $C_\xi=0$ in 
(\ref{eqn:HSigmareln}), consistent with 
assumption \ref{assm:1stlaw}, converts this free energy difference
to the generalized entropy for the subregion.  It is instructive to 
consider this step in the absence of an observer, in which case 
the observer Hamiltonian drops from (\ref{eqn:HSigmareln}) and their stress
tensor drops from the constraint.  This leads to the
relationship 
\beq
S_\text{rel}(\Phi||\Psi) = -\left\langle\frac{A}{4G_N}\right\rangle_\Phi
-S_\Phi^\text{QFT} + \left\langle\frac{A}{4G_N}\right\rangle_\Psi
+S_\Psi^\text{QFT} = -S_\text{gen}(\Phi) + S_\text{gen}(\Psi)
\eeq
Since the relative entropy is generically finite when
$|\Phi\rangle$ and $|\Psi\rangle$ are cyclic-separating, this argument demonstrates that the 
{\it difference} in generalized
entropies between two states is finite.  One simply needs to ensure
that the regularization scheme employed to define the entropy and 
area operator is consistent with the localized gravitational
constraint, so that $\vev{\int_\Sigma C_\xi}_{\Phi,\Psi} = 0$.
To conclude finiteness of $S_\text{gen}(\Phi)$ itself, one would need 
to be able to argue that generalized entropy is finite 
in the KMS state $|\Psi\rangle$.  In some sense, $S_\text{gen}(\Psi) 
=\infty$ in the $G_N\rightarrow 0$ limit, since it is 
dominated by $\frac{A}{4G_N}$ with the area remaining finite. 
This suggests that full finiteness of $S_\text{gen}$ is a nonperturbative
statement, that we only expect to see when working with the 
finite $G_N$ quantum gravity description.\footnote{See
\cite{Gesteau:2023hbq} for a discussion of finiteness
of $S_\text{gen}$ in the context of asymptotically
isometric codes \cite{Faulkner:2022ada}.}  Nevertheless,
finiteness of generalized entropy differences is still 
a nontrivial statement, since it is not immediately clear
that divergences will cancel out in the generalized entropy expression
for perturbations that change the area.  
Invoking the local gravitational constraint 
allows this cancellation to be derived in the perturbative theory.  

Restoring the observer to the discussion does not 
affect the argument for finiteness of generalized entropy differences.  It makes a finite
contribution to the matter entropy and affects the area through
its backreaction on the geometry, but neither of these effects produce new 
UV divergences.  What the observer adds is the ability to interpret the 
generalized entropy 
as an entropy of a von Neumann algebra.  This provides a 
statistical interpretation of the generalized entropy, although not quite 
a state-counting interpretation due to the fact that the von Neumann
algebra is type $\ttwo$.  Nevertheless, it opens the door to a number of further
investigations into properties of generalized entropy in semiclassical 
geometries.  It could even be viewed as the correct entropic quantity
to consider in semiclassical gravity, which reduces to the generalized
entropy for classes of states in which the observer is weakly entangled. 

An immediate topic to investigate would be to determine how 
the type $\ttwo$ entropy behaves for generic classical-quantum states,
lifting the semiclassical assumption that was employed to arrive at 
equation (\ref{eqn:SrhoSrelobs}).  The exact expressions for the density
matrices given in (\ref{eqn:rhoPhioutline}) and
(\ref{eqn:rhoPhi'outline})
provide 
a first step, but it is still a nontrivial problem to compute
the logarithm due to noncommutativity between $h_\Psi$ and $\Delta_{\Phi|\Psi}$. 

A more ambitious goal would be to frame the quantum focusing
conjecture \cite{Bousso:2015mna} in terms of  type $\ttwo$ gravitational algebras,
and to seek a proof from properties of the entropy under algebra inclusions.  
Our work takes a first step on this problem by giving a prescription for 
constructing the algebra for generic subregions.  Hence, in principle we can
meaningfully discuss how the entropy responds to changes in the subregion
induced by evolution along a causal horizon.  The fact that the entropy 
of the type $\ttwo$ algebras is automatically finite sidesteps tricky issues
related to renormalization and finiteness of the generalized entropy; since the 
type $\ttwo$ entropy limits to a generalized entropy difference, it can be 
taken as a proper definition of the  renormalized generalized entropy.  
A more nontrivial task would be to consistently relate the 
 additive ambiguities in type $\ttwo$ entropies between different subregions.  
This is related to the question of how the algebra for a subregion relates to the algebra for a larger region containing it.  In particular, the KMS state 
for the subregion would not have any obvious relation to the KMS state for the
larger region, and hence some work is required to determine how the algebras 
are related.  The fact that crossed products with respect to 
different states are simply related by Connes cocycles
\cite{Connes1973, Witten2021}
suggests that a natural
relation  could 
be achieved for crossed product algebras 
associated with a subregion and a proper 
subspace thereof.
Clearly the type $\ttwo$ gravitational algebras  constructed here 
provide a new set of tools for investigating the quantum focusing conjecture 
and other semiclassical entropy relations. 

Another direction of inquiry relates to higher curvature corrections to the 
generalized entropy.  For Killing horizons, these are given by the Wald
entropy
\cite{wald1993black,Iyer1994}, but for generic entangling surfaces 
there can be additional contributions from the extrinsic curvature,
as appear in the Dong entropy
\cite{Dong:2013qoa, Camps2013}.  Most derivations 
of the higher curvature entropy functionals rely on 
Euclidean methods,\footnote{Although see
\cite{Wall2015,Bhattacharyya2017,Hollands2022} 
for derivations based on the classical higher curvature 
second law of black hole mechanics. } so 
a Lorentzian derivation  in terms of von Neumann algebras would be enlightening.  
As we have emphasized, the crossed product construction of gravitational
algebras is a consequence of diffeomorphism invariance, and hence would be 
valid for any diffeomorphism-invariant theory of gravity.
As reviewed in appendix \ref{app:CPS}, 
the constraint operator can always be unambiguously defined,
since it is constructed directly from the higher curvature equations 
of motion. 
The main subtlety
in obtaining the entropy formula is the correct determination of the 
subregion gravitational Hamiltonian $H_\xi^\Sigma$, 
which is sensitive to how ambiguities in the covariant 
phase space are resolved.  These ambiguities propagate into the entropy 
functional, and hence the main problem to address is how to treat them
consistently in the present context.  We expect the recent advances 
in the theory of covariant phase space with boundaries could provide
insights into this question 
\cite{Harlow2019, Chandrasekaran2020, Chandrasekaran2021}.

Finally, we noted in section \ref{sec:encond} that the occurrence of a
type $\ttwo_1$ algebra for bounded subregions in gravity implies a version
of Jacobson's entanglement equilibrium hypothesis
\cite{Jacobson2015}.  
The main difference is that in place of the vacuum state, 
the local KMS state $|\Psi\rangle$ determines the maximal entropy
state for the subregion.  This observation may help resolve some puzzles
arising in the original construction, which considered small causal
diamonds in maximally symmetric backgrounds.  In particular, when working
with nonconformal matter fields, it was found that the entropy
of the vacuum did not seem to behave correctly in conformal perturbation
theory to be consistent with Jacobson's original hypothesis
\cite{Casini2016a,Speranza2016}.  
It is possible that a more careful treatment
of this problem using the algebraic techniques developed here 
and thinking carefully about the appropriate KMS state for the causal diamond
may lead to a resolution of these puzzles.
Note that the type $\ttwo$ entropy formula provides a missing piece of the 
entanglement equilibrium hypothesis, namely an independent definition
of the entropy, which is provided by equation
(\ref{eqn:SrhoSrelobs}).  Assuming the equality
of this formula with the subregion generalized entropy would yield 
a derivation of the local subregion constraints, exactly analogous
to the derivations of the bulk Einstein equations from the holographic entropy
formula \cite{Lashkari:2013koa, Faulkner:2013ica,
Swingle:2014uza, Faulkner:2017tkh, Lewkowycz:2018sgn}. 

\subsection{Holographic applications} \label{sec:holoapp}

Although motivated by large-$N$ limits in holography,
our construction of gravitational subregion algebras 
was performed directly in the bulk, employing purely
quantum gravitational arguments.  
However, given that holography provides concrete, UV-complete
models of bulk quantum gravity in terms of a dual CFT, 
determining 
how the bulk type $\ttwo$ algebras
identified here arise in the CFT is a natural direction to pursue.  
A major component of such a top-down derivation 
of the gravitational algebra would be the explicit construction of
the type $\tthr_1$ algebra of bulk quantum fields, i.e.\
determining how to justify assumption \ref{assm:aqft} using a large-$N$
limit.  Although a challenging problem, inroads have already been 
made in some recent works \cite{Leutheusser:2022bgi, Faulkner:2022ada, 
Bahiru:2022oas, Bahiru:2023zlc},  
and there is a wide literature  devoted to 
bulk reconstruction in holography that has addressed aspects of this problem 
as well \cite{hamilton2006holographic, Faulkner:2013ana, Jafferis:2015del, Dong:2016eik, cotler2019entanglement, Faulkner:2017vdd}.
An explicit construction of such an algebra in the CFT would also shed
light on how to view the observer degree of freedom, 
which is somewhat enigmatic from the bulk quantum gravitational
perspective (see section \ref{sec:observer} for further discussion).

One area where the holographic picture already provides insight is 
for subregions whose entangling surfaces extend out to infinity. For example,
the AdS-Rindler wedge, or more general entanglement wedges, are all of this 
flavor.  Entanglement wedge duality 
\cite{Czech:2012bh, Jafferis:2015del,
Dong:2016eik}
implies that the dual algebra  
consists of all local CFT operators
in the causally complete boundary region  that forms the asymptotic
boundary of the  entanglement wedge. As a local algebra of a quantum
field theory, we know that this algebra should be type $\tthr_1$ for any value
of $N$.
Focusing on the case of AdS-Rindler, which is an entanglement wedge
with boost symmetry, 
if one were to try directly applying the arguments in this 
paper, one would find that 
the asymptotic observer Hamiltonian
should correspond to the one-sided asymptotic boost 
Hamiltonian.  This is problematic because this 
operator corresponds to a one-sided boost Hamiltonian in the 
dual CFT, which is ill-defined in the continuum.  
Hence, 
it is not clear that imposing the gravitational constraints
could be interpreted as a legitimate crossed-product 
construction in this context.  
We expect 
similar arguments to apply for more general subregions
whose entangling surfaces extend to infinity, and whose
corresponding generalized entropy is infinite at finite $G_N$.
A modification of this idea involving a bulk
radial cutoff was recently considered in \cite{Bahiru:2022mwh}, and the cutoff
algebra was argued to be type $\ttwo_\infty$.  We expect however that 
a consistent picture of the boundary algebra in the continuum limit should exist
that both incorporates the bulk gravitational constraints and also
reproduces the type $\tthr_1$ structure.

Entanglement wedges relate to another potential application of 
our construction to holography: understanding the quantum
extremal surface prescription \cite{Engelhardt:2014gca}.
Quantum extremal surfaces provide the nonperturbative
quantum-corrected generalization of the Ryu-Taka\-ya\-nagi formula,
and are determined by extremizing the generalized entropy over all
choices of entangling surfaces in the bulk. All derivations of this 
formula to date involve Euclidean path integral methods, which, although
highly useful, obscure the algebraic origin of the entropy being computed.
It would be of great interest to derive the QES formula in 
a Lorentzian formulation involving the subregion gravitational algebras
constructed in this work.  As we have seen, the type $\ttwo$ entropy
agrees with the generalized entropy, up to a state-independent (but possibly
subregion-dependent) constant,
and so it would be interesting to understand the extremization 
procedure in terms of some property of these von Neumann algebras.
This motivates a broader investigation into the
structure of the
full net of von Neumann algebras associated with quantum gravitational
subregions. 

Note that in order to fix our region in the first place we had to enforce a constraint, that a notion of volume for the region (see section \ref{sec:fixregion}) was constant. The generalized entropy we find should then be understood as the entropy subject to this constraint. This is reminiscent of the coarse-grained entropies of~\cite{engelhardt2019coarse}. Perhaps, in analogy with the behavior of entropy in thermodynamics, by relaxing this constraint we can land on not just the generalized entropy, but an extremization of the generalized entropy over the shape of the entangling surface. This could give a purely Lorentzian derivation of the QES formula.

Finally, we note a possible application of this work to tensor
networks in holography.  This connection is motivated by the observation
that any type $\ttwo_\infty$ von Neumann algebra can be realized 
as a tensor product of a type $\ttwo_1$ and a type $\tone_\infty$ 
von Neumann algebra (see e.g. \cite[section 7.2]{sorce2023notes}).  We argued in section \ref{sec:asymp}
that subregions that include complete asymptotic boundaries yield
algebras of type $\ttwo_\infty$, while those associated with a bounded 
subregion were argued in section \ref{sec:encond} to be type $\ttwo_1$.  
This suggests that we could view the asymptotic subregion as consisting 
of an infinite lattice of local subregions, each associated with a 
type $\ttwo_1$ algebra. 
Operators acting on short distance scales would be represented on an
individual type $\ttwo_1$ algebra, and operators that mix individual
lattice sites would correspond to operators acting on the type $\tone_\infty$
tensor factor, and together these would generate the full
type $\ttwo_\infty$ asymptotic algebra.  
It would be natural to choose the size of the local subregions 
in the lattice to be on the order of the AdS length and construct
the lattice as a hyperbolic tiling as in the HaPPY code
\cite{Pastawski:2015qua}.
This could provide a tensor network model that incorporates sub-AdS locality,
in that the type $\ttwo_1$ factors would describe features of the gravitational
algebra at short distance scales.

\subsection{Geometric modular flow conjecture}

The major workhorse behind the results presented 
here is assumption \ref{assm:mod} concerning the 
existence of a state on the algebra $\aqft$ 
whose modular flow is geometric in the vicinity
of Cauchy surface $\Sigma$.  
It allows the 
algebra $\ac$ presented in section \ref{sec:crossconstr}
to be identified as a crossed product of $\aqft$ with 
respect to its modular automorphism group.
Furthermore, the existence of a KMS state on $\aqft$ for this 
flow is the key input that, in conjunction
with assumption \ref{assm:1stlaw}, allows the 
relative entropy  
$S_\text{rel}(\Phi||\Psi)$ appearing in the type $\ttwo$ entropy
formula (\ref{eqn:SrhoSrelobs}) to be expressed in terms of 
a generalized entropy, and  therefore yields an explanation
for the cancellation of UV divergences in generalized 
entropy differences.  
We motivated this assumption in section \ref{sec:modham}
from the intuitive picture that modular flows should 
approach the local Rindler boost near the entangling 
surface, and also provided examples in which the associated
KMS state can be explicitly constructed.

Given the prominent role it plays in this work, 
an important direction for future investigation would be to explore the 
validity of assumption \ref{assm:mod} in greater detail.  
One direction of inquiry would be to determine further geometric
constraints on the properties of the flow needed to produce a KMS
state.  For example, it would be interesting to determine how quickly
the flow must approach the local Rindler boost 
near the entangling surface, 
and  whether this depends on geometric properties 
of $\partial\Sigma$ such as its extrinsic curvature.  
We also expect the flow to be highly constrained along the null boundary
of the subregion $\sr$, and it would be interesting to explore 
these constraints in further detail.  
One property to investigate is the behavior
of the surface gravity along the null surface, 
away from the entangling surface.  Equation
\ref{eqn:nabxi} gives one definition of surface gravity for the 
null surface, but a number of other definitions exist that generically
do not agree when the surface is not a Killing horizon
\cite{Jacobson1993, Belin2022}.
These surface gravities depend on the behavior
of the generator $\xi^a$ on or near the null surface,
and hence we expect that conditions relating them 
to the geometry of the null boundary should arise
for modular flows.
A particular example of this kind of condition was realized in the construction of a KMS state for deformed CFT in section \ref{sec:CHM}.

Because assumption \ref{assm:mod} applies in the 
$\gc\rightarrow 0$ limit in which gravity decouples, 
it could be investigated purely from the perspective
of nongravitational algebraic quantum field theory, treating 
the matter fields and free spin-2 gravitons separately.   
Hence, one might hope to be able to construct
a proof of (or counterexample to) 
the conjecture in situations where 
we have some control, such as renormalizable
quantum field theories in Minkowski space.  
We gave one argument in section \ref{sec:modham}
involving canceling nonlocal terms in the vacuum
modular Hamiltonian using operators from $\aqft$
and invoking the converse of the cocycle
derivative theorem (see appendix \ref{sec:convcoc}).  It is possible
this could provide a framework for constructing a general
proof, although one would have to carefully analyze
that the cancellation of nonlocal terms 
can be achieved, perhaps in a limiting sense, using only elements
from $\aqft$.

\subsection{Interpretation of the observer} \label{sec:observer}

The introduction of an observer degree of freedom into the 
local subregion
algebra played a crucial role in arriving at a nontrivial
type $\ttwo$ gravitational algebra.  It serves as an anchor
to which operators in the subregion can be gravitationally
dressed, thereby providing a means to satisfy the quasilocal constraints
of diffeomorphism invariance.
Even when working with an unbounded region where,
instead of an observer, the asymptotic boundary and ADM Hamiltonian
provide the anchor for dressing,
the fact that any representation
of a type $\ttwo$ algebra must have a nontrivial commutant invariably 
leads to the conclusion that there is an observer degree of freedom
in the complementary subregion, as discussed in section 
\ref{sec:asymp}.
Nevertheless, the observer was introduced by hand in the crossed
product construction, and it remains an open question how this 
degree of freedom emerges from either a bulk quantum gravitational
or holographic description.  

A number of recent ideas have been proposed related to the problem
of observers in quantum gravity.  CLPW suggest that the observer
could emerge as a degree of freedom within the appropriate ``code subspace''
of the full quantum gravitational Hilbert space in which 
the subregion algebra is well-defined \cite{Chandrasekaran2022a}.  
This code subspace should 
roughly be identified as a class of states in which bulk effective field 
theory provides a good description.
A similar proposal advocates for using features of the state
defining the background geometry as a means to dress operators
\cite{Bahiru:2022oas,Bahiru:2023zlc}, and hence in the context
of subregion von Neumann algebras 
one could interpret the observer 
as being constructed from these features of the background.
Susskind has offered a related perspective, arguing that the 
observer can emerge as a fluctuation of the de Sitter static patch
degrees of freedom, and also connected the existence of the 
observer to 
the need to gauge-fix the time-translation symmetry in the effective
theory \cite{Susskind:2023rxm}.
These ideas all share the property of obtaining the observer 
from an intrinsic degree of freedom of the complete theory.

An alternative viewpoint is that the observer could arise as 
an external degree of freedom that is consistently coupled to 
the gravitational theory.  For example, 
a holographic model for a bulk observer as a probe
black hole was described in
\cite{Jafferis:2020ora,deBoer:2022zps}.  The black hole
arises via entangling the CFT with a reference system, and hence
one could interpret the reference system as  the extra
degree of freedom associated with the observer.  
The idea of entangling with a reference also plays a prominent
role in recent works on black hole evaporation
and the information problem \cite{Almheiri:2019psf, Penington:2019npb,
Almheiri:2019hni, Almheiri:2019qdq, Penington:2019kki, Almheiri:2020cfm}.
A possibly related idea comes from the recent construction
of a constrained instanton for gravity restricted to a subregion
with a fixed spatial volume constraint \cite{Jacobson:2022jir}.  
The constraint is implemented with a Lagrange multiplier in
the gravitational action (a la~\cite{Cotler:2020lxj}), which resembles
the extra observer degree of freedom needed to obtain a nontrivial subregion algebra.  Note also that the fixed volume constraint
appeared in our construction 
for certain choices of subregions  
in relation to the problem of specifying the location
of the entangling surface considered in section 
\ref{sec:fixregion}.  It is therefore possible
that a direct relation to the constrained instanton
construction  could be found.

A possible idea for an intrinsic model of the observer
within the bulk quantum gravitational theory relates to 
our discussion of the constraints in section~\ref{sec:constraints}. There it was emphasized that it is important to consider
the gravitational constraints at first nonlinear order 
in the gravitational coupling $\gc$, which nevertheless
leads to a constraint between quantum fields and the observer
that is visible even at $\mathcal{O}(\gc^0)$.  It is possible that this 
constraint could be implemented in the linearized theory
by quantizing a single nonlinear graviton mode exactly, and that
this extra mode might plausibly play the role of the observer.  
Such a mode would be nonlocal, 
related to the difference in times experienced
by the quantum fields in the two subregions.  A fruitful starting
point to explore this idea would be lower dimensional gravitational
models such as JT gravity coupled to matter.  There, the recently
derived gravitational algebra found in \cite{Harlow:2021dfp}
may yield the desired nonlinear mode to implement this idea.

A final perspective on the observer is provided by the quasilocal
constraint relation (\ref{eqn:HSigmareln}), 
which, after imposing $C_\xi^{\text{mat}+\text{obs}}$
can be rearranged to express the observer Hamiltonian
in terms of the area and one-sided boost generator,
\beq
H_\text{obs} = -\frac{\kappa A}{8\pi G} - H_\xi^\Sigma.
\eeq
The area operator on its own is singular in the quantum
theory, as is the one-sided Hamiltonian $H_\xi^\Sigma$.  
However, the above relation suggests that their sum
defines a UV-finite operator in semiclassical gravity 
that behaves like a local energy contribution within 
the subregion.  Using this relation, one could 
interpret the observer as a smoothed-out version
of the area operator, which fails to commute with 
local fields within the subregion when quantum gravitational
effects are taken into account. 
See \cite{Witten2021, Chandrasekaran2022b} 
for related comments concerning this regulated
version of the area operator.

\subsection{Gravitational edge modes} \label{sec:edgemodes}
An intriguing proposal for modeling the observer as an
external degree of freedom is provided by gravitational
edge modes.
Introducing a subregion boundary into a theory
with gauge symmetry can cause some gauge transformations
to become physical symmetries, whose charges define
edge modes.
There has been much speculation that 
entanglement between these edge mode degrees of freedom
could provide an interpretation
of the area term appearing in the generalized
entropy formula.  This is motivated by 
Donnelly's formula for entropy in nonabelian gauge
theories, which expresses the entropy as a sum
of a bulk entropy and the expectation
value of an operator defined at the entangling surface
\cite{Donnelly:2014gva}.  
This entropy formula is derived in the context of 
the extended Hilbert space, in which additional degrees
of freedom are added to the physical Hilbert space in 
order to obtain a factorization across spatial boundaries.
The physical Hilbert space is recovered by the entangling
product described in 
\cite{Donnelly2016}, which implements the gauge
constraints on the extended Hilbert space.  
There has been much work devoted to extending this construction
to gravitational theories, see
\cite{Balachandran:1994up, Carlip:1995cd, Donnelly2016, Speranza2017, Freidel:2020xyx, 
Freidel:2020svx, 
Chandrasekaran2020, Donnelly:2020xgu, Ciambelli:2021vnn, Ciambelli2021,
Speranza2022, Donnelly:2022kfs}.

The similarity to the crossed product construction considered
here is readily apparent.\footnote{See
\cite{Shaghoulian2023} for a related discussion
on the connection between observers and edge modes.} 
The observer appears as an
additional degree of freedom needed in order to define
the subregion algebra, whose clock is charged under the
boost transformation that evolves the subregion forward in time.
As discussed by CLPW \cite{Chandrasekaran2022a}, 
a nontrivial algebra consistent with the gravitational
constraint
also requires
an observer in the complementary region, and the
extended kinematical Hilbert space is a tensor product
of both observers' Hilbert spaces and the Hilbert
space of quantum fields.  Imposing the constraint
yields a physical Hilbert space for the subregion via the 
crossed product, directly analogous to the constraint
imposed for the entangling product for edge modes.  
The crossed product has a further advantage
of not needing to assume that the quantum field Hilbert
space factorizes, and therefore can be viewed as a 
continuum version of the entangling product of \cite{Donnelly2016}.

Interpreting the observer as a gravitational edge mode
has a bearing on the question of how many degrees of freedom
should be associated with the observer.  The simple model
employed here defined the observer in terms of a single
particle Hilbert space.  However, such a description is likely
too simplistic for a realistic theory.  As pointed out 
by CLPW \cite{Chandrasekaran2022a}, one needs to at least 
include a frame for the observer to allow nontrivial angular
dependence of the algebra in the de Sitter example.  
In the context of gravitational edge modes, a natural
choice is provided by the infinite-dimensional symmetry
algebra that arises from subregion charges, of which
the area is a single generator
\cite{Donnelly2016, Speranza2017, Ciambelli2021}.
Taking this idea to its logical
conclusion, one should model the observer as a representation
of the edge mode symmetry algebra, and consider the crossed 
product with respect to this much larger group.  We have been
informed that this connection between observers, edge modes, and 
crossed products will be explored in upcoming work by Freidel
and Gesteau \cite{Freidel2023}.

\subsection{Constraints, diffeomorphism invariance, and dressing}

We have found that imposing a single gravitational
constraint associated with the constant boost about the 
entangling surface is enough to arrive at 
type $\ttwo$ algebras for gravitational subregions.
However, in gravity, all compactly supported
diffeomorphisms are 
gauge transformations, and each is associated with an
independent constraint that must be imposed on the Hilbert
space and observable algebra.  This immediately 
raises the question of what became of this infinite
set of other constraints in the construction
of the subregion algebra.

To a large extent, these constraints have been
subsumed by assumption \ref{assm:aqft}, asserting
the existence of commuting subregion
algebras $\aqft$, $\aqft'$ constructed to 
all orders in the $\gc$ expansion.
This assumption entails the construction of operators with 
appropriate gravitational dressings in order to commute
with the gravitational constraints order by order 
in the $\gc$ expansion.  We described schematically
how this procedure might work in section \ref{sec:outline} by
dressing operators to the entangling surface in order to
satisfy microcausality,
but a more detailed investigation 
is warranted, perhaps along the lines
of the constructions considered in 
\cite{Donnelly:2015hta,
Donnelly:2016rvo, Giddings2022, Goeller2022}.

As discussed in section \ref{sec:outline}, the boost constraint 
is explicitly imposed on these algebras because we do not
expect quasilocal dressing within the subregion
to produce fully gauge-invariant operators.  This can
be understood from the perspective of partial gauge-fixing.
It is possible to fix the gauge within each subregion relative
to the entangling surface without violating 
microcausality, but a complete gauge fixing also requires
the gauges in the separate subregions to be related to
each other.  
The boost constraint  of assumption
\ref{assm:constr} synchronizes the global time variables
defined by the observer in each subregion.  However,
it seems likely that this is insufficient for
constructing a complete gauge fixing.  
Any diffeomorphism
that acts as an outer automorphism of the algebras 
$\aqft$ and $\aqft'$ could potentially lead to a nontrivial constraint
that is not solved by the quasilocal construction of the 
algebras.  These transformations include arbitrary diffeomorphisms
of the entangling surface, as well as position-dependent boosts.
A proposal for handling these constraints would be 
to give the observer additional degrees of freedom and 
implement a crossed product with respect to these
transformations, as suggested by the gravitational edge
mode picture discussed in section \ref{sec:edgemodes}.

One can also consider constraints associated with 
diffeomorphisms that deform the subregions, such 
as a time translation of the entangling surface to its 
future.  Satisfying these constraints seems to relate
 to the problem of specifying the subregion boundary
in a diffeo\-morphism-invariant manner.  
This question was briefly addressed in section \ref{sec:fixregion},
which proposed setting the leading order change in 
the subregion gravitational Hamiltonian $H_\xi^\Sigma$ to
zero as a gauge-fixing condition.  
This alone does not fully determine the subregion,
and we mentioned an additional idea for 
dynamically fixing the entangling surface $\partial\Sigma$
 by extremizing 
a functional $\frac{\kappa A}{8\pi G_N} + \mathcal{V}[\xi]$,
where the geometric functional 
$\mathcal{V}[\xi]$ is related to the subregion
Hamiltonian $H_\xi^\Sigma$.  When $\mathcal{V}[\xi] = 0$,
this procedure reduces to the Ryu-Takayanagi prescription
\cite{Ryu:2006bv,Ryu:2006ef,Hubeny:2007xt}, and when $\mathcal{V}[\xi]
 = -k\frac{\kappa V}{8\pi G_N}$, it can lead to bounded subregions
which in maximally symmetric spaces reduce to causal diamonds.  
It seems likely that a wide variety of subregions could be given
 diffeomorphism-invariant specifications by judiciously
choosing the functional $\mathcal{V}[\xi]$.

Finally, a common feature of diffeomorphism-invariant theories
is a lack of a preferred notion of time evolution.  
In our construction, a version of this arises in an arbitrariness
in the choice of the boost-generating vector field $\xi^a$
away from the boundary of the subregion.  Although one 
should expect this choice to not affect the resulting
gravitational algebra, it has a marked effect on
various formulas since different choices of vector fields
will produce different flows on the algebras 
$\aqft$, $\aqft'$, and result in different KMS states.
Luckily, so long as the vectors agree near the entangling 
surface, these flows will be related by Connes cocycles (assuming
the validity of the geometric modular flow conjecture 
discussed in section \ref{sec:modham}),
and the resulting crossed product algebras will
be unitarily equivalent \cite{Connes1973, Witten2021}.  
This is an instantiation 
of the background-independence of the gravitational crossed product
construction proposed by Witten \cite{Witten2021} and 
is reminiscent of the state-independent notion 
of thermal time proposed by Connes and Rovelli
\cite{Connes:1994hv}.  Exploring 
how to leverage this 
unitary equivalence to obtain unambiguous results for the 
gravitational subregion algebra and entropies
would be an interesting future direction to pursue.

\subsection*{Acknowledgments}
We thank Tom Faulkner, Laurent Freidel, Ted Jacobson, Moshe Rozali, and Mark Van Raamsdonk for helpful discussions. This work was initiated at the Galileo Galilei Institute workshop ``Reconstructing the Gravitational Hologram with Quantum Information.'' KJ thanks the Kavli Institute for Theoretical Physics for their hospitality while this work was being completed. His research was supported in part by an NSERC Discovery Grant and by the National Science Foundation under Grant No. NSF PHY-1748958. JS is supported by the Templeton Foundation via the Black Hole Initiative. AJS is supported by the Air Force Office of Scientific Research under award number FA9550-19-1-036.

\appendix

\section{Covariant phase space and constraints} \label{app:CPS}

The gravitational constraints are a key player in the argument
leading to local type $\ttwo$ von Neumann algebras, being the 
subjects of assumptions \ref{assm:constr} and \ref{assm:1stlaw}
given in section \ref{sec:assumptions}, and discussed
at length in section \ref{sec:constraints}.  This appendix gives 
an account of how these constraints arise in the canonical
theory as a consequence of diffeomorphism gauge symmetry.  
The covariant phase space formalism \cite{Crnkovic1987,
Lee:1990nz, Iyer1994}
is particularly
well suited for addressing this point, since it is 
a canonical formulation of the theory that preserves
the manifest diffeomorphism symmetry present in the Lagrangian
(see \cite{Harlow2019,Chandrasekaran2020,Chandrasekaran2021} for recent reviews).
The treatment of the constraints given here is closely 
related to the presentation in appendix B of \cite{Donnelly:2016rvo}.

The starting point is the Lagrangian $L[\phi]$, 
a differential form of maximal degree in spacetime that is 
a functional of the dynamical fields, collectively
denoted $\phi$, which include the metric $g_{ab}$ and 
other matter fields $\psi$. Varying the 
Lagrangian with respect to the dynamical
field produces the relations
\beq\label{eqn:deltaL}
\delta L = E_\phi \cdot \delta \phi
+ d\theta,
\eeq
where $E_\phi = 0$
define the metric and matter field equations, 
and $\theta = \theta[\phi;\delta\phi]$ is the symplectic
potential current. 

Diffeomorphisms are generate by vector fields $\xi^a$ and 
act on the dynamical fields by Lie derivatives,
\beq\label{eqn:delxiL}
\delta_\xi\phi = \lie_\xi\phi.
\eeq
Diffeomorphism-covariance of the Lagrangian implies that
\beq
\delta_\xi L = \lie_\xi L = di_\xi L,
\eeq
where $d$ denotes the exterior derivative and $i_\xi$ 
denotes contraction of the vector field $\xi^a$ into
the differential form.  Due to this equation, one can 
define the Noether current 
\beq
J_\xi = \theta[\phi;\lie_\xi\phi] - i_\xi L,
\eeq
whose exterior derivative is determined 
by (\ref{eqn:deltaL}) and (\ref{eqn:delxiL}) to be
\beq
dJ_\xi = -E_\phi \cdot \lie_\xi \phi,
\eeq
so that $J_\xi$ is conserved on-shell.  Since this equation
holds identically for arbitrary vectors $\xi^a$, the right hand 
side can be decomposed uniquely as 
\beq \label{eqn:Eliexiphi}
-E_\phi\cdot \lie_\xi \phi = dC_\xi + N_\xi
\eeq
where $C_\xi$ and $N_\xi$ depend algebraically on $\xi^a$
and not on its derivatives.  Because $N_\xi = d(J_\xi - C_\xi)$
is exact for arbitrary $\xi^a$, it must be the 
case that it vanishes.  The relations
\beq
N_\xi = 0
\eeq
are known as the Noether identities, and arise as a consequence 
of Noether's second theorem applied to local diffeomorphism
symmetry.  Therefore, $J_\xi - C_\xi$ is an identically
closed form for arbitrary $\xi^a$, and 
hence must be exact \cite{Wald1990}.  Thus there must 
exist a Noether potential $Q_\xi$ so that the Noether current
is given by
\beq
J_\xi = C_\xi + d Q_\xi.
\eeq

According to equation (\ref{eqn:Eliexiphi}), $C_\xi$ consists
of specific combinations of the equations of motion, and these 
combinations define the constraints of the theory.  
Their role as constraints, as opposed to standard dynamical
field equations, follows from the fact that they involve
fewer time derivatives than the other equations of motion,
and therefore restrict the initial data as opposed to determining
dynamics 
\cite{Seifert2007, Jacobson2011}.  

In the canonical formulation, the constraints play the role
of generators of gauge transformations.  This is seen by 
constructing a symplectic current 
$\omega = \delta\theta$,\footnote{We employ notation where 
$\delta$ denotes an exterior derivative on the space
of field configurations.  Hence, $\delta\theta$ should be thought
of as an antisymmetrized variation $\delta_2\theta[\delta_1\phi]
-\delta_1\theta[\delta_2\phi]$.}
whose integral over a complete Cauchy surface $\csfull$ 
determines the symplectic form
\beq \label{eqn:Omega}
\Omega = \int_{\csfull}\omega.
\eeq
Generically there can be additional boundary terms 
appearing in the expression for $\Omega$ (see the 
recent treatments in  \cite{Harlow2019,Chandrasekaran2020,Chandrasekaran2021})
but we will not display these explicitly; we will comment
on how these terms affect some later expressions below.
When evaluated on a diffeomorphism transformation,
the standard Iyer-Wald 
identities \cite{Iyer1994} produce the relation
\beq \label{eqn:Hameqn}
\Omega[\delta\phi, \lie_\xi\phi]
=
\int_{\csfull}\delta C_\xi + \int_{\partial\csfull}
(\delta Q_\xi - i_\xi\theta)
+\int_{\csfull} i_\xi E_\phi\cdot\delta\phi
\eeq
Upon accounting for the boundary terms in the symplectic 
form and the action and imposing necessary boundary conditions
\cite{Harlow2019, Iyer1994},
the second integral can be written as a total variation 
$\int_{\partial\csfull} \delta B_\xi$,
assuming that the vector field $\xi^a$ preserves
the chosen boundary conditions.  For asymptotic
boundaries, this generally requires
that $\xi^a$ approach an asymptotic Killing 
vector.  Since the last 
integral involves the field equations, the Hamiltonian
$\Hxig$
for this transformation is identified with
\beq\label{eqn:Hxigbulk+bdy}
\Hxig = \int_{\csfull} C_\xi + \int_{\partial\csfull} B_\xi,
\eeq
which generates the dynamics by Hamilton's equation
of motion,
\beq\label{eqn:Hameqnmotion}
\delta \Hxig = \Omega[\delta\phi,\lie_\xi\phi]
-\int_{\csfull} i_\xi E_\phi \cdot\delta\phi.
\eeq
Equation (\ref{eqn:Hxigbulk+bdy}) is the expression of the 
well-known statement that in gravity, the Hamiltonian
is given by an integral of the constraints, up to a 
boundary term.  

Equation (\ref{eqn:Hameqnmotion})  
shows that $\Hxig$ can be viewed as the generator
of the field transformation $\lie_\xi\phi$, as it implies the 
Poisson bracket relation
\beq
\{\phi,\Hxig\} = \lie_\xi\phi.
\eeq
When $\xi^a$ is compactly supported, the boundary term in 
(\ref{eqn:Hxigbulk+bdy}) vanishes, and the Hamiltonian
becomes purely an integral of the constraint.  This compactly 
supported diffeomorphism is a gauge transformation
of the theory, and hence we arrive at the statement
that the constraints generate gauge transformations.

To be more explicit about the form of the constraints,
we now specialize the theory to general relativity
minimally coupled to matter.  The dynamical fields
consist of the metric $g_{ab}$ and a collection
of matter fields $\psi$.  The Lagrangian splits into a sum
of a gravitational and matter contribution, $L=
L^g + L^m$, with
\beq
L^g = \frac{1}{16\pi G} \epsilon(R-2\Lambda)
\eeq
with $\epsilon$ the spacetime volume form, $R$ the 
Ricci scalar, and $\Lambda$ the cosmological constant.
The constraint is determined by equation 
(\ref{eqn:Eliexiphi}), and its precise form
depends on the tensor structure of the matter fields
$\psi$.  
It is always given by a term involving the 
Einstein equation, plus possible additional terms involving 
matter equations of motion,
\beq\label{eqn:Cxiexpr}
C_\xi = \epsilon_{a\ldots}\left(\frac{1}{8\pi G}(G\indices{^a_b}+\Lambda
\delta\indices{^a_b}) -T\indices{^a_b} \right) \xi^b +
E_\psi\text{-terms}.
\eeq
Exact expressions for the $E_\psi$ terms are given in
 \cite{Seifert2007, Jacobson2011, Faulkner:2013ica}.

In addition to matter fields, the crossed product 
construction
outlined in section \ref{sec:outline}
involves including
an observer degree of freedom into the theory,
who must couple to gravity universally via their
energy-momentum.  The simplest way to achieve this 
is to add an auxiliary phase space 
associated with the observer, which is simply taken to 
be the standard phase space on $\mathbb{R}^2$, with position
and momentum coordinates $(q,p)$.  As discussed in 
section \ref{sec:constraints}, the observer is modeled as a clock,
with $q$ interpreted as the observer's energy and the momentum
$p$ interpreted as the clock's time.  The observer's symplectic
form is simply given by
\beq
\Omega_\text{obs} = \delta q\wedge \delta p
\eeq
Since $p$ is the clock variable, defined to measure time 
along the flow of the specific generator $\xi^a$ considered
in the crossed product construction, this diffeomorphism acts on the observer variables 
by
\beq
\delta_\xi q = 0,\qquad \delta_\xi p = 1.
\eeq
Hence,
\beq
\Omega_\text{obs}(\delta(q,p), \delta_\xi(q,p)) = \delta q = 
\delta H_\text{obs}.
\eeq
Note that the constraint (\ref{eqn:Cxiexpr}) will be modified
to also include the observer's stress tensor in addition
to the matter stress tensor.  
Taking the extended symplectic form $\Omega_\text{ext}$ 
to be the sum of 
$\Omega_\text{obs}$ and the gravitational symplectic
form $\Omega$ defined in (\ref{eqn:Omega}), we find that
the Hamiltonian is modified to be
\beq
\Hxig + H_\text{obs} = \int_{\csfull}C^{\text{mat}+\text{obs}}_\xi +
\int_{\partial\csfull} B_\xi,
\eeq
where $C^{\text{mat}+\text{obs}}_\xi$ now includes the contribution
of the observer's energy momentum tensor.  

Finally, we can also discuss the construction of the local phase 
space within a subregion.  For this, we integrate
the symplectic current only over the partial Cauchy surface $\Sigma$,
\beq
\Omega^\sr = \int_\Sigma \omega.
\eeq
The Iyer-Wald identity now yields the relation for evaluating
this subregion symplectic form on a diffeomorphism
\beq
\Omega^\Sigma[\delta\phi,\lie_\xi\phi]
=\int_\Sigma \delta C_\xi +\int_{\partial\Sigma}\delta Q_\xi
+\int_\Sigma i_\xi E_\phi\cdot \delta\phi
\eeq
where the term $i_\xi\theta$ does not contribute
at $\partial\Sigma$ since $\xi^a$ is taken to vanish there.  
This equation defines a subregion gravitational Hamiltonian
generating the local flow according to
\beq
H_\xi^\Sigma = \int_\Sigma C_\xi + \int_{\partial\Sigma} Q_\xi,
\eeq
up to an additive constant, which we absorb into the 
definition of the subregion Hamiltonian $H_\xi^\Sigma$.
In general relativity, the Noether potential is given by
\cite{wald1993black, Iyer1994}
\beq
-\frac{1}{16\pi G} \epsilon_{ab\ldots}\nabla^a\xi^b
\eeq
At $\partial\Sigma$, the spacetime volume form decomposes
as $\epsilon = - n\wedge \mu$, with $n_{ab}$ the binormal
2-form normalized by $n^{ab} n_{ab} = -2$ and $\mu$ the 
volume form on $\partial\Sigma$.  By  choosing $\xi^a$
at $\partial\Sigma$ to satisfy
\beq
\nabla^a\xi^b \overset{\partial\Sigma}{=} \kappa n^{ab},
\eeq
with the surface gravity $\kappa$ constant, 
the Noether potential integrated over the boundary
becomes 
\beq
\int_{\partial\Sigma} Q_\xi= -\frac{\kappa A}{8\pi G}.
\eeq
The localized constraint including the contribution of the 
observer can then be written as 
\beq \label{eqn:Sigma1stlaw}
H_\xi^\Sigma + H_\text{obs}+ \frac{\kappa A}{8\pi G} 
= \int_{\partial\Sigma}C_\xi^{\text{mat}+\text{obs}}
\eeq
which is employed in (\ref{eqn:HSigmareln}).  A similar argument
applied to the complementary region $\sr'$ leads to 
equation (\ref{eqn:HbarSigmareln}).  

Finally, we briefly comment on possible ambiguities
that can arise in the definition of the gravitational
Hamiltonian and entropy functionals.  These 
arise from the ability to shift the Lagrangian 
and symplectic  current 
by boundary terms $L\rightarrow L+d\ell$, $\theta\rightarrow\theta
+d\beta$.  At physical boundaries, these ambiguities
are resolved by boundary conditions and demanding that the 
full action (including boundary terms) is stationary 
on-shell 
\cite{Compere:2008us, Harlow2019}.
For the subregion phase space, the ambiguities can be resolved by
considering the form of the boundary conditions one would 
impose for a subregion variational principle, even if these 
boundary conditions are not explicitly imposed
\cite{Chandrasekaran2020, Chandrasekaran2021}.  
Many of these ambiguities are not relevant
for the vector field $\xi^a$ considered here,
since it vanishes at $\partial\Sigma$.  However,
specifically in higher curvature gravitational theories, the
correct entropy functional is expected to be given by the Dong
entropy, which differs from the Wald entropy constructed 
from the covariant Noether potential by extrinsic curvature
terms \cite{Dong:2013qoa, Camps2013}. These corrections can be viewed as 
a choice of ambiguities for the covariant phase space 
\cite{Wall2015}, and 
it was suggested in \cite{Harlow2019}
that they might be determined by a 
more detailed analysis of the boundary
conditions for the subregion
phase space.

Note however that for any choice of ambiguity terms, equation
(\ref{eqn:Sigma1stlaw}) 
will continue to hold, as will the analogous equation
in the complementary region.  This is because the constraints
on the right hand side are constructed directly from the equations
of motion, which do not depend on how the ambiguities are resolved.
Hence, although the higher curvature gravitational Hamiltonian
$\Hxig$ depends on the choice of ambiguity terms, the entropy functional
does as well in such a way that the combination appearing 
in (\ref{eqn:Sigma1stlaw}) is independent of this choice.
It would be interesting to determine how these ambiguities
enter into the construction of the crossed product algebra
and generalized entropy in higher curvature theories.

\section{Types, crossed products, and their use} \label{app:algebraic-background}

This appendix gathers some pedagogical background on the aspects of von Neumann type theory and crossed product theory that are relevant for the present work.
More detail about the type classification of von Neumann algebras can be found in the review \cite{sorce2023notes}, and more detail about crossed products can be found in the review \cite{van1978continuous}.

A von Neumann algebra $\alg$ is a set of bounded operators on a Hilbert space $\hs$ that contains the identity and is closed under addition, multiplication, scalar multiplication, adjoints, and a particular kind of limit called a ``weak limit.''
$\alg$ is said to be a ``factor'' if its center is trivial, meaning that the only elements of $\alg$ which commute with all of $\alg$ are the scalar multiples of the identity.

Von Neumann factors are classified into types based on whether they contain operators that can be treated as density matrices.
Usually, a density matrix is defined as a positive operator with unit trace ($\rho \geq 0$ and $\tr(\rho) = 1$), though more generally one sometimes considers unnormalized density matrices where the trace is only required to be finite and nonzero.
Not every von Neumann algebra has density matrices.
For example, if $\hs$ is infinite-dimensional and $\alg$ is the set of scalar multiples of the identity, then $\alg$ contains no positive operators with finite, nonzero trace.
However, there is a sense in which certain von Neumann algebras admit \textit{effective} or \textit{renormalizable} density matrices.
To define these, we enlarge our definition of what we mean by a ``trace.''
Rather than taking ``trace'' to mean the specific Hilbert space trace defined on positive operators by $\tr(P) = \sum_{j} \bra{e_j}P\ket{e_j}$ for an orthonormal basis $\ket{e_j},$ we take a trace to be any map on $\alg$ that has all of the important physical properties enjoyed by the Hilbert space trace.

More precisely, a trace is defined to be a map $\tau$ from the space of positive operators in $\alg$ to the set $[0, \infty]$ satisfying certain linearity and cyclicity conditions.
Just like the Hilbert space trace, it takes values in the extended positive reals $[0, \infty]$ (some operators have infinite Hilbert space trace) and is naively defined only on positive operators (because not every non-positive operator has a well defined, basis-independent Hilbert space trace), but can be extended to take finite values in $\mathbb{C}$ for any ``trace class'' operator $T$ satisfying $\tau(\sqrt{T^{\dagger} T}) < \infty$.
The linearity and cyclicity conditions are spelled out explicitly in \cite[definition 6.1]{sorce2023notes}.
A \textit{physical} trace --- called ``faithful, normal, and semifinite'' by mathematicians --- is a trace $\tau$ that has good physical properties: nonzero operators have nonzero trace, $\tau$ is continuous in a sense appropriate for functions valued in $[0, \infty]$, and $\tau$ assigns finite trace to at least one nonzero operator.
See \cite[definition 6.2]{sorce2023notes} for details.

The questions of existence and uniqueness for physical traces on a von Neumann algebra can be thought of as issues of \textit{renormalizability} and \textit{scheme dependence}.
If a physical trace exists, we can think of it as a renormalization scheme for treating certain infinite-Hilbert-space-trace operators as effective density matrices.
If there are multiple inequivalent physical traces, these represent inequivalent renormalization schemes.
It has been shown (see e.g. theorem 2.31 of \cite{TakesakiI}) that on a von Neumann factor $\alg,$ any two physical traces must be related by a multiplicative constant, $\tau = c \tau',$ so for factors the scheme dependence of renormalization is extremely restricted.
Among von Neumann factors, we say a factor is type III if it admits no physical trace --- i.e., the only faithful and normal trace sends every nonzero positive operator to infinity --- and type I or II if it admits a physical trace.
The classification is completed by saying that a factor is of type I if it has an orthogonal projector with minimal trace, and type II if it has orthogonal projectors of arbitrarily small trace.
Physically, type I factors should be thought of as containing pure states, since they have projectors that are ``effectively rank-one'' in that they cannot be subdivided further; type II factors should be thought of as containing mixed states, but no pure states.
A type II factor is said to be of type II$_1$ if the identity operator has finite trace, in which case the identity can be treated as the density matrix for a ``maximally mixed'' state, and plays the role of a state of maximum entropy.
A type II factor is said to be of type II$_{\infty}$ if the identity has infinite trace, in which case there is no state of maximum entropy.

The relevance of all this for the present paper is that it was argued in \cite{fredenhagen1985modular, buchholz1995scaling} that within a single superselection sector of a quantum field theory with an ultraviolet fixed point, the algebras of operators localized to subregions are von Neumann factors of type III$_1.$
The subscript in ``type III$_1$'' relates to a further classification of type III factors due to Connes \cite{Connes1973}, and means that every modular operator has spectrum supported on the full positive reals $[0, \infty).$
While type III algebras admit no renormalization schemes, Takesaki proved \cite[corollary 9.7]{Takesaki1973} that the crossed product of a type III$_1$ factor by any of its modular operators is a type II factor (more specifically, a type II$_{\infty}$ factor).
Since type II factors admit essentially unique renormalization schemes, the crossed product of a type III$_1$ factor by one of its modular operators is a useful tool for renormalizing infinite quantities.
This tool is actually computationally practical: Takesaki gave an implicit characterization of the physical trace on a crossed product algebra in \cite[lemma 8.2]{Takesaki1973}, which was later written as a concrete formula by Witten in \cite{Witten2021}.

The traditional construction of a crossed product by a modular operator is given as follows.
Let $\alg$ be a von Neumann algebra acting on the Hilbert space $\hs,$ and let $L^2(\mathbb{R})$ be an auxiliary Hilbert space, so that the total space under consideration is $\hs \otimes L^2(\mathbb{R}).$
Let $\Delta_{\Psi}$ be the modular operator for some state $\ket{\Psi}$ (see appendix \ref{app:modth}).
The Hilbert space $\hs \otimes L^2(\mathbb{R})$ can be thought of as the space of square-integrable functions from $\mathbb{R}$ into $\hs,$ by decomposing a vector $\ket{\psi} \in \hs \otimes L^2(\mathbb{R})$ in terms of the unnormalized momentum basis as\footnote{Note that in much of the mathematical literature, the position basis is used, which leads to an equivalent description of the crossed product. We use the momentum basis to match the conventions of \cite{Chandrasekaran2022a}.}
\begin{equation}
    \ket{\psi} = \int dp \ket{\psi(p)} \otimes \ket{p}.
\end{equation}
In this way, we can identify $\ket{\psi}$ with the function $p \mapsto \ket{\psi(p)}.$
The crossed product $\alg \rtimes_{\Delta} \mathbb{R}$ is defined as the smallest von Neumann algebra containing two kinds of operators: (i) the translation operators that act on $L^2(\mathbb{R})$ as $\ket{p} \mapsto \ket{p+ t}$, and (ii) the twirled operators that act on $\ket{\psi(p)}$ as $\ket{\psi(p)} \mapsto \Delta_{\Psi}^{- i p} \msf{a} \Delta_{\Psi}^{i p} \ket{\psi(p)}$ for some $\msf{a} \in \alg.$
In terms of the position and momentum operators $\hat{q}, \hat{p}$ acting on $L^2(\mathbb{R}),$ and the modular Hamiltonian $h_{\Psi} = - \log \Delta_{\Psi}$ acting on $\hs,$ these two kinds of operators can be written
\begin{alignat}{2}
    \text{(i)} : & \quad e^{i \hat{q} t} & \quad t \in \mathbb{R}, \\
    \text{(ii)} : & \quad e^{i h_{\Psi} \hat{p}} \msf{a} e^{- i h_{\Psi} \hat{p}} & \quad \msf{a} \in \alg.
\end{alignat}
Thus we may write the crossed product as
\begin{equation} \label{eqn:standard-crossed-product}
    \alg \rtimes_{\Delta} \mathbb{R}
        = \{ e^{i h_{\Psi} \hat{p}} \msf{a} e^{- i h_{\Psi} \hat{p}}, e^{i \hat{q} t}\, | \, \msf{a} \in \alg, t \in \mathbb{R}\}''
\end{equation}
where $S''$ denotes the \textit{double commutant} of the set $S$, which is known to be equal to the smallest von Neumann algebra containing all elements of $S$ \cite{v1930algebra}.
Conjugating by the unitary operator $e^{-i h_{\Psi} \hat{p}}$ leads to an equivalent definition of the crossed product algebra with respect to the following set of generators:
\begin{equation} \label{eqn:dressing-crossed-product}
    \alg \rtimes_{\Delta} \mathbb{R}
        \cong \{ \msf{a} \otimes 1, \Delta_{\Psi}^{i t} \otimes e^{i \hat{q} t}\, | \, \msf{a} \in \alg, t \in \mathbb{R}\}''.
\end{equation}
This representation of the crossed product gives an intuitive picture for its physical meaning: the crossed product consists of operators in $\alg$ dressed by modular flow, with momentum-space translations in the auxiliary register $L^2(\mathbb{R})$ being used to keep track of modular time.

A powerful result known as the \textit{commutation theorem} for crossed products tells us that the crossed product algebra given in equation \eqref{eqn:standard-crossed-product} is exactly the subalgebra of $\alg \otimes \mathcal{B}(L^2(\mathbb{R}))$ fixed under the flow generated by $h_{\Psi} + \hat{q}$.
That is, we have (see e.g. \cite[chapter I, theorem 3.11]{van1978continuous})
\begin{equation}
    \alg \rtimes_{\Delta} \mathbb{R}
        = \{\wh{\msf{a}} \in \alg \otimes \mathcal{B}(L^2(\mathbb{R}))\, | \, e^{i (h_{\Psi} + \hat{q}) t} \wh{\msf{a}} e^{- i (h_{\Psi} + \hat{q}) t} = \wh{\msf{a}} \text{ for all } t \in \mathbb{R}\}.
\end{equation}
This is why, in the main text, crossed products arise as a consequence of imposing gauge symmetry; if $h_{\Psi} + \hat{q}$ is the generator of some gauge symmetry, then gauge invariance requires restricting to the crossed product subalgebra.

It is known \cite[corollary 9.7]{Takesaki1973} that if $\alg$ is a type III$_1$ factor, then $\alg \rtimes_{\Delta} \mathbb{R}$ is a type II$_{\infty}$ factor.
As emphasized above, this means that it has a preferred physical trace that is unique up to rescaling.
A concrete formula for the trace can be obtained by unpacking the proof of lemma 8.2 in \cite{Takesaki1973}.
In the present notation if $\ket{\Psi}$ is the state whose modular flow is used to define the crossed product, and $\ket{0}_p$ is the unnormalizable zero-momentum state, then the formal state $|\wh{\Psi}\rangle = \ket{\Psi} \otimes \ket{0}_p$ has special properties.
In particular, its modular flow $\wh{\Delta}_{\Psi}^{it}$ satisfies\footnote{Sign differences in this expression relative to expressions in \cite{Takesaki1973} are due to our convention that modular flow acts as $\msf{a} \mapsto \Delta^{-i t} \msf{a} \Delta^{it},$ as opposed to Takesaki's convention $\msf{a} \mapsto \Delta^{it} \msf{a} \Delta^{-it}.$}
\begin{equation}
    \wh{\Delta}_{\Psi}^{-it}\, \wh{\msf{a}}\, \wh{\Delta}_{\Psi}^{i t}
        = e^{-i \hat{q} t} \wh{\msf{a}} e^{i \hat{q} t}
\end{equation}
for all $\wh{\msf{a}} \in \alg \rtimes_{\Delta} \mathbb{R}.$
This shows that the modular flow associated to the state $|\tilde{\Psi}\rangle$ is inner.
Takesaki then shows that if any algebraic state $\varphi$ has inner modular flow of the form $\msf{c}^{it}$ for some positive operator $\msf{c}$ affiliated with the algebra, then the map $\tau(\msf{a}) = \varphi(\msf{c} \msf{a})$ is a physical trace.
In the present context, where $\varphi$ is the algebraic state on the crossed product defined by taking expectation values in $\ket{\Psi} \otimes \ket{0},$ this implies that the map
\begin{equation}
    \atr(\wh{\msf{a}})
        = (\bra{\Psi} \otimes \bra{0}) e^{- \hat{q}} \wh{\msf{a}} (\ket{\Psi}\otimes \ket{0})
\end{equation}
is a physical trace on $\alg \rtimes_{\Delta} \mathbb{R}.$
We have adopted the notation $\atr$ for the physical trace on a crossed product, as this is the notation used in the main text.

In the main text, the crossed product is not taken exactly by a modular flow, but rather by a modular flow rescaled by an inverse temperature $\beta$ determined by the surface gravity of a vector field $\xi^a.$
This requires rescaling $h_{\Psi}$ to $h_{\Psi}/\beta$.
The effect of this is to give the crossed product algebra as
\begin{equation}
	\alg \rtimes_{\Delta} \mathbb{R}
	= \{ e^{i \hat{p} \hb} \msf{a} e^{- i \hat{p} \hb}, e^{i \hat{q} t}\, | \, \msf{a} \in \alg, t \in \mathbb{R}\}'',
\end{equation}
as in equation \eqref{eq:crossed-product-main-text}, with the isomorphic representation
\begin{equation}
	\alg \rtimes_{\Delta} \mathbb{R}
	= \{ \msf{a} \otimes \mathbbm{1}, e^{- i \hb t} \otimes e^{i \hat{q} t} \, | \, \msf{a} \in \alg, t \in \mathbb{R}\}''.
\end{equation}
The commutation theorem tells us that $\alg \rtimes_{\Delta} \mathbb{R}$ is exactly the flow-fixed algebra
\begin{equation}
    \alg \rtimes_{\Delta} \mathbb{R}
        = \{ \wh{\msf{a}} \in \alg \otimes \mathcal{B}(L^2(\mathbb{R}))\, |\,
            e^{i \left( \hb + \hat{q} \right) t} \wh{\msf{a}} e^{- i \left( \hb + \hat{q} \right) t} = \wh{\msf{a}} \text{ for all } t \in \mathbb{R}\},
\end{equation}
as claimed in section \ref{sec:outline}.
Finally, the physical trace is
\begin{equation} \label{eq:appendix-trace-formula}
    \atr\left( \wh{\msf{a}} \right)
        = 2 \pi \beta (\bra{\Psi} \otimes \bra{0}_p) e^{- \beta \hat{q}} \wh{\msf{a}} (\ket{\Psi} \otimes \ket{0}_p),
\end{equation}
where $\ket{0}_p$ is the unnormalizable zero-momentum state and we have introduced an overall constant $2 \pi \beta$ to make certain equations in the main text simpler.
In particular, when we produce a type II$_1$ algebra in section \ref{sec:encond}, the identity in that algebra has unit trace with respect to this normalization.

\section{Modular theory} \label{app:modth}

Many of the results in this work rely on properties of modular
automorphism groups and Tomita-Takesaki theory for von Neumann
algebras.  In this appendix, we will collect the main results that 
are used throughout the paper, to serve as a quick reference for notation
and definitions.  The discussion here will be informal; 
precise mathematical statements and proofs of the results quoted here
can be found in the cited references.  
See \cite{Witten:2018zxz} for an accessible introduction
and more detailed explanations.
Further mathematical detail can be found in \cite{sunder2012invitation, Petz1993}, with a complete treatment in \cite{TakesakiII}.

Given a von Neumann algebra $\alg$ acting on a Hilbert space $\hs$ and 
a cyclic and separating vector $|\Psi\rangle \in \hs$, the Tomita
operator $S_\Psi$ is an antilinear operator defined by the relation
\beq
S_\Psi \msf{a} |\Psi\rangle = \msf{a}^\dagger|\Psi\rangle, \qquad \forall
\msf{a}\in\alg.
\eeq
It admits a polar decomposition
\beq
S_\Psi = J_\Psi \Delta_\Psi^{\frac12},
\eeq
where the modular conjugation $J_\Psi$ is an antiunitary operator satisfying $J_{\Psi}^2 = 1$ and the modular operator
$\Delta_\Psi=S_\Psi^\dagger S_\Psi$ is Hermitian and positive-definite.
Both operators leave the state $|\Psi\rangle$ invariant, $J_\Psi|\Psi\rangle = \Delta_\Psi|\Psi\rangle = |\Psi\rangle$,
and furthermore modular conjugation sends $\Delta_\Psi$ to its inverse,
\beq
J_\Psi \Delta_\Psi J_\Psi = \Delta_\Psi^{-1}.
\eeq
The Tomita operator $S_{\Psi}'$ for the commutant algebra $\alg'$ admits a polar decomposition $S_{\Psi}' = J_{\Psi} \Delta_{\Psi}^{-1/2},$ so the modular conjugation for $\alg'$ is the same as for $\alg$ and the modular operator is $\Delta_{\Psi}^{-1}.$
Furthermore, if $\msf{a}$ is in $\alg,$ then $J_{\Psi} \msf{a} J_{\Psi}$ is in $\alg'.$

From its definition, one finds that $\Delta_\Psi$ satisfies
\beq\label{eqn:DelPsiflip}
\langle \Psi|\msf{a}\msf{b}|\Psi\rangle = \langle\Psi|\msf{b}\Delta_\Psi
\msf{a}|\Psi\rangle, \qquad \forall \msf{a},\msf{b}\in\alg.
\eeq
The modular Hamiltonian $h_\Psi$ is defined by 
\beq
h_\Psi = -\log\Delta_\Psi,
\eeq
and is used to define the modular flow of operators in $\alg$,
\beq
\msf{a}_s = e^{ish_\Psi} \msf{a}e^{-is h_\Psi}.
\eeq
The flowed operator $\msf{a}_s$ remains in $\alg$ for all real values of $s$, and
hence modular flow defines an automorphism of $\alg$.  When $\alg$ is type
$\tthr_1$, this automorphism is outer for all values of $s$, implying that 
$h_\Psi$ cannot be expressed as a sum of an element of $\alg$ and an element
of $\alg'$.

Although $\Delta_\Psi$ does not factorize
when $\alg$ is type $\tthr$, it is often helpful to keep in mind the formula 
for this operator when $\alg$ is type $\tone$ and the Hilbert space factorizes
$\hs = \hs_\alg \otimes \hs_{\alg'}$.  In this case, any state $|\Psi\rangle$
defines a density matrix $\rho_\Psi$ for $\alg$ acting on $\hs_\alg$ and a density matrix
$\rho_\Psi'$
for $\alg'$ acting on $\hs_{\alg'}$.  In terms of these, the modular operator
is given by
\beq \label{eqn:factorized-Delta-type-I}
\Delta_\Psi = \rho_\Psi\otimes (\rho_\Psi')^{-1}
\eeq
and the modular Hamiltonian can be expressed as a sum
\beq
h_\Psi = h_\Psi^\alg - h_\Psi^{\alg'},
\eeq
with $h_{\Psi}^{\alg} = - \log \rho_{\Psi}$ and $h_{\Psi}^{\alg'} = - \log \rho_{\Psi}'.$
This formal split of the modular Hamiltonian is employed in sections
\ref{sec:modham} and \ref{sec:crossprod}, especially in deriving the generalized entropy
formula (\ref{eqn:SrhoSgen}), but when $\alg$ is type $\tthr$, the one-sided 
modular Hamiltonians $h_\Psi^\alg$, $h_\Psi^{\alg'}$ are singular
operators.

Given another cyclic and separating state $|\Phi\rangle$, we can
define the relative Tomita operator $S_{\Phi|\Psi}$ by the relation\footnote{There
are differing conventions for relative operators throughout the literature,
where the Tomita operator in this equation is sometimes denoted $S_{\Psi|\Phi}$.  
Here we follow the conventions employed in CLPW \cite{Chandrasekaran2022a},
which also agrees with those of \cite{Lashkari:2019ixo, Haag1992},
but is opposite the conventions of \cite{Witten:2018zxz, Ceyhan:2018zfg}.}
\beq
S_{\Phi|\Psi}\msf{a}|\Psi\rangle = \msf{a}^\dagger|\Phi\rangle.
\eeq
The polar decomposition $S_{\Phi|\Psi} = J_{\Phi|\Psi}
\Delta_{\Phi|\Psi}^{\frac12}$ defines the antiunitary relative
modular conjugation $J_{\Phi|\Psi}$ and relative modular operator
$\Delta_{\Phi|\Psi}$.
From the fact that $S_{\Phi|\Psi} S_{\Psi|\Phi} = \mathbbm{1}$, one finds
that $J_{\Phi|\Psi} J_{\Psi|\Phi} = \mathbbm{1}$, implying
\beq \label{eqn:rel-mod-adjoint-flip}
J_{\Phi|\Psi}^\dagger = J_{\Psi|\Phi},
\eeq
and also 
\beq \label{eqn:relmod-inverse}
J_{\Phi|\Psi}\Delta_{\Phi|\Psi}^{\frac12} J_{\Psi|\Phi} =
\Delta_{\Psi|\Phi}^{-\frac12}.
\eeq
Similar manipulations yield the following relations for the relative Tomita operator $S_{\Phi|\Psi}'$ for the commutant $\alg',$ with polar decomposition $J_{\Phi|\Psi}' (\Delta_{\Phi|\Psi}')^{\frac12}$:
\begin{align} \label{eqn:rel-mod-commutant-identities}
    S_{\Phi|\Psi}^{\dagger}
         = S_{\Phi|\Psi}', \qquad
    J_{\Phi|\Psi}'
         = J_{\Phi|\Psi}^{\dagger}, \qquad
    (\Delta_{\Phi|\Psi}')^{1/2}
         = \Delta_{\Psi|\Phi}^{-1/2}.
\end{align}
The relative modular operator may be expressed as
$\Delta_{\Phi|\Psi}=S_{\Phi|\Psi}^\dagger S_{\Phi|\Psi}$, which leads
to a relation analogous to (\ref{eqn:DelPsiflip}),
\beq\label{eqn:relmodflip}
\langle\Phi |\msf{a}\msf{b}|\Phi\rangle = \langle\Psi|\msf{b}
\Delta_{\Phi|\Psi}\msf{a}|\Psi\rangle.
\eeq
The logarithm of $\Delta_{\Phi|\Psi}$ determines the relative
modular Hamiltonian
\beq
h_{\Phi|\Psi} = -\log\Delta_{\Phi|\Psi}.
\eeq
In terms of the formal type $\tone$ factorization, the relative modular 
operator and modular Hamiltonian can be expressed
\begin{align}
\Delta_{\Phi|\Psi} &= \rho_\Phi\otimes (\rho_\Psi')^{-1} \\
h_{\Phi|\Psi} &= h_\Phi^\alg - h_\Psi^{\alg'}.
\end{align}

An important set of operators that appear when relating
modular flows of different states are the Connes cocycles $u_{\Phi|\Psi}(s)$,
$u'_{\Psi|\Phi}(s)$, defined by 
\cite{Connes1973, Haag1992, TakesakiII}
\begin{align}
u_{\Phi|\Psi}(s) &= \Delta_{\Phi|\Psi}^{is}\Delta_\Psi^{-is} = \Delta_\Phi^{is}
\Delta_{\Psi|\Phi}^{-is} \\
u_{\Psi|\Phi}'(s) &= \Delta_{\Phi|\Psi}^{-is}\Delta_{\Phi}^{is} = \Delta_\Psi^{-is}
\Delta_{\Psi|\Phi}^{is}
\label{eqn:u'}
\end{align}
The equivalence of these two definitions of the Connes cocycles can be checked
using the type $\tone$ density matrix expressions for the modular operators,
which also reveals that $u_{\Phi|\Psi}(s)$ is an element of $\alg$ and 
$u_{\Psi|\Phi}'(s)$ is an element of $\alg'$.  These statements remain
true even in the case of type $\tthr$ algebras, despite the fact that 
the factorization of the modular operators is not valid in this situation
(see \cite{Lashkari:2019ixo} for a recent review of the proof for generic
von Neumann algebras).
From these definitions, one also verifies that  evolution with 
respect to different modular Hamiltonians $h_\Phi$ and $h_\Psi$ is 
related by
\beq
e^{-ish_\Phi} = u_{\Phi|\Psi}(s) e^{-is h_\Psi} u_{\Psi|\Phi}'(s),
\eeq
implying that any two modular flows on $\alg$ are related by an inner
automorphism, and similarly for $\alg'$.  This is the content
of the cocycle derivative theorem  in the theory
of von Neumann algebras \cite{Connes1973}\cite[Theorem 3.3]{TakesakiII}.  

A useful class of states employed in this 
work are canonical purifications.  
Given a fixed vector $|\Psi\rangle \in \hs$ and some state $\omega$ on $\alg$, 
there is a unique canonical purification $|\Phi\rangle$ that reproduces
the expectation values of $\omega$ on elements of $\alg$.  The vector $|\Phi\rangle$
is 
an element of the {\it canonical cone} associated
with $|\Psi\rangle$, 
and has the property of possessing the same modular conjugation
as $|\Psi\rangle$, i.e.\ $J_\Phi  = J_\Psi$, and
these are equal to the relative modular conjugations $J_{\Psi|\Phi} = 
J_{\Phi|\Psi} = J_\Psi$ \cite{Araki1974, Haag1992}. 
The vector $|\Phi\rangle$ satisfies the properties
\beq
J_\Psi|\Phi\rangle = |\Phi\rangle, \qquad |\Phi\rangle = \Delta_{\Phi|\Psi}^{\frac12}
|\Psi\rangle.
\eeq
Because all of the modular conjugations are equal, one also finds
\beq
J_\Psi\Delta_\Phi J_\Psi = \Delta_\Phi^{-1}, \qquad J_\Psi \Delta_{\Phi|\Psi}
J_\Psi = \Delta_{\Psi|\Phi}^{-1}.
\eeq
When $|\Phi\rangle$ 
is a cyclic-separating vector that is not in the canonical cone of 
$|\Psi\rangle$, 
it can always be written in the form $|\Phi\rangle
= \msf{u}'|\Phi_c\rangle$, where $|\Phi_c\rangle$ is in the canonical
cone and $\msf{u}'\in\alg'$ is unitary \cite{Araki1974}.  
Thus, when computing expectation values or entropies for the algebra
$\alg$, one is always free to assume that the vector for the state
under consideration is a canonical purification.

An explicit formula for $\msf{u}'$ can be obtained by noting
that the relative Tomita operators for $|\Phi\rangle$ and $|\Phi_c\rangle
$ with respect to $|\Psi\rangle$ are related by 
\beq
S_{\Phi|\Psi} = \msf{u}'S_{\Phi_c|\Psi},
\eeq
which follows directly from the definitions of the 
relative Tomita operators: 
$S_{\Phi|\Psi}\msf{a}|\Psi\rangle
= \msf{a}^\dagger |\Phi\rangle = \msf{u}' a^\dagger|\Phi_c\rangle
= \msf{u}'S_{\Phi_c|\Psi}\msf{a}|\Psi\rangle$.  By the uniqueness
of the polar decomposition and the fact that $J_{\Phi_c|\Psi} = J_\Psi$, 
we then find that 
\beq
\Delta_{\Phi|\Psi} = \Delta_{\Phi_c|\Psi},\qquad J_{\Phi|\Psi} = \msf{u}'
J_\Psi,
\eeq
and hence 
\beq \label{eqn:uprime-explicit}
\msf{u}' = J_{\Phi|\Psi} J_\Psi.
\eeq
Note that other modular operators are related to their
canonically purified versions according to
\beq \label{eqn:DelPsiPhicanon}
\Delta_{\Psi|\Phi} = \msf{u}'\Delta_{\Psi|\Phi_c}
(\msf{u}')^\dagger, \qquad 
\Delta_{\Phi} = \msf{u}'\Delta_{\Phi_c}
(\msf{u}')^\dagger.
\eeq

We close with a summary of certain useful identities relating products of modular operators that are affiliated with $\alg$ or $\alg'.$
As explained in e.g. \cite[remark 5.3.10]{pedersen1979c}, an unbounded operator is said to be affiliated with $\alg$ if it commutes with every operator in $\alg'$; equivalently, if every bounded function of the operator is in $\alg.$
It is easy to see that the operator $S'_{\Psi} S'_{\Psi|\Phi}$ is affiliated with $\alg$, as for any $\msf{a}', \msf{b}'$ in $\alg',$ we have
\begin{equation}
    S_{\Psi}' S_{\Psi|\Phi}' \msf{a}' \ket{\msf{b}' \Phi}
        = \msf{a}' \ket{\msf{b}' \Psi}
        = \msf{a}' S'_{\Psi} S'_{\Psi|\Phi} \ket{\msf{b}' \Phi},
\end{equation}
so the operator $[S'_{\Psi} S'_{\Psi|\Phi}, \msf{a}']$ vanishes on all states of the form $\ket{\msf{b}' \Phi}.$ These states are dense in $\hs$ by the assumption that $\ket{\Phi}$ is cyclic and separating, so we have $[S_{\Psi}' S_{\Psi|\Phi}', \msf{a}'] = 0.$
Writing the operator $S_{\Psi}' S_{\Psi|\Phi}'$ in terms of its polar decomposition and using equations \eqref{eqn:rel-mod-adjoint-flip}, \eqref{eqn:relmod-inverse}, and \eqref{eqn:rel-mod-commutant-identities} gives
\begin{equation}
    S_{\Psi}' S_{\Psi|\Phi}'
        = J_{\Psi} \Delta_{\Psi}^{-\frac12} \Delta_{\Psi|\Phi}^{\frac12} J_{\Phi | \Psi}.
\end{equation}
Since $S_{\Psi}' S_{\Psi|\Phi}'$ is affiliated with $\alg,$ conjugating this operator by $J_{\Psi}$ produces an operator affiliated with $\alg'.$ So the operator
\begin{equation}
    J_{\Psi}S_{\Psi}' S_{\Psi|\Phi}' J_{\Psi}
        = \Delta_{\Psi}^{-\frac12} \Delta_{\Psi|\Phi}^{\frac12} J_{\Phi | \Psi} J_{\Psi}
\end{equation}
is affiliated with $\alg'.$
But we already know via equation \eqref{eqn:uprime-explicit} that $J_{\Phi | \Psi} J_{\Psi}$ is in $\alg'$, so $\Delta_{\Psi}^{-\frac12} \Delta_{\Psi|\Phi}^{\frac12}$ must be affiliated with $\alg'$ as well.

By repeating the logic of the above paragraph but switching the roles of $\alg$ and $\alg',$ using the identities in equation \eqref{eqn:rel-mod-commutant-identities}, and taking adjoints or inverses when convenient, we see that the following operators are affiliated with $\alg$ and $\alg'.$
\begin{equation} \label{eqn:affiliated-operators}
  \begin{alignedat}{3}
   \ccol{\underline{Affiliated with $\alg$}} & \qquad\quad & \ccol{\underline{Affiliated with $\alg'$}} \\
    \ccol{$S_{\Psi|\Phi} S_{\Psi}$} & \qquad\quad & \ccol{$S_{\Psi} S_{\Psi|\Phi}$} \\
    \ccol{$J_{\Psi|\Phi} J_{\Psi}$} & \qquad\quad & \ccol{$J_{\Phi|\Psi} J_{\Psi}$} \\
    \ccol{$\Delta_{\Phi|\Psi}^{\frac12} \Delta_{\Psi}^{-\frac12}$} & \qquad\quad & \ccol{$\Delta_{\Psi|\Phi}^{\frac12} \Delta_{\Psi}^{-\frac12} $}
  \end{alignedat}
\end{equation}
These operators satisfy certain useful identities.
It is easy to verify the expression
\begin{equation}\label{eqn:S-equals-S}
    S_{\Psi} S_{\Psi|\Phi} = S_{\Psi|\Phi} S_{\Phi}
\end{equation}
by checking that these operators have the same action on the dense set of states of the form $\ket{\msf{a} \Phi}.$
Cocycle manipulations presented in e.g. \cite[appendix C]{Araki1982}\cite[appendix A]{Ceyhan:2018zfg}
verify the identity
\begin{equation} \label{eqn:modular-conjugation-cocycle}
    J_{\Phi|\Psi} J_{\Psi} = J_{\Phi} J_{\Phi|\Psi}.
\end{equation}
Finally, expanding equation \eqref{eqn:S-equals-S} in terms of polar decompositions and applying equations \eqref{eqn:rel-mod-adjoint-flip}, \eqref{eqn:relmod-inverse}, \eqref{eqn:rel-mod-commutant-identities} gives
\begin{equation}
    J_{\Psi} \Delta_{\Psi}^{\frac12} \Delta_{\Phi|\Psi}^{-\frac12} J_{\Psi|\Phi} = J_{\Psi|\Phi} \Delta_{\Psi|\Phi}^{\frac12} \Delta_{\Phi}^{-\frac12} J_{\Phi}.
\end{equation}
Left-multiplying by $J_{\Psi}$ and right-multiplying by $J_{\Phi|\Psi}$ gives
\begin{equation}
    \Delta_{\Psi}^{\frac12} \Delta_{\Phi|\Psi}^{-\frac12} = J_{\Psi} J_{\Psi|\Phi} \Delta_{\Psi|\Phi}^{\frac12} \Delta_{\Phi}^{-\frac12} J_{\Phi} J_{\Phi|\Psi}.
\end{equation}
By equation \eqref{eqn:affiliated-operators}, $J_{\Phi} J_{\Phi|\Psi}$ is affiliated with $\alg'$ and $\Delta_{\Psi|\Phi} \Delta_{\Phi}^{-1/2}$ is affiliated with $\alg,$ so these commute, and we have
\begin{equation}
    \Delta_{\Psi}^{\frac12} \Delta_{\Phi|\Psi}^{-\frac12} = J_{\Psi} J_{\Psi|\Phi} J_{\Phi} J_{\Phi|\Psi} \Delta_{\Psi|\Phi}^{\frac12} \Delta_{\Phi}^{-\frac12}.
\end{equation}
Applying the identity \eqref{eqn:modular-conjugation-cocycle} then gives
\begin{equation}
    \Delta_{\Psi}^{\frac12} \Delta_{\Phi|\Psi}^{-\frac12} = \Delta_{\Psi|\Phi}^{\frac12} \Delta_{\Phi}^{-\frac12}.
\end{equation}
Repeating these arguments with slight variations, occasionally substituting $\Psi \leftrightarrow \Phi$, gives the following table of identities.
\begin{equation} \label{eqn:affiliated-operators-cocycles}
  \begin{alignedat}{3}
   \ccol{\underline{Affiliated with $\alg$}} & \qquad\quad & \ccol{\underline{Affiliated with $\alg'$}} \\
    \ccol{$S_{\Psi|\Phi} S_{\Psi} = S_{\Phi} S_{\Psi|\Phi}$} & \qquad\quad & \ccol{$S_{\Psi} S_{\Psi|\Phi} = S_{\Psi|\Phi} S_{\Phi}$} \\
    \ccol{$J_{\Psi|\Phi} J_{\Psi} = J_{\Phi} J_{\Psi|\Phi}$} & \qquad\quad & \ccol{$J_{\Phi|\Psi} J_{\Psi} = J_{\Phi} J_{\Phi|\Psi}$} \\
    \ccol{$\Delta_{\Phi|\Psi}^{\frac12} \Delta_{\Psi}^{-\frac12} = \Delta_{\Phi}^{\frac12} \Delta_{\Psi|\Phi}^{-\frac12}$} & \qquad\quad & \ccol{$\Delta_{\Psi|\Phi}^{\frac12} \Delta_{\Psi}^{-\frac12} = \Delta_{\Phi}^{\frac12} \Delta_{\Phi|\Psi}^{-\frac12}$}
  \end{alignedat}
\end{equation}

\section{Converse of the cocycle derivative theorem}
\label{sec:convcoc}

One of the main justifications for assumption \ref{assm:mod}
provided in section \ref{sec:modham} is the fact that 
any operator $h_{\msf{a}\msf{b'}}$ 
related to a modular Hamiltonian $h_0$ 
as in equation (\ref{eqn:innerrelated}) is itself the modular
Hamiltonian of some state.  To show this, one invokes the 
converse of the cocycle derivative theorem
\cite{Connes1973, TakesakiII}, a version of which
can be stated as follows.  Suppose that $h_1$ is the generator
of a flow on a von Neumann algebra $\alg$ and on its commutant $\alg'$
that is related to a modular flow generated by $h_0$ according 
to
\beq \label{eqn:innerequiv}
e^{ish_1} = \msf{u}(s) e^{ish_0} \msf{u}'(s)
\eeq
where $\msf{u}(s)\in\alg$ and $\msf{u}'(s)\in\alg'$ for 
all $s\in\mathbb{R}$.  Then $h_1$ is a modular Hamiltonian
of some state.  The fact that the exponential on the left
hand side of (\ref{eqn:innerequiv}) involves an 
$s$-independent generator $h_1$ implies that $\msf{u}(s)$,
$\msf{u}'(s)$ satisfy certain cocycle conditions, which we 
can derive by evaluating $e^{i(s+t)h_1}$:
\begin{align}
\msf{u}(s+t)e^{i(s+t)h_0} \msf{u}'(s+t) &=
e^{is h_1} e^{ith_1} 
\nonumber 
\\
&=
\msf{u}(s)e^{ish_0} \msf{u}'(s) \msf{u}(t)e^{ith_0}\msf{u}'(t)
\nonumber 
\\
&=
\big[ \msf{u}(s) e^{ish_0}\msf{u}(t)e^{-is h_0}\big]\cdot
e^{i(s+t)h_0} \cdot\big[e^{-it h_0} \msf{u}'(s)e^{ith_0} \msf{u}'(t)\big].
\end{align}
where we used $[\msf{u}'(s),\msf{u}(t)] = 0$.
Hence, $\msf{u}(s)$ and $\msf{u}'(s)$ satisfy the cocycle 
conditions
\begin{align}
\msf{u}(s+t) &= \msf{u}(s) e^{ish_0}\msf{u}(t)e^{-ish_0} 
\label{eqn:ucoc}\\
\msf{u}'(s+t) &= e^{-ith_0}\msf{u}'(s)e^{ith_0}\msf{u}'(t),
\label{eqn:u'coc}
\end{align}
which are the necessary conditions in order to apply
the converse of the cocycle derivative theorem
as given in \cite[Theorem 3.8]{TakesakiII}.

This cocycle identity and the relation (\ref{eqn:innerequiv})
imply the existence of relative Hamiltonians $h_{1|0}$ and $h_{0|1}$
by the relations
\begin{align}
e^{ish_{1|0}} &= \msf{u}(s)e^{ish_0} = e^{ish_1}(\msf{u}'(s))^\dagger 
\\
e^{ish_{0|1}} &= \msf{u}(s)^\dagger e^{ish_1} = e^{ish_0}\msf{u}'(s).
\end{align}
To see that these equations define $s$-independent
operators $h_{0|1}$ and $h_{1|0}$, we can compute the 
Baker-Campbell-Hausdorff expansion of the two expressions
for the relative Hamiltonian.  Writing $\msf{u}(s) = e^{is W(s)}$,
$\msf{u}'(s) = e^{isW'(s)}$ with $W(s)\in\alg$ and $W'(s)\in\alg'$
both Hermitian, this determines two expansions 
for $h_{1|0}$,
\begin{align}
is h_{1|0} = \log(e^{isW(s)} e^{is h_0}) &= \log(e^{ish_1} e^{-isW'(s)})
\nonumber
\\
is(W(s)+h_0) + \frac{(is)^2}{2} [W(s),h_0] +\ldots
&= is(h_1-W'(s)) - \frac{(is)^2}{2}[h_1,W'(s)] +\ldots
\end{align}
On the left hand side, all terms are elements of $\alg$ except 
for $is h_0$, since $h_0$ generates an automorphism of $\alg$
so that commutators $[h_0,W(s)]$, $[h_0,[h_0,W(s)]]$ etc.\
are all elements of $\alg$.  Similarly, on the right
hand side, the only term that is not an element 
of $\alg'$ is $ish_1$.  Assuming that $\alg$ has no center,
this implies that all terms beyond the linear term in $s$ must
cancel on each side of the above equation, meaning that 
\beq
h_{1|0} = W(0)+ h_0 = h_1-W'(0),
\eeq
which is $s$-independent.  A similar argument holds for 
$h_{0|1}$.  

This argument also reveals that the Hamiltonians are related by 
the equation
\beq
h_1 = W(0) + h_0 + W'(0),
\eeq
which was the condition quoted in section \ref{sec:modham}
for $h_1$ to be a modular Hamiltonian.  Hence, for a general
Hamiltonian of the form
\beq
h_{\msf{a}\msf{b}'} = \msf{a} + h_0 +\msf{b}'
\eeq
with $\msf{a}$, $\msf{b}'$ Hermitian, 
one can construct the relative Hamiltonian
\beq
h_{\msf{a}\msf{b'}|0} = \msf{a} + h_0,
\eeq
and from it construct the cocycle
\beq
\msf{u}(s) = e^{is h_{\msf{a}\msf{b}'|0}}e^{-is h_0}.
\eeq
With this definition, one immediately verifies that 
$\msf{u}(s)$ satisfies the cocycle identity
(\ref{eqn:ucoc}).  The other relative Hamiltonian 
can be defined by 
\beq
h_{0|\msf{a}\msf{b}'} = h_0 + \msf{b}',
\eeq
and the cocycle
\beq
\msf{u}'(s) = e^{-ish_0} e^{is h_{0|\msf{a}\msf{b}'}}
\eeq
will satisfy (\ref{eqn:u'coc}).  One can then apply the converse 
of the cocycle derivative theorem to conclude that $h_{\msf{a}\msf{b}'}$
is a modular Hamiltonian of some state, as claimed in 
section \ref{sec:modham}.

Finally, we comment on the technical requirements for the application
of this theorem.  Theorem 3.8 of \cite{TakesakiII} 
applies to a flow on a von Neumann algebra $\alg$ that is 
related to a modular flow by a cocycle satisfying 
the condition (\ref{eqn:ucoc}).  This modular flow is 
with respect to a faithful, semi-finite, normal weight $\varphi$
on the algebra $\alg$.   Being a weight, as opposed to a state,
means that $\varphi$ may assign infinite expectation value to 
some operators in $\alg$.  The semi-finite requirement
means that sufficiently many operators in $\alg$ 
have finite expectation values, in the sense that these
operators generate the full algebra $\alg$.  Faithful 
refers to the fact that no nonzero positive operator in $\alg$ is 
assigned zero expectation value.  Finally, the most important
requirement is that $\varphi$ is normal, which is a continuity 
requirement on the expectation values that $\varphi$ assigns
to algebra elements. 
For bounded subregions, one should think of normality as a 
condition that the entanglement structure of quantum fields 
near the entangling surface agrees with the local vacuum.

\section{Computation of modular operators} \label{app:modcomp}

An important technical result in the present
work is the exact set of expressions (\ref{eqn:rhoPhioutline}), 
(\ref{eqn:rhoPhi'outline})
for the 
 modular operator $\Delta_{\wh{\Phi}}$ of a
 classical-quantum state $|\wh{\Phi}\rangle$ 
for the crossed product algebra $\ac$
constructed in section 
\ref{sec:crossprod}.  In this appendix,
we derive these expressions for the modular operator,
as well as expressions for a related class of 
twirled classical-quantum states of the form
$|\tilde{\Phi}\rangle = e^{i\hat{p}\hb}|\Phi,f\rangle$.  
These expressions will be compared to those
obtained in \cite{Chandrasekaran2022a, Chandrasekaran2022b}
under a semiclassical approximation on the observer
wavefunction $f$ for the classical-quantum state.
We will see explicitly that the modular operator factorizes into a piece affiliated with $\ac$ and a piece affiliated with $(\ac)',$ in agreement with classic theorems showing that this must always be the case in a type II factor (see e.g. \cite[chapter V.2.4]{Haag1992}).
We will interpret the piece of the modular operator affiliated with $\ac$ as the density matrix for the state $|\wh{\Phi}\rangle$ in the algebra $\ac$, and show that this agrees with a natural definition of density matrices related to the algebraic traces defined in appendix \ref{app:algebraic-background}.

We begin with some notation and conventions. 
We will work with both a position $|y\rangle$
and momentum $|s\rangle$ basis
for wavefunctions in $\hs_\text{obs} = L^2(\mathbb{R})$.
These satisfy
\begin{align}
\begin{split}
\hat{q}|y\rangle &= y|y\rangle, \quad 
\langle y'|y\rangle = \delta(y'-y), \quad 
e^{ia\hat{p}}|y\rangle = |y-a\rangle
\\
\hat{p}|s\rangle &=s|s\rangle, \quad
\langle s'|s\rangle = \delta(s'-s), \quad
e^{ib\hat{q}}|s\rangle = |s+b\rangle \\
\langle y |s\rangle &= \frac{e^{isy}}{\sqrt{2\pi}}
\end{split}
\end{align}
A given state $|f\rangle$ in $\hs_\text{obs}$ 
is expressed in 
these bases as
\beq
|f\rangle = \int dy f(y)|y\rangle = \int ds \tilde{f}(s)|s\rangle
\eeq
so that $f(y)$ and $\tilde{f}(s)$ are Fourier
transforms of each other, with the convention
\begin{align}
\begin{split}
\tilde f(s) &= \frac{1}{\sqrt{2\pi}}\int_{-\infty}^\infty
dy f(y)e^{-isy} \\
f(y)&= \frac{1}{\sqrt{2\pi}}\int_{-\infty}^\infty
ds \tilde{f}(s)e^{isy}.
\end{split}
\end{align}
We also keep in mind the Fourier representation of the 
delta function,
\beq
2\pi \delta(s) = \int_{-\infty}^\infty dy e^{-isy}.
\eeq

Given any operator $\msf{a}\in\aqft$,
its modular flow with respect to the state $|\Psi\rangle$
will be denoted
\beq
\msf{a}_s \equiv e^{ish_\Psi} \msf{a}e^{-ish_\Psi} 
=\Delta_\Psi^{-is} \msf{a} \Delta_\Psi^{is}.
\eeq
To determine the modular operator $\Delta_{\wh{\Phi}}$
of the state $|\wh\Phi\rangle = |\Phi,f\rangle$
for the crossed product algebra $\ac$,
we will explicitly solve the relation
\beq \label{eqn:PhiabPhi}
\langle \wh{\Phi}|\,\ho a\,\ho b\,|\wh{\Phi}\rangle
=
\langle \wh{\Phi}|\,\ho b\,\Delta_{\wh\Phi}\,\ho a\,|
\wh\Phi\rangle
\eeq
for generic elements $\ho a,\ho b\in\ac$.  We take
these operators to be 
\begin{align}
\begin{split}
\ho{a} &= 
\msf{a}_{\frac{\hat{p}}{\beta}} e^{iu\hat{q}}
=
e^{i\hat{p}\hb} \msf{a} e^{-i\hat{p}\hb}
e^{iu\hat{q}} \\
\ho{b} &=
\msf{b}_{\frac{\hat{p}}{\beta}} e^{iv\hat{q}}
=
e^{i\hat{p}\hb} \msf{b} e^{-i\hat{p}\hb}
e^{iv\hat{q}},
\end{split}
\end{align}
which additively span the algebra $\ac$.  We will assume that $|\Phi\rangle$
lies in the canonical cone of $|\Psi\rangle$ (see appendix
\ref{app:modth}) throughout the derivation, since the modular operator
for more general states is easily obtained by conjugating with $\msf{u}'
= J_{\Phi|\Psi} J_\Psi$.  

We start by evaluating the left hand side of (\ref{eqn:PhiabPhi})
using the Fourier representation of the wavefunction:
\begin{align}
\langle \wh\Phi|\,\ho a\,\ho b\,|\wh\Phi\rangle
&=
\langle \Phi,f|\,\ho a\,\ho b\, |\Phi,f\rangle
=
\int ds' ds \tilde{f}(s')^*\tilde{f}(s)
\langle\Phi|\langle s'| \msf{a}_{\frac{\hat p}{\beta}}
e^{iu\hat q}\msf{b}_{\frac{\hat p}{\beta}} e^{iv\hat q} |s\rangle|\Phi\rangle
\nonumber 
\\
&=
\int ds' ds \tilde{f}(s')^*\tilde{f}(s)
\langle \Phi|\msf{a}_{\frac{s'}{\beta}}
\langle s'-u|s+v\rangle \msf{b}_{\frac{s+v}{\beta}}|\Phi\rangle
\nonumber
\\
&=
\int ds \tilde{f}(s+u+v)^*\tilde f(s)
\langle \Phi|\msf{a}_{\frac{s+u+v}{\beta}}\msf{b}_{\frac{s+v}{\beta}} |\Phi\rangle
\nonumber 
\\
&=
\int ds \tilde{f}(s+u)^*\tilde f(s-v)\langle\Phi |\msf{a}_{\frac{s+u}{\beta}}
\msf{b}_{\frac{s}{\beta}}|\Phi\rangle
\nonumber 
\\
&=
\int ds \tilde f(s+u)^*\tilde f(s-v)\langle\Psi|\msf{b}_{\frac{s}{\beta}}
\Delta_{\Phi|\Psi} \msf{a}_{\frac{s+u}{\beta}} |\Psi\rangle
\nonumber
\\
&=
\int ds\frac{dy'dy}{2\pi} f^*(y') f(y) e^{iy'(s+u)} e^{-iy(s-v)}
\langle \Psi | \msf{b} e^{-is\hb}\Delta_{\Phi|\Psi} e^{is\hb}
\msf{a}_{\frac{u}{\beta}}|\Psi\rangle
\nonumber
\\
&=
\int ds dy' dy f^*(y') f(y) \langle y' | u\rangle\langle -v|y\rangle
e^{iy's} e^{-iys} 
\langle \Psi | \msf{b} e^{-is\hb}\Delta_{\Phi|\Psi} e^{is\hb}
\msf{a}_{\frac{u}{\beta}}|\Psi\rangle
\nonumber
\\
&=
\langle\Psi,-v|\msf{b} \int ds e^{-is(\hat{q}+\hb)}
|f\rangle \Delta_{\Phi|\Psi}\langle f| e^{is(\hat{q}+\hb)}
\msf{a}_{\frac{u}{\beta}}|\Psi,u\rangle
\nonumber 
\\
&=
\langle \Psi,-v|\msf{b}e^{i\hat{p}\hb}
\int ds e^{-is\hat{q}} e^{-i\hat{p}\hb}
|f\rangle \Delta_{\Phi|\Psi}\langle f|
e^{i\hat{p}\hb} e^{is\hat{q}} e^{-i\hat{p}\hb} \msf{a}_{\frac{u}{\beta}}
|\Psi,u\rangle
\label{eqn:PabP1}
\end{align}
In the fifth line, we have applied the identity (\ref{eqn:relmodflip})
to flip the order of $\msf{a}_{\frac{s+u}{\beta}}$ 
and $\msf{b}_{\frac{s}{\beta}}$. 
To carry out the $s$-integral in this expression,
it helps to introduce an eigenbasis $|\omega\rangle$
for $h_\Psi$ satisfying $h_\Psi|\omega\rangle = \omega|\omega\rangle$.
Computing the matrix elements of the $s$-integral operator in this 
basis yields
\begin{align}
&\langle\omega'|\int dse^{-is\hat{q}} e^{-i\hat{p}\hb}
|f\rangle \Delta_{\Phi|\Psi}\langle f|
e^{i\hat{p}\hb} e^{is\hat{q}}
|\omega\rangle 
\nonumber
\\
=&\;
\langle\omega'|
\int ds dy' dy e^{-is\hat{q}} f\left(\hat{q}-\hb\right)
e^{-i\hat{p}\hb}|y'\rangle\Delta_{\Phi|\Psi}\langle y|
e^{i\hat{p}\hb}
f^*\left(\hat{q}-\hb\right) e^{is\hat{q}} |\omega\rangle
\nonumber 
\\
=&\;
\langle\omega'| f\left(\hat{q}-\hb\right)\int ds dy' dy 
e^{-is\hat{q}}|y'+\frac{\omega'}{\beta}\rangle
\Delta_{\Phi|\Psi}\langle y+\frac{\omega}{\beta}|e^{is\hat{q}} 
f^*\left(\hat{q}-\hb\right)|\omega\rangle
\nonumber
\\
=&\;
\langle\omega'|2\pi f\left(\hat{q}-\hb\right)
\int dy' dy \delta(y'+\frac{\omega'}{\beta} - y - \frac{\omega}{\beta})
|y'+\frac{\omega'}{\beta}\rangle
\Delta_{\Phi|\Psi}\langle y+\frac{\omega}{\beta}|
f^*\left(\hat{q}-\hb\right)|\omega\rangle
\nonumber 
\\
=&\;
\langle \omega'|2\pi f\left(\hat{q}-\hb\right) \int dy |y\rangle\Delta_{\Phi|\Psi}\langle y|
f^*\left(\hat{q}-\hb\right)|\omega\rangle
\nonumber
\\
=&\;
\langle\omega'| 2\pi f\left(\hat{q}-\hb\right)\Delta_{\Phi|\Psi} f^*\left(\hat{q}-\hb\right)|\omega\rangle
\end{align}
Plugging this into (\ref{eqn:PabP1}) then yields
\begin{align}
\langle \wh\Phi|\,\ho a\,\ho b\,|\wh\Phi\rangle
&=
2\pi \langle\Psi,-v|\msf{b}e^{i\hat{p}\hb}f\left(\hat{q}-\hb\right)\Delta_{\Phi|\Psi}
f^*\left(\hat{q}-\hb\right)e^{-i\hat{p}\hb}\msf{a}_{\frac{u}{\beta}}|\Psi,u\rangle
\label{eqn:Phiabfinal}
\end{align}

To evaluate the right hand side of (\ref{eqn:PhiabPhi}),
we make use of the assumption that $|\Phi\rangle$ lies in the 
canonical cone of $|\Psi\rangle$, so that $|\Phi\rangle =
\Delta_{\Phi|\Psi}^{\frac12}|\Psi\rangle$
(see appendix \ref{app:modth}).  
Then we compute
\begin{align}
\langle\wh\Phi|\,\ho b\, \Delta_{\wh\Phi}\,\ho a \,|\wh\Phi\rangle
&=
\langle\Phi,f|\msf{b}_{\frac{\hat{p}}{\beta}}
e^{iv\hat{q}}\Delta_{\wh\Phi}\msf{a}_{\frac{\hat{p}}{\beta}} e^{iu\hat{q}}
|\Phi,f\rangle
\nonumber 
\\
&=
\int ds'ds\tilde{f}(s')^*\tilde f(s) \langle\Phi|\msf{b}_{\frac{s'}{\beta}}
\langle s'-v|\Delta_{\wh\Phi}|s+u\rangle
\msf{a}_{\frac{s+u}{\beta}}|\Phi\rangle
\nonumber 
\\
&=
\int ds' ds\tilde f(s')^*\tilde f(s)\langle\Phi,-v|
\msf{b}_{\frac{s'}{\beta}} e^{-is'\hat{q}}\Delta_{\wh\Phi}e^{is\hat{q}}
\msf{a}_{\frac{s+u}{\beta}}|\Phi,u\rangle
\nonumber
\\
&=
\int ds' ds\tilde f(s')^*\tilde f(s)\langle\Psi,-v|
\Delta_{\Phi|\Psi}^{\frac12}\Delta_\Phi^{-\frac12}
\msf{b}_{\frac{s'}{\beta}}
e^{-is'\hat{q}}\Delta_{\wh\Phi}e^{is\hat{q}}\msf{a}_{\frac{s+u}{\beta}}
\Delta_\Phi^{-\frac12}\Delta_{\Phi|\Psi}^{\frac12}|\Psi,u\rangle
\nonumber
\\
&=
\int ds'ds \tilde f(s')^*\tilde f(s)\langle\Psi,-v|
\msf{b} e^{-is'(\hat{q}+\hb)}\Delta_\Psi^{\frac12}\Delta_{\Psi|\Phi}^{-\frac12}
\Delta_{\wh{\Phi}}
\Delta_{\Psi|\Phi}^{-\frac12}\Delta_{\Psi}^{\frac12}
e^{is(\hat{q}+\hb)} \msf{a}_{\frac{u}{\beta}}
|\Psi,u\rangle
\nonumber 
\\
&=
2\pi\langle\Psi,-v|\msf{b} f^*\left(\hat{q}+\hb\right)\Delta_{\Psi}^{\frac12}
\Delta_{\Psi|\Phi}^{-\frac12} \Delta_{\wh\Phi}\Delta_{\Psi|\Phi}^{-\frac12}
\Delta_\Psi^{\frac12} f\left(\hat{q}+\hb\right) \msf{a}_{\frac{u}{\beta}}
|\Psi,u\rangle
\label{eqn:bDela}
\end{align}
In the fifth line, we have used equation \eqref{eqn:affiliated-operators-cocycles}, which gives 
$\Delta_{\Phi|\Psi}^{\frac12}\Delta_{\Phi}^{-\frac12}
=\Delta_{\Psi}^\frac12\Delta_{\Psi|\Phi}^{-\frac12}$ and tells us that 
this operator is in $\aqft'$, so it can be commuted past $\msf{b}$.
We also apply similar manipulations to the operator 
$\Delta_{\Phi}^{-\frac12}\Delta_{\Phi|\Psi}^{\frac12}$.

Equating (\ref{eqn:bDela}) 
with (\ref{eqn:Phiabfinal}) gives an equation for $\Delta_{\wh\Phi}$
that is straightforwardly solved,
\begin{align} \label{eqn:modular-operator-appendix-eqn}
\Delta_{\wh{\Phi}} 
&=
\Delta_{\Psi|\Phi}^{\frac12}\frac{1}{f^*\left(\hat{q}+\hb\right)}
\Delta_\Psi^{-\frac12}
e^{i\hat{p}\hb} f\left(\hat{q}-\hb\right)\Delta_{\Phi|\Psi}f^*\left(\hat{q}-\hb\right)
e^{-i\hat{p}\hb}\Delta_\Psi^{-\frac12}\frac{1}{f\left(\hat{q}+\hb\right)}
\Delta_{\Psi|\Phi}^{\frac12}
\nonumber 
\\
&=
\Delta_{\Psi|\Phi}^{\frac12}\frac{e^{-\frac{\beta\hat{q}}{2}}}{f^*\left(\hat{q}+\hb\right)}
\cdot
\left[
e^{i\hat{p}\hb} f\left(\hat{q}-\hb\right)e^{\beta\hat{q}/2} \Delta_{\Phi|\Psi} e^{\beta \hat{q}/2}
f^*\left(\hat{q}-\hb\right)e^{-i\hat{p}\hb}
\right]
\cdot
\frac{e^{-\frac{\beta\hat{q}}{2}}}{f\left(\hat{q}+\hb\right)}
\Delta_{\Psi|\Phi}^{\frac12}.
\end{align}
One can verify using equation \eqref{eqn:affiliated-operators-cocycles} that the terms outside of the brackets are elements 
of $(\ac)'$, while the quantity inside
the bracket is in $\ac$,
as discussed in section \ref{sec:moddensity}.
We therefore find that the modular operator factorizes 
into density matrices
$\Delta_{\wh{\Phi}} = \rho_{\wh{\Phi}}(\rho_{\wh{\Phi}}')^{-1}$. 
Finally, to lift 
the requirement that $|\Phi\rangle$ is in the canonical
cone of $|\Psi\rangle$, we simply conjugate 
$\Delta_{\wh{\Phi}}$ by $\msf{u'} = J_{\Phi|\Psi} J_\Psi$,
as explained in section \ref{sec:moddensity}.
Since $\Delta_{\Phi|\Psi}$ is the same as in the canonically
purified modular operator, but $\Delta_{\Psi|\Phi}$
is related to the canonically purified version
according to equation (\ref{eqn:DelPsiPhicanon}), 
the 
density matrices are then given by
\begin{align} \label{eqn:appendix-E-DM}
\rho_{\wh{\Phi}} 
&= 
\frac{1}{\beta}
e^{i\hat{p}\hb} f\left(\hat{q}-\hb\right)e^{\beta\hat{q}/2} \Delta_{\Phi|\Psi} e^{\beta \hat{q}/2}
f^*\left(\hat{q}-\hb\right)e^{-i\hat{p}\hb}
\\
\rho_{\wh{\Phi}}' 
&=\frac{1}{\beta}
\Delta_{\Psi|\Phi}^{-\frac12}
J_{\Phi|\Psi}J_\Psi
e^{\frac{\beta\hat{q}}{2}}
\left|f\left(\hat{q}+\hb\right)\right|^2
e^{\frac{\beta\hat{q}}{2}}
J_\Psi J_{\Psi|\Phi}
\Delta_{\Psi|\Phi}^{-\frac12}
\end{align}

These expressions can be compared to the density matrices 
obtained in \cite{Chandrasekaran2022a, Chandrasekaran2022b},
computed for a similar class of states.  To make the comparison,
we first note that these works employed a unitarily equivalent algebra
obtained from $\ac$ by conjugating with respect to 
$e^{-i\hat{p}\hb}$.  Denoting this algebra by
$\alg_\text{cr} = e^{-i\hat{p}\hb}\ac e^{i\hat{p}\hb}$, 
we see that it is given by
\beq
    \alg_\text{cr} = \big\{ \msf{a}, e^{i \left(\hat{q}-\hb\right) t}|\msf{a}\in\aqft, t \in \mathbb{R} \big\}''.
\eeq
Under this transformation, the classical-quantum
state $|\wh{\Phi}\rangle$ 
maps to $|\Phi_\text{cr}\rangle = e^{-i\hat{p}\hb}
|\Phi,f\rangle$, which we refer to as a twirled  state.
These states are somewhat more natural for the algebra $\alg_\text{cr}$
than ones that do not involve twirling since they are states on which it is easy to implement the positive energy projection: since 
this projection becomes $\Theta(\hat{q}+\hb)$ after the conjugation,
the projected states are just twirled states with $f(q<0)=0$.
The modular operator $\Delta_{\Phi_\text{cr}}$ of this state 
on the algebra $\alg_\text{cr}$ can be obtained 
immediately from $\Delta_{\wh{\Phi}}$ by conjugation.  
This produces the density matrices for the twirled state
\begin{align}
\rho_{\Phi_\text{cr}} &= \frac1\beta f\left(\hat{q}-\hb\right) e^{\beta\hat{q}/2}
\Delta_{\Phi|\Psi} e^{\beta \hat{q}/2} f^*\left(\hat{q}-\hb\right)
\label{eqn:rhotw}
\\
\rho_{\Phi_\text{cr}}' &=\frac1\beta e^{-i\hat{p}\hb}
\Delta_{\Psi|\Phi}^{-\frac12} 
J_{\Phi|\Psi} J_\Psi
e^{\frac{\beta\hat{q}}{2}} \left|f\left(\hat{q}+\hb\right)\right|^2 
e^{\frac{\beta\hat{q}}{2}}
J_{\Psi}J_{\Psi|\Phi}
\Delta_{\Psi|\Phi}^{-\frac12}
e^{i\hat{p}\hb}.
\label{eqn:rho'tw}
\end{align}
Interestingly, the expression for $\rho_{\Phi_\text{cr}}$ agrees with the 
form of the density matrix derived in \cite{Chandrasekaran2022a, 
Chandrasekaran2022b} (up to the order of $f$ and $f^*$) after 
accounting for the definition $\hat{x} = -\hat{q}$ and making the
appropriate changes to the wavefunction $f$, which is a function of 
$x$ instead of $q$ in \cite{Chandrasekaran2022a, 
Chandrasekaran2022b}.  This is somewhat surprising since 
their density matrix was computed for the {\it untwirled} state
$|\Phi,f\rangle$ on $\alg_\text{cr}$.  
However, they also employed semiclassical assumptions on the wavefunction
$f$ when deriving the density matrix, which 
appears to make the state largely insensitive to the twirling 
operation.

The exact density matrix for the untwirled state $|\Phi_\text{cq}\rangle=
|\Phi,f\rangle$
on the algebra $\alg_\text{cr}$ can be derived following a similar 
sequence of steps as employed above.  The steps are almost identical
if one instead computes the modular operator for the commutant 
algebra $\alg_\text{cr}'$,
\beq
\alg_\text{cr}' = \{ e^{-i\hat{p}\hb} \msf{a}' e^{i\hat{p}\hb},e^{i\hat{q} t}
|\msf{a}'\in\aqft', t \in \mathbb{R}\}'',
\eeq
and then uses the relation $\Delta_{\Phi_\text{cq}}' =
\Delta_{\Phi_\text{cq}}^{-1}$.  In 
this case, one should express a generic 
state $|\Phi\rangle$ in terms of a canonical purification
$|\Phi_c\rangle$ 
according to $|\Phi\rangle = \msf{u} |\Phi_c\rangle$, 
with  $\msf{u} = J_{\Psi|\Phi}J_\Psi \in
\aqft$.  This results in the following
density matrices for this state:
\begin{align}
\rho_{\Phi_\text{cq}} 
&= 
\frac{1}{\beta}
\Delta_{\Phi|\Psi}^{\frac12}
J_{\Psi|\Phi} J_\Psi
e^{\beta\hat{q}/2}
\left|f\left(\hat{q}-\hb\right)\right|^2 e^{\beta\hat{q}/2}
J_\Psi J_{\Phi|\Psi}
\Delta_{\Phi|\Psi}^{\frac12}
\label{eqn:rhocq} 
\\
\rho_{\Phi_\text{cq}}' 
&=
\frac{1}{\beta} e^{-i\hat{p}\hb}f\left(\hat{q}+\hb\right)e^{\beta\hat{q}/2}
\Delta_{\Psi|\Phi}^{-1} e^{\beta\hat{q}/2}
f^*\left(\hat{q}+\hb\right)e^{i\hat{p}\hb}.
\label{eqn:rhocq'}
\end{align}
The expression for $\rho_{\Phi_\text{cq}}$ agrees with that given 
in \cite{Chandrasekaran2022a, Chandrasekaran2022b}
upon application of the semiclassical approximation, which implies
$\left[\Delta_{\Phi|\Psi}^{\frac12}, f\left(\hat{q}-\hb\right)\right]\approx 0 \approx
\left[J_{\Psi|\Phi}J_\Psi,f\left(\hat{q}-\hb\right)\right]$.

It is also worth comparing the expressions for the twirled state
density matrices (\ref{eqn:rhotw}), (\ref{eqn:rho'tw})
to those of the untwirled states (\ref{eqn:rhocq}), (\ref{eqn:rhocq'}).  
One notices that the expression 
of the density matrix for $\alg_\text{cr}$ in the twirled state
is similar in structure to the density matrix for $\alg_\text{cr}'$
in the untwirled state, and vice-versa.  
This just reflects the fact that classical-quantum states for $\alg_\text{cr}$
behave more like twirled classical-quantum states for $\alg_\text{cr}'$.
Of course, neither state is truly classical-quantum, since there is always
entanglement between the observer and the quantum field degrees of 
freedom within the crossed product algebras.
Finally, note that the untwirled state
$|\Phi_\text{cq}\rangle$ for the algebra $\alg_\text{cr}$
naturally maps to a twirled classical quantum
state $|\tilde{\Phi}\rangle = e^{i\hat{p}\hb}
|\Phi,f\rangle$ for the original crossed 
product algebra $\ac$.  The density matrices
for this state are obtained by conjugating 
(\ref{eqn:rhocq}) and (\ref{eqn:rhocq'})
by $e^{i\hat{p}\hb}$, resulting in
\begin{align}
\rho_{\tilde\Phi} 
&=
\frac{1}{\beta}e^{i\hat{p}\hb}
\Delta_{\Phi|\Psi}^{\frac12}
J_{\Psi|\Phi} J_\Psi
e^{\beta\hat{q}/2}
\left|f\left(\hat{q}-\hb\right)\right|^2 e^{\beta\hat{q}/2}
J_{\Psi}J_{\Phi|\Psi}
\Delta_{\Phi|\Psi}^{\frac12}
 e^{-i\hat{p}\hb}
\\
\rho_{\tilde\Phi}'
&=
\frac{1}{\beta} f\left(\hat{q}+\hb\right)e^{\beta\hat{q}/2}
\Delta_{\Psi|\Phi}^{-1} e^{\beta\hat{q}/2}
f^*\left(\hat{q}+\hb\right).
\end{align}

To finish off, it is interesting to see how the characterization of the density matrix $\rho_{\wh{\Phi}}$ from equation \eqref{eqn:appendix-E-DM}, as the piece of the modular operator $\Delta_{\wh{\Phi}}$ affiliated with $\ac$, compares with the usual definition of a density matrix in quantum mechanics.
In ordinary quantum mechanics, given a tensor-factorized Hilbert space $\hs = \hs_{\alg} \otimes \hs_{\alg'}$, the density matrix $\rho_{\psi}$ for a state $\ket{\psi} \in \hs$ is defined as the unique positive operator in $\mathcal{B}(\hs_{\alg})$ satisfying
\begin{equation}
    \bra{\psi} \msf{a} \otimes 1 \ket{\psi}
        = \tr_{\hs_{\alg}}(\rho_{\psi} \msf{a})
\end{equation}
for all $\msf{a} \in \mathcal{B}(\hs_{\alg}).$
As explained in appendix \ref{app:algebraic-background}, while the Hilbert space trace is always infinite on a type II von Neumann factor, there is a renormalized notion of the trace that is uniquely defined up to rescaling.
On the crossed product, we denote this trace by $\atr,$ and we showed in appendix \ref{app:algebraic-background} that it is given by the formula
\begin{equation}
    \atr(\wh{\msf{a}})
        = 2 \pi \beta (\bra{\Psi} \otimes \bra{0}_p) e^{-\beta \hat{q}} \wh{\msf{a}} (\ket{\Psi} \otimes \ket{0}_p),
\end{equation}
where $\ket{0}_p$ is the unnormalized zero-momentum state and $\wh{\msf{a}}$ is an operator in $\ac.$
It is then natural to define the density matrix $\rho_{\wh{\Phi}}$ for the state $|\wh{\Phi}\rangle,$ should one exist, as a positive operator affiliated with $\ac$ satisfying
\begin{equation} \label{eq:density-matrix-def}
    \langle \wh{\Phi} | \wh{\msf{a}} | \wh{\Phi} \rangle
        = \atr(\rho_{\wh{\Phi}} \wh{\msf{a}})
\end{equation}
for any $\wh{\msf{a}}$ in $\ac.$
The properties of the renormalized trace discussed in appendix \ref{app:algebraic-background} --- in particular, its ``faithfulness'' --- imply that if this operator exists, then it is unique.
In fact, there is a theorem showing that this operator always exists (see e.g. \cite[theorem 5.3.11]{pedersen1979c}), but since we already have an expression for $\rho_{\wh{\Phi}}$ computed via the modular operator $\Delta_{\wh{\Phi}},$ it suffices to plug that expression into equation \eqref{eq:density-matrix-def} and verify that the identity is satisfied.

Since every operator in $\ac$ is a limit of finite linear combinations of operators of the form $e^{i \hat{p} \hb} \msf{a} e^{- i \hat{p} \hb} g(\hat{q}),$ it suffices to check the identity \eqref{eq:density-matrix-def} for operators $\wh{\msf{a}}$ of this form. Using the expression for the density matrix $\rho_{\wh{\Phi}}$ coming from equation \eqref{eqn:modular-operator-appendix-eqn}, we have
\begin{align}
    & \atr(\rho_{\wh{\Phi}} e^{i \hat{p} \hb} \msf{a} e^{- i \hat{p} \hb} g(\hat{q})) \nonumber \\
    & \quad = 2 \pi \langle \Psi| \langle 0|_p e^{- \beta \hat{q}} e^{i \hat{p} \hb} f\left(\hat{q} - \hb \right) e^{\beta \hat{q}/2} \Delta_{\Phi|\Psi} e^{\beta \hat{q}/2} f^*\left( \hat{q} - \hb \right) \msf{a} e^{- i \hat{p} \hb} g(\hat{q}) \ket{\Psi} \ket{0}_p.
\end{align}
The leftmost factor of $e^{i \hat{p} \hb}$ can be commuted through $e^{\beta \hat{q}}$ at the expense of translating $\hat{q}$ by $\hb,$ but since we have $h_{\Psi} \ket{\Psi} = 0,$ all terms involving $h_{\Psi}$ appearing at the left side of the expression trivialize.
This introduces the simplification
\begin{align}
    & \atr(\rho_{\wh{\Phi}} e^{i \hat{p} \hb} \msf{a} e^{- i \hat{p} \hb} g(\hat{q})) \nonumber \\
    & \quad = 2 \pi \langle \Psi| \langle 0|_p e^{- \beta \hat{q}} f\left(\hat{q} \right) e^{\beta \hat{q}/2} \Delta_{\Phi|\Psi} e^{\beta \hat{q}/2} f^*\left( \hat{q} - \hb \right) \msf{a} e^{- i \hat{p} \hb} g(\hat{q}) \ket{\Psi} \ket{0}_p,
\end{align}
and we may now commute both $e^{\beta \hat{q}/2}$ terms to the left side of the expression to obtain the further simplification
\begin{align}
    \atr(\rho_{\wh{\Phi}} e^{i \hat{p} \hb} \msf{a} e^{- i \hat{p} \hb} g(\hat{q}))
    & = 2 \pi \langle \Psi| \langle 0|_p f\left(\hat{q} \right) \Delta_{\Phi|\Psi} f^*\left( \hat{q} - \hb \right) \msf{a} e^{- i \hat{p} \hb} g(\hat{q}) \ket{\Psi} \ket{0}_p.
\end{align}
We will now employ the helpful identity that if $\ket{s}$ is a momentum eigenstate, then we have
\begin{equation} \label{eqn:f-momentum-identity}
    f(\hat{q}) \ket{s}
        = \int dq\, \frac{e^{i s q}}{\sqrt{2 \pi}} f(q) \ket{q}
        = \frac{e^{i s \hat{q}}}{\sqrt{2 \pi}} \ket{f}.
\end{equation}
Using this identity, and also pulling factors of $e^{\pm i \hat{p} \hb}$ out of $f^*\left(\hat{q} - \hb \right),$ we may write the trace expression as
\begin{align}
    \atr(\rho_{\wh{\Phi}} e^{i \hat{p} \hb} \msf{a} e^{- i \hat{p} \hb} g(\hat{q}))
    & = \langle \Psi| \langle f^*| \Delta_{\Phi|\Psi} e^{-i \hat{p} \hb} f^*\left( \hat{q}\right) e^{i \hat{p} \hb} \msf{a} e^{- i \hat{p} \hb} \ket{\Psi} \ket{g}.
\end{align}
We now insert two resolutions of the identity in the momentum basis, $1 = \int ds\, \ketbra{s} = \int ds'\, \ketbra{s'},$ to obtain the expression
\begin{align}
    & \atr(\rho_{\wh{\Phi}} e^{i \hat{p} \hb} \msf{a} e^{- i \hat{p} \hb} g(\hat{q})) \nonumber \\
    & \quad = \int ds\, ds'\, \langle f^*| s \rangle  \braket{s'}{g} \bra{s}f^*\left( \hat{q}\right) \ket{s'} \langle \Psi| \Delta_{\Phi|\Psi} e^{i (s'-s) \hb} \msf{a} e^{- i s' \hb} \ket{\Psi}.
\end{align}
Since we have $e^{i s \hb} \ket{\Psi} = \ket{\Psi},$ we are free to insert this operator at the right end of the expression, and obtain
\begin{align}
    & \atr(\rho_{\wh{\Phi}} e^{i \hat{p} \hb} \msf{a} e^{- i \hat{p} \hb} g(\hat{q})) \nonumber \\
    & \quad = \int ds\, ds'\, \langle f^*| s \rangle  \braket{s'}{g} \bra{s}f^*\left( \hat{q}\right) \ket{s'} \langle \Psi| \Delta_{\Phi|\Psi} e^{i (s'-s) \hb} \msf{a} e^{- i (s'-s) \hb} \ket{\Psi}.
\end{align}
Because $e^{i (s' - s) \hb} \msf{a} e^{-i (s' - s) \hb}$ is in $\alg,$ we may apply the relative modular operator identity \eqref{eqn:relmodflip} to obtain
\begin{align}
    & \atr(\rho_{\wh{\Phi}} e^{i \hat{p} \hb} \msf{a} e^{- i \hat{p} \hb} g(\hat{q})) \nonumber \\
    & \quad = \int ds\, ds'\, \langle f^*| s \rangle  \braket{s'}{g} \bra{s}f^*\left( \hat{q}\right) \ket{s'} \langle \Phi| e^{i (s'-s) \hb} \msf{a} e^{- i (s'-s) \hb} \ket{\Phi}.
\end{align}
We now make the change of variables $s' \mapsto s' + s$ to obtain
\begin{align}
    & \atr(\rho_{\wh{\Phi}} e^{i \hat{p} \hb} \msf{a} e^{- i \hat{p} \hb} g(\hat{q})) \nonumber \\
    & \quad = \int ds\, ds'\, \langle f^*| s \rangle  \braket{s'+s}{g} \bra{s}f^*\left( \hat{q}\right) \ket{s'+s} \langle \Phi| e^{i s' \hb} \msf{a} e^{- i s' \hb} \ket{\Phi}.
\end{align}
We now apply equation \eqref{eqn:f-momentum-identity} to the term $\bra{s} f^*(\hat{q}) \ket{s'+s}$ to obtain
\begin{align}
    & \atr(\rho_{\wh{\Phi}} e^{i \hat{p} \hb} \msf{a} e^{- i \hat{p} \hb} g(\hat{q})) \nonumber \\
    & \quad = \frac{1}{\sqrt{2 \pi}} \int ds\, ds'\, \langle f^*| s \rangle  \braket{s'+s}{g} \bra{f}e^{- i s \hat{q}} \ket{s'+s} \langle \Phi| e^{i s' \hb} \msf{a} e^{- i s' \hb} \ket{\Phi}.
\end{align}
We may then apply the identity $\ket{s'+s} = e^{i s' \hat{q}} \ket{s}$ to obtain the expression
\begin{align}
    & \atr(\rho_{\wh{\Phi}} e^{i \hat{p} \hb} \msf{a} e^{- i \hat{p} \hb} g(\hat{q})) \nonumber \\
    & \quad = \frac{1}{\sqrt{2 \pi}} \int ds\, ds'\, \langle f^*| s \rangle  \bra{s} e^{- i s' \hat{q}}\ket{g} \braket{f}{s'} \langle \Phi| e^{i s' \hb} \msf{a} e^{- i s' \hb} \ket{\Phi}.
\end{align}
The parameter $s$ now appears only as the resolution of the identity $\int ds\, \ketbra{s} = 1,$ so we may perform that integral to obtain
\begin{align}
    \atr(\rho_{\wh{\Phi}} e^{i \hat{p} \hb} \msf{a} e^{- i \hat{p} \hb} g(\hat{q}))
    & = \frac{1}{\sqrt{2 \pi}} \int ds'\, \langle f^*| e^{- i s' \hat{q}}\ket{g} \braket{f}{s'} \langle \Phi| e^{i s' \hb} \msf{a} e^{- i s' \hb} \ket{\Phi}.
\end{align}
By inserting a complete position basis, it is easy to verify the identity
\begin{equation}
\bra{f^*} e^{- i s' \hat{q}} \ket{g} = \bra{g^*} e^{- i s' \hat{q}} \ket{f},
\end{equation}
and we may then use the identity \eqref{eqn:f-momentum-identity} in the form $\frac{1}{\sqrt{2 \pi}} e^{i s' \hat{q}} \ket{g^*} = g^*(\hat{q}) \ket{s'}$ to obtain
\begin{align}
    \atr(\rho_{\wh{\Phi}} e^{i \hat{p} \hb} \msf{a} e^{- i \hat{p} \hb} g(\hat{q}))
    & = \int ds'\, \bra{s'} g(\hat{q}) \ket{f} \braket{f}{s'} \langle \Phi| e^{i s' \hb} \msf{a} e^{- i s' \hb} \ket{\Phi}.
\end{align}
After some rearranging, we may replace $e^{\pm i s' \hb}$ with $e^{\pm i \hat{p} \hb},$ and integrate over the complete momentum basis $\int ds'\, \ketbra{s'} = 1$ to obtain
\begin{align}
    \atr(\rho_{\wh{\Phi}} e^{i \hat{p} \hb} \msf{a} e^{- i \hat{p} \hb} g(\hat{q}))
    & = \bra{\Phi} \bra{f} e^{i \hat{p} \hb} \msf{a} e^{- i \hat{p} \hb} g(\hat{q}) \ket{\Phi} \ket{f},
\end{align}
as desired.

As a final comment, as emphasized in section \ref{sec:moddensity}, we note that in the expression \eqref{eqn:modular-operator-appendix-eqn} for the modular operator, we could have freely multiplied the affiliated-with-$\ac$ part by a scalar function of $|\wh{\Phi}\rangle$, and multiplied the affiliated-with-$(\ac)'$ part by the reciprocal of that function.
This reflects a state-dependent ambiguity in the normalization of the density matrices $\rho_{\wh{\Phi}}$ and $\rho_{\wh{\Phi}}'.$
In the main text, we resolved this to a state-independent normalization ambiguity by requiring $\atr(\rho_{\wh{\Phi}}) = 1,$ so that the only lingering ambiguity comes from the normalization of the trace.
In terms of the calculation presented above for the identity $\atr(\rho_{\wh{\Phi}} \wh{\msf{a}}) = \langle \wh{\Phi}| \wh{\msf{a}} | \wh{\Phi} \rangle$, the normalization condition can be thought of as the special case where $\wh{\msf{a}}$ is the identity operator.

\bibliography{cdrefs}

\bibliographystyle{JHEPthesis}

\end{document}